\input harvmac
\input epsf.tex
\def\figin{\epsfcheck\figin}\def\figins{\epsfcheck\figins}
\def\epsfcheck{\ifx\epsfbox\UnDeFiSIeD
\message{(NO epsf.tex, FIGURES WILL BE IGNORED)}
\gdef\figin##1{\vskip2in}\gdef\figins##1{\hskip.5in}
\else\message{(FIGURES WILL BE INCLUDED)}%
\gdef\figin##1{##1}\gdef\figins##1{##1}\fi}
\def\DefWarn#1{}
\def\figinsert{\goodbreak\midinsert}
\def\ifig#1#2#3{\DefWarn#1\xdef#1{fig.~\the\figno}
\writedef{#1\leftbracket fig.\noexpand~\the\figno}%
\figinsert\figin{\centerline{#3}}\medskip\centerline{\vbox{\baselineskip12pt
\advance\hsize by -1truein\noindent\footnotefont{\bf Fig.~\the\figno:} #2}}
\bigskip\endinsert\global\advance\figno by1}

\def\cf{{\cal F}}
\def\G{\Gamma}
\def\khalf{ {k \over 2} }
\def\half{ {1 \over 2}}



\lref\teschnerrecent{J.~Teschner,
``Crossing symmetry in the $H_3^+$ WZNW model,
Phys.\ Lett.\ B {\bf 521}, 127 (2001)
[arXiv:hep-th/0108121].
}

\lref\teschnersecond{ J.~Teschner,
``Operator product expansion and factorization in the
$H_3^+$ WZNW model,''
Nucl.\ Phys.\ B 571 (2000) 555;
{\tt hep-th/9906215}.
}

\lref\freedman{ D.~Z.~Freedman, S.~D.~Mathur, A.~Matusis and L.~Rastelli,
``Correlation functions in the CFT$_d$/AdS$_{d+1}$ correspondence,''
Nucl.\ Phys.\ B 546 (1999) 96; {\tt
hep-th/9804058}.
}

\lref\first{
J.~Maldacena and H.~Ooguri,
``Strings in $AdS_3$ and $SL(2,R)$ WZW model. Part 1: the spectrum,''
J.\ Math.\ Phys. 42 (2001) 2929
({\sl Special issue on M Theory}); {\tt hep-th/0001053}.
}

\lref\second{J.~Maldacena, H.~Ooguri and J.~Son,
``Strings in $AdS_3$ and the $SL(2,R)$ WZW model.
Part 2: Euclidean black hole,''
J.\ Math.\ Phys.  42 (2001) 2961
({\sl Special issue on M Theory}); {\tt
hep-th/0005183}.
}

\lref\wittenthermal{E.~Witten,
``Anti-de Sitter space, thermal phase
transition, and confinement in  gauge theories,''
Adv.\ Theor.\ Math.\ Phys.\  2 (1989) 505;
{\tt hep-th/9803131}.
}

\lref\eva{E.~Silverstein and E.~Witten,
``Criteria for conformal invariance of $(0,2)$ models,''
Nucl.\ Phys.\ B 444 (1995) 161;
{\tt hep-th/9503212}.
}

\lref\sw{N.~Seiberg and E.~Witten,
``The D1/D5 system and singular CFT,''
JHEP 9904 (1999) 017;
{\tt hep-th/9903224}.
}

\lref\beah{O.~Aharony and M.~Berkooz,
``IR dynamics of $d = 2, N = (4,4)$ gauge theories and DLCQ of
'little  string theories',''
JHEP 9910 (1999) 030;
{\tt hep-th/9909101}.
 }

\lref\HosomichiBM{
K.~Hosomichi, K.~Okuyama and Y.~Satoh,
``Free field approach to string theory on $AdS_3$,''
Nucl.\ Phys.\ B 598 (2001) 451;
{\tt hep-th/0009107}.
}
\lref\HosomichiFM{
K.~Hosomichi and Y.~Satoh,
``Operator product expansion in string theory on $AdS_3$,''
{\tt hep-th/0105283}.
}%
\lref\GiribetFY{
G.~Giribet and C.~Nunez,
``Aspects of the free field description of string theory on AdS$_3$,''
JHEP  0006  (2000) 033;
{\tt hep-th/0006070}.
}
\lref\giribet{
G.~Giribet and C.~Nunez,
``Correlators in AdS$_3$ string theory,''
JHEP 0106  (2001) 010;
{\tt hep-th/0105200}.
}

\lref\SatohBI{
Y.~Satoh,
``Three-point functions and operator product
expansion in the $SL(2)$ conformal field theory,''
{\tt hep-th/0109059}.
}

\lref\wittensurfaces{C.~R.~Graham and E.~Witten,
``Conformal anomaly of submanifold observables in AdS/CFT
correspondence,''
Nucl.\ Phys.\ B 546 (1999) 52;
{\tt hep-th/9901021}.
}

\lref\son{J.~Son,
``String theory on $AdS_3 / Z_N$,''
{\tt hep-th/0107131}.
}

\lref\zam{V.~Fateev, A.~Zamolodchikov,
and Al.~Zamolodchikov, unpublished notes.}

\lref\teschnerone{J.~Teschner,
``On structure constants and fusion rules in the $SL(2,C)/SU(2)$ WZNW  model,''
Nucl.\ Phys.\ B 546 (1999) 390;
{\tt hep-th/9712256}.
}

\lref\gklast{
A.~Giveon and D.~Kutasov,
``Notes on $AdS_3$,''
{\tt hep-th/0106004}.
}
\lref\deboer{J. de \thinspace Boer, H. Ooguri,
H. Robins, and J. Tannenhauser, ``String theory in $AdS_3$,''
JHEP 9812 (1998) 026; {\tt hep-th/9812046}.}

\lref\shenkerseiberg{N.~Seiberg and S.~Shenker,
``A note on background (in)dependence,''
Phys.\ Rev.\ D 45 (1992) 4581;
{\tt hep-th/9201017}.
}
\lref\seiberg{N.~Seiberg,
``Notes on quantum Liouville theory and quantum gravity,''
Prog.\ Theor.\ Phys.\ Suppl.\  102 (1990) 319,
Proceedings of 1990 Yukawa International Seminar,
{\it Common Trends in Mathematics and Quantum Field Theories},
Kyoto, Japan.
 }
\lref\ks{
D.~Kutasov and N.~Seiberg,
``More comments on string theory on $AdS_3$,''
JHEP 9904 (1999) 008;
{\tt hep-th/9903219}.
}
\lref\teschnernew{
J.~Teschner,
``Crossing symmetry in the $H_3^+$ WZNW model,''
{\tt hep-th/0108121}.
}

\lref\jeremy{
J.~Maldacena, J.~Michelson and A.~Strominger,
``Anti-de Sitter fragmentation,''
JHEP 9902  (1999) 011;
{\tt hep-th/9812073}.
}

\lref\gawedzki{
K.~Gawedzki,
``Noncompact WZW conformal field theories,''
Proceedings of the NATO Advanced Study Institute,
{\it New Symmetry Principles in Quantum Field Theory,}
Cargese, 1991, p. 247, $eds$. J. Fr\"ohlich,
G. 't Hooft, A. Jaffe, G. Mack, P.K. Mitter, and R. Stora,
Plenum Press 1992; {\tt hep-th/9110076}.
}

\lref\btz{
M.~Banados, C.~Teitelboim and J.~Zanelli,
``The Black hole in three-dimensional space-time,''
Phys.\ Rev.\ Lett.\  69 (1992) 1849;
{\tt hep-th/9204099}.
}

\lref\mvgv{
S.~Mukhi and C.~Vafa,
``Two-dimensional black hole as a topological coset model of c = 1 string theory,''
Nucl.\ Phys.\ B 407 (1993) 667;
{\tt hep-th/9301083};
D.~Ghoshal and C.~Vafa,
``c = 1 string as the topological theory of the conifold,''
Nucl.\ Phys.\ B 453  (1995) 121;
{\tt hep-th/9506122}.
}

\lref\gk{
A.~Giveon and D.~Kutasov,
``Little string theory in a double scaling limit,''
JHEP 9910 (1999) 034;
{\tt hep-th/9909110},
A.~Giveon and D.~Kutasov,
``Comments on double scaled little string theory,''
JHEP 0001 (2000)  023;
{\tt hep-th/9911039}.
}


{\Title{\vbox{
\hbox{CALT-68-2360, CITUSC/01-042}
\hbox{\tt hep-th/0111180}}}
{\vbox{
\centerline{Strings in $AdS_3$ and the $SL(2,R)$ WZW Model.}
\vskip .1in \centerline{Part 3: Correlation Functions}}}
\vskip .3in
\centerline{Juan Maldacena$^{1,2}$ and Hirosi Ooguri$^3$}
\vskip .4in
}

\centerline{$^1$ Jefferson Physical Laboratory, Harvard University,
Cambridge, MA 02138}
\vskip .1in
\centerline{$^2$ Institute for Advanced Study, Princeton, NJ 08540}
\vskip .1in
\centerline{$^3$ California Institute of Technology 452-48,
Pasadena, CA 91125}

\vskip .4in

We consider correlation functions for string theory on
$AdS_3$. We analyze their singularities and we provide a
physical interpretation for them. We explain which
worldsheet  correlation
functions have a sensible physical interpretation in
terms of the boundary theory.
We consider the operator product expansion of  the four
point function and we find that it factorizes
only if a certain condition is obeyed. We explain that
this is the correct physical result.
We compute correlation functions involving spectral flowed
operators and we derive a constraint on the amount of
winding violation.

\vfill
\eject

\newsec{Introduction}

This is the third instalment of our series of papers on
the $SL(2,R)$ WZW model and its relation to string theory
in $AdS_3$, three-dimensional anti-de Sitter space.
In the first two papers,
\refs{\first, \second}, we have determined the structure of the Hilbert
space of the WZW model, computed the spectrum of physical
states of the string theory, and studied the one loop amplitude.
In this paper, we will discuss the
correlation functions of the model.

The $SL(2,R)$ WZW model has many important
applications in string theory and related subjects.
It has  close connections to the Liouville theory of
two-dimensional gravity (for a review, see for example, \seiberg )
and  three-dimensional Einstein
gravity \ref\wittenthreed{E. Witten,
``(2+1)-dimensional gravity as an exactly soluble system,''
Nucl. Phys. B311 (1988) 46.}. It is used to describe  string theory in
two-dimensional black hole geometries \ref\wittentwod{
E. Witten, ``On string theory and black holes,'' Phys. Rev.
D44 (1991) 314.}. Its quotients  are  an important
ingredient in understanding string theory in the background of
Neveu-Schwarz 5-branes \ref\chs{C. G. Callan,
J. A. Harvey, and A. Strominger,
``Supersymmetric string solitons,'' {\tt hep-th/9112030}.},
and they  capture aspects of strings
propagating near singularities of Calabi-Yau
spaces \lref\ov{H. Ooguri and C. Vafa, ``Two-dimensional black hole
and singularity of CY manifolds,'' Nucl. Phys. B463 (1996) 55,
{\tt hep-th/9511164}.}
\refs{\mvgv,\ov, \gk}. One can also construct
a black hole geometry in three dimensions by taking
a quotient of the $SL(2,R)$ group manifold \btz .
Moreover, sigma-models with non-compact
target spaces such as $SL(2,R)$ have various
applications to condensed matter physics \ref\zir{M. Zirnbauer,
``Riemannian symmetric superspaces and their origin in
random matrix theory,'' J. Math. Phys. 37 (1996) 4986;
{\tt math-phys/9808012}.}. For these reasons,
the model has been studied extensively for more than a
decade.\foot{For a list of historical references see
the bibliography in \first.}
Recently the model has becomes particularly important
in connection with the $AdS$/CFT correspondence
\lref\malda{J.~Maldacena,
``The large N limit of superconformal field theories and supergravity,''
Adv.\ Theor.\ Math.\ Phys.\  2 (1998) 231;
{\tt hep-th/9711200}.
} \lref\magoo{For a review of the subject, see for example,
O.~Aharony, S.~S.~Gubser, J.~Maldacena, H.~Ooguri and Y.~Oz,
``Large N field theories, string theory and gravity,''
Phys.\ Rept.\  323 (2000) 183;
{\tt hep-th/9905111}.
} \refs{\malda,\magoo}
since it describes the worldsheet of a string propagating
in $AdS_3$ with a background NS-NS $B$-field.
According to the correspondence, Type IIB superstring theory
on $AdS_3 \times S^3 \times M_4$ is dual to the supersymmetric
non-linear sigma model in two dimensions whose target space
is the moduli space of Yang-Mills instantons on
$M_4$ \magoo \ref\maldastrom{
J.~Maldacena and A.~Strominger,
``$AdS_3$ black holes and a stringy exclusion principle,''
JHEP 9812 (1998) 005;
{\tt hep-th/9804085}.
}. Here $M_4$ is a four-dimensional Ricci flat K\"ahler manifold,
which can be either a torus $T^4$ or a K3 surface.
So far this has been the only case where we have been
able to explore the  correspondence beyond
the supergravity approximation with complete control
over the worldsheet theory.

Besides the $AdS$/CFT correspondence, understanding
string theory in $AdS_3$ is interesting since $AdS_3$
is the simplest example of a curved spacetime where the
metric component $g_{00}$ is non-trivial. Using this
model, one can discuss various questions which involve
the concept of time in string theory.
This will give us important lessons on how to deal with
string theory in geometries which involve time
in more complicated ways. In this connection, there
had been a long standing puzzle, first raised in
\lref\noghost{J.~Balog,
L.~O'Raifeartaigh, P.~Forgacs, and A.~Wipf,
``Consistency of string
  propagation on curved space-time: an
$SU(1,1)$ based counterexample,''
  Nucl.\ Phys. B325 (1989) 225.}
\lref\dpl{
L.~J.~Dixon, M.~E.~Peskin, and J.~Lykken,
``$N=2$ superconformal
symmetry and $SO(2,1)$ current algebra,''
Nucl.\ Phys. B325 (1989) 329.} \refs{\noghost,\dpl},
about whether the no-ghost
theorem holds for string in $AdS_3$.
The proof of no-ghost theorem in this case is more
involved than in Minkowski
space since the time variable in $AdS_3$
does not decouple from the rest of the degrees
of freedom on the worldsheet.
The task was further complicated by the
fact that $AdS_3$ is a non-compact space and
the worldsheet CFT is not rational.
Thus it was difficult to decipher the spectrum
of the worldsheet theory.

This problem was solved in \refs{\first,\second}.
In \first, we proposed the
spectrum of the WZW model and gave a complete
proof of the no ghost theorem base on the proposed
spectrum. This proposal itself
was verified in \second\ by exact computation of
the one loop free energy for string on
$AdS_3 \times {\cal M}$, where
${\cal M}$ is a compact space represented
by a unitary conformal field theory on the
worldsheet. Although
the one loop free energy receives contributions only
from physical states of the string theory, we can
deduce the full spectrum of the $SL(2,R)$ WZW
model from the dependence of the partition function
on the spectrum of the internal CFT representing
${\cal M}$, which can be arbitrary as far as
it has the appropriate central charge.
Thus the result of \second\ can be regarded
as a string theory
proof of the full spectrum proposed in \first.

The spectrum of the $SL(2,R)$ WZW model established in
\refs{\first , \second} is as follows.
Since the model has the symmetry generated by
the $SL(2,R) \times SL(2,R)$ current algebra, the
Hilbert space ${\cal H}$ is decomposed into its representations
as
\eqn\hilbertspace{
{\cal H} = \oplus_{w=-\infty}^\infty
 \left[ \left(\int_{1\over 2}^{{k-1\over 2}}
 dj {\cal D}_j^w \otimes {\cal D}_j^w\right)
 \oplus \left(\int_{{1\over 2}+i{\bf R}}
dj \int_0^1 d\alpha C_{j,\alpha}^w
\otimes C_{j,\alpha}^w\right) \right].}
Here ${\cal D}_j^w$ is an irreducible representation
of the $SL(2,R)$ current algebra generated from
the highest weight state $|j; w\rangle$ defined
by
\eqn\hwsdisc{
 \eqalign{ &J_{n+w}^+|j; w\rangle = 0,
~~J_{n-w-1}^-|j; w\rangle = 0, ~~J_n^3
 |j; w\rangle = 0, ~~~(n=1,2,\cdots) \cr
&J_0^3|j; w\rangle = \left(j+{k \over 2}w\right)
|j;w\rangle, \cr
&\left[- (J_0^3 -\khalf w)^2  + {1\over 2}\left(J_w^+ J_{-w}^-
+ J_{-w}^- J_w^+ \right)\right] |j; w\rangle
= -j(j-1) |j; w\rangle,
}}
and ${\cal C}_j^w$ is generated from the state
$|j, \alpha; w\rangle$ obeying
\eqn\hwscont{\eqalign{
 &J_{n\pm w}^\pm |j, \alpha ; w\rangle = 0, ~~J_n^3
 |j, \alpha; w\rangle = 0, ~~~(n=1,2,\cdots) \cr
 &J_0^3 |j, \alpha;w\rangle
 = \left(\alpha+{k \over 2}w\right)|j, \alpha; w \rangle,\cr
&\left[- (J_0^3- \khalf w)^2  + {1\over 2}\left(J_w^+ J_{-w}^-
+ J_{-w}^- J_w^+ \right)\right] |j, \alpha; w\rangle
= -j(j-1) |j, \alpha; w\rangle.}}
The representations with $w=0$ are conventional ones,
where $|j; 0 \rangle$ and $| j, \alpha; 0\rangle$
are annihilated by the positive frequency modes
of the currents $J_n^{\pm,3}$ ($n \geq 1$).
These representations ${\cal D}_j^0$ and ${\cal C}_{j,\alpha}^0$
are called the discrete and continuous
representations respectively.\foot{We call ${\cal D}_j^{0}$
a discrete representation even though the spectrum of $j$
in the Hilbert space \hilbertspace\ of the WZW model
is continuous. It would have been discrete
if the target space were the single cover of the $SL(2,R)$
group manifold.
In order to avoid  closed timelike curves, we take
the target space to be
the universal cover of $SL(2,R)$,
in which case the spectrum of $j$ is continuous.
We still call these representations discrete since
their $J_0^3$ eigenvalues are related to the values of
the Casimir operator, $-j(j-1)$, while the $J^3_0$ eigenvalue
for continuous representations  is not related to their values
of the Casimir operator. }
The representations with $w\neq 0$ are related
to the ones with $w=0$ by the spectral flow automorphism
of the current algebra, $J_n^{3,\pm} \rightarrow
\tilde{J}_n^{3,\pm}$, defined by
\eqn\introflow{
  \eqalign{  & \tilde{J}^\pm_n=
J^\pm_{n\pm w}, \cr
             & \tilde{J}^3_n = J_n^3- {k \over 2}
w \delta_{n,0}.}}
In the standard WZW model, based on a compact Lie group,
 spectral flow does not generate new types of representations;
it simply maps a conventional  representation into
another, where the highest weight state of one representation
turns into a current algebra descendant of another. In the case
of $SL(2,R)$, representations with
different amounts of $w$ are not equivalent.

In \first, it was shown that \hilbertspace\ leads to
the physical spectrum of string in $AdS_3$ without ghosts
and that the spectrum agrees with various aspects of the dual CFT$_2$
on the boundary of the target space. In particular,
it is shown that the spectral flow images
of the continuous representations lead to physical states with
continuous energy spectrum of the form,
\eqn\firstlongstring{
 J_0^3 = {k \over 4} w + {1\over w}\left(
 {s^2 + {1 \over 4}\over k-2} +N+ h - 1 \right),}
where $s$ is a continuous parameter for the states, $N$ is
the amount of the current algebra excitations before
we take the spectral flow,
and $h$ is the conformal weight of the state in the internal CFT
representing the compact directions in the target space.
These states are called ``long strings''  with winding number $w$,
and their continuous spectrum is related to the presence
of  non-compact directions in the target space of the dual CFT$_2$
\refs{\jeremy,\sw}.
The continuous parameter $s$ is identified with the momentum
in the non-compact directions. The continuous representations
with $w=0$ give no physical states except for the tachyon,
which is projected out in superstring. On the other hand,
the discrete representations and their spectral flow
images give the so called ``short strings,'' whose physical spectra
are discrete.

In this paper, we compute amplitudes of these physical states
of the string theory and interpret  them as  correlation functions
of the dual CFT$_2$. We show that the string theory
amplitudes satisfy various properties expected
for correlation functions of the dual CFT$_2$.

The dual CFT$_2$ is unitary with a Hamiltonian
of positive definite spectrum, and the density of
states grows much slower than the exponential
of the energy.\foot{The Cardy formula states that
the density of states of conformal field theory
on a unit circle grows as $\exp\left( 2\pi \sqrt{c E/6}\right)$,
where $E$ is the energy and $c$ is the central charge
of the theory.}
Therefore one should be able to
analytically continue the time variable of CFT$_2$
to Euclidean time. Correspondingly the $AdS_3$ geometry
can be analytically continued to the three-dimensional
hyperbolic space $H_3$, whose boundary is $S^2$.
The worldsheet of the string on $H_3$ is described by
the $SL(2,C)/SU(2)$ coset model. We would like to stress
that the $SL(2,R)$ WZW model and the $SL(2,C)/SU(2)$
coset model are quite distinct even though their  actions
are formally related by analytical continuations of
field variables. For example,
the Hilbert spaces of the two models are completely
different since all the
states in the Hilbert space \hilbertspace\
of the $SL(2,R)$ WZW model, except for the continuous representations
with $w=0$, correspond to non-normalizable
states in the $SL(2,C)/SU(2)$ model. This means that
all the physical states in string theory, except for
the tachyon, are represented
by non-normalizable states in the $SL(2,C)/SU(2)$ model.
It is in the context of string theory computations of
physical observables
that one can establish connections between the
two worldsheet models.
We will discuss in detail how this connection
works when we use string theory to compute correlation
functions of the dual CFT$_2$ on the boundary of the
target space.

Correlation functions of the $SL(2,C)/SU(2)$ model have
been derived in \refs{\teschnerone,\zam,\teschnersecond}
for operators corresponding to normalizable states and
some non-normalizable states simply related to them by analytic
continuation.
Although the
correlation functions for normalizable states are completely
normal, those for non-normalizable states  contain singularities
of various kinds. Thus we need to understand the origins of these
singularities and to learn how to deal with them.

For clarity, we separate our discussion into
two parts. First we will discuss the origins of these
singularities purely from the point of view of the
worldsheet theory. We will show how functional integrals
of the $SL(2,C)/SU(2)$ model generate these singularities.
We find that some of these singularities can be understood in the
point particle limit while others come from
 large ``worldsheet instantons.''

After explaining all the singularities from the worldsheet
point of view, we turn to string theory computations
and interpret these singularities from the point of view
of the target spacetime physics. Some of the singularities
are interpreted as due to operator mixings, and others
originate from the existence of the non-compact directions
in the target space of the dual CFT$_2$. In addition to
the singularities in the worldsheet correlation functions,
the integral over the moduli space of string worldsheets can
generate additional singularities of stringy nature.
In Minkowski space, singularities are all at boundaries
of moduli spaces ($e.g.$, when two vertex operators collide
with each other or when the worldsheet degenerates) and
divergences coming from them are interpreted as due
to the  propagation of intermediate physical states.
For strings in $AdS_3$, we find that amplitudes
can have singularities in the middle of moduli space. We
have already encountered such phenomena in one loop free
energy computation in \second, and they are attributed to
the existence of the long string states in the physical
spectrum. We will find related singularities in our computation
of four point correlation functions.

By taking into account these singularities on the worldsheet
moduli space, we prove the factorization of four point
correlation functions in the target space. We show that
the four point correlation function,
obtained by  integrating over the moduli space of the worldsheet,
is expressed as a sum of products of three point functions
summed over possible intermediate physical states.
The structure of the factorization agrees with the
physical Hilbert space of string given in \refs{\first,\second}.
We also check that normalization factors for
intermediate states come  out  precisely as expected.
The resulting factorization formula
shows a partial conservation of the total ``winding number''
$w$ of string.\foot{As explained in \first , $w$ is in general
 a label of the type of
representation and is not the actual winding number of the string,
although, for some states,
it could coincide with the winding number
of the string in the angular direction of $AdS_3$. }
 We will explain its origin from  the worldsheet
$SL(2,R)$ current algebra symmetry and the structure
of the two and three point functions.
In the course of this, we will
clarify various issues about the analytic continuation relating
the $SL(2,C)/SU(2)$ model and the $SL(2,R)$ model.

We find that, in certain situations, the four point
functions fail to factorize into a sum of products of
three point functions with physical intermediate states.
We show that this failure of the factorization happens
exactly when it is expected from the point
of view of the boundary CFT$_2$. Namely, the four point
functions factorize only when they should.

This paper is organized as follows. In section 2, we
review correlation functions of the $SL(2,C)/SU(2)$
coset model derived in \refs{\teschnerone,\zam} and explain
the worldsheet origin of their singularities.
In section 3,
we turn to the string theory computation and discuss the
target space interpretation  of the singularities
in two and three point correlation functions.
In section 4, we give a detailed treatment of four
point correlation functions. On the worldsheet,
a four point function of the $SL(2,C)/SU(2)$ model
is expressed as an integral over solutions to the
Knizhnik-Zamolodchikov equation \teschnersecond . We integrate
the amplitude over the worldsheet moduli, which in
this case is the cross ratio of the four points on $S^2$, and
obtain the target space four point correlation function.
We examine factorization properties of the resulting
correlation function. We explain when it factorizes and why it
sometimes fails to factorize.
In section 5, we compute two and three point
functions of states with non-zero winding numbers.
We also explain the origin of the constraint on the
winding number violation.
In section 6, we use the result of section 5 to show that
the factorization of the four point function works
with precisely the correct coefficients.

In appendix A, we derive the target space two point function
of short string with $w=0$. The normalization of the target space
two point function is different from that of the worldsheet
two point function. The target space normalization is precisely
the one that shows up in the factorization of the four point
amplitudes. In Appendix B, we derive some properties
of conformal blocks of four point functions.
In Appendix C, we derive a formula for integrals of
hypergeometric functions used in section 4.
In appendix D,  we derive a  constraint on winding number violation
from the $SL(2,R)$ current algebra symmetry of the theory.
In Appendix E, we introduce another definition of the spectral
flowed operator, working directly in the coordinate basis (rather
than in the momentum basis) on the
boundary of the target space. We compute
two and three point functions containing the spectral flowed
operators using this definition.

Some aspects of correlation functions
of the $SL(2,C)/SU(2)$ model have also been discussed in
\refs{\HosomichiBM,
\HosomichiFM,\gklast,\GiribetFY,
\giribet,\SatohBI}.

\newsec{General remarks about the $SL(2,C)/SU(2)$ model}

In this section, we study properties of the sigma model whose target
space is Euclidean $AdS_3$ or three dimensional hyperbolic space,
which is denoted by $H_3$. This sigma model is a building block
for the construction of string theory in $H_3 \times {\cal M}$,
where ${\cal M}$ is an internal space represented by some
unitary conformal field theory.
It is also used to compute string amplitudes for
Lorentzian signature $AdS_3$. A precise prescription for
computing string amplitudes will be given in section 3.
Before discussing the string theory interpretation,
let us clarify some properties of the sigma model itself.

The hyperbolic space $H_3$ can be realized as a right-coset space
$SL(2,C)/SU(2)$ \gawedzki . Accordingly the conformal field theory
with the target space $H_3$ and a non-zero NS-NS 2-form
field $B_{\mu\nu}$ can be constructed as a coset of the
$SL(2,C)$ WZW model by the right action of $SU(2)$.
The action of the $SL(2,C)/SU(2)$ model
can be expressed in terms of the Poincare coordinates
$(\phi, \gamma, \bar{\gamma})$
and the global coordinates $(\rho, \theta, \varphi)$ of
$H_3$ as
\eqn\lagrangian{\eqalign{
S =& {k \over \pi} \int d^2z
\left( \partial \phi \bar{\partial} \phi
 + e^{2 \phi} \partial \bar \gamma \bar \partial
 \gamma\right)\cr
 = & {k \over \pi} \int d^2z
\Big[ \partial \rho \bar{\partial} \rho
+ \sinh^2 \rho ( \partial \theta \bar{\partial}\theta
  +\sin^2 \theta \partial \varphi \bar{\partial} \varphi ) + \cr
&~~~~~~~~~~~~~~
\left. + i \left( {1 \over 2} \sinh 2\rho - \rho \right)
  \sin\theta (\partial \theta \bar{\partial} \varphi
 - \bar{\partial}\theta \partial \varphi ) \right].}
}
We are considering the Euclidean worldsheet with
$\partial = \partial_z$, $etc$.
Near the boundary, $\rho \rightarrow \infty$,
the action becomes
\eqn\largerhoaction{\eqalign{
 S \sim &{k \over \pi} \int d^2z \left[
\partial \rho \bar{\partial}\rho
+ {1 \over 4} e^{2\rho}
  (\partial \theta - i \sin\theta \partial \varphi)
  (\bar{\partial} \theta + i \sin\theta
   \bar{\partial} \varphi)  - \right. \cr
&~~~~~~~~~~~~~~~~~~ -i \left. \rho \sin\theta(
\partial \theta \bar{\partial} \varphi
 - \bar{\partial} \theta \partial \varphi) \right].}}
Because of the second term in the right-hand side,
contributions from large values of $\rho$ are
suppressed in the functional integral as $\sim
\exp(-\alpha e^{2\rho})$; the coefficient
$\alpha$ is positive semi-definite, and
it vanishes only when
$(\theta,\varphi)$ is
a holomorphic map from the worldsheet to $S^2$
obeying
\eqn\holcondition{
 \bar{\partial} \theta + i \sin \theta \bar{\partial} \varphi
=0.}
Even for $\alpha=0$, if the map is non-trivial,
the last term in \largerhoaction\ may grow linearly
in $\rho$. For constant $\rho$ and $(\theta,\varphi)$
obeying \holcondition, the action
goes as $S \sim 2 kn \rho$ where $n$ is the number
of times the worldsheet wraps the $S^2$.

The action on the Euclidean worldsheet is real-valued
since the $B$ field is pure imaginary.\foot{This is
so that the $B$ field becomes real-valued after
analytically continuing the target space to
the Lorentzian signature $AdS_3$.} The action is
positive definite, and all normalizable operators have positive
conformal weights.
Thus one expects Euclidean functional integrals
to behave reasonably well  in this model. The only novelty
is the fact that the target space $H_3$ of this sigma
model is non-compact, but it is just as in the case of
a free scalar field taking values in ${\bf R}$, which
is also non-compact.

The  interpretation of this model on a Lorentzian
worldsheet is more subtle. Because of the $B$ field,
the action \lagrangian \ is not invariant under
reflection of the Euclidean time, and it becomes
complex valued after analytically continuing to
the Lorentzian worldsheet. Thus the Hilbert space
of the $SL(2,C)/SU(2)$ model
on the Lorentzian worldsheet may not have a positive
definite inner product; in fact, an action of
the $SL(2,C)$ current $J_{-n}^3$ generates
negative norm states. As we mentioned in the above
paragraph, the model
on the Euclidean worldsheet appears to be completely
normal --- just that it does not have an analytic
continuation to a normal field theory on a
Lorentzian worldsheet.\foot{This is somewhat
mirror to the situation of the $SL(2,R)$ WZW model.
The $SL(2,R)$ model makes sense in the Lorentzian
worldsheet as discussed in \refs{\ks,\first}.
However we cannot
analytically continue
to the Euclidean worldsheet since the Hamiltonian
of the model is not positive definite. Note that here we are
talking about
 analytically continuing the worldsheet without analytically
continuing the spacetime.}

What is the space of states of this conformal field theory?
In the semiclassical approximation, which is valid when
$k$ in the action \largerhoaction\ is large, states
are given by normalizable functions on the target space.
More precisely, since the target space $H_3$ is non-compact,
we allow functions to be continuum normalizable. Because of
the $SL(2,C)$ isometry of $H_3$,
the space of continuum normalizable functions is decomposed into
a sum of irreducible unitary representations of $SL(2,C)$.
The representations are parametrized
by $j = {1\over 2} + is$ with $s$ being a real number, and
the Casimir of each representation is given by $-j(j-1)$.
The Casimir is proportional to the eigenvalue of the Laplacian
on $H_3$.
Corresponding to each of these states, there is
an operator in the $SL(2,C)/SU(2)$ model,
which is also called
normalizable. They can be conveniently written as \lref\zf{
A. B. Zamolodchikov and V. A. Fateev, ``Operator
algebra and correlation functions
in the two-dimensional Wess-Zumino $SU(2) \times SU(2)$
chiral model,'' Sov.\ J.\ Nucl.\ Phys.  43 (1986) 657.}
\refs{\zf,\teschnerone},
\eqn\operators{
\Phi_j(x , \bar x; z,\bar z)  = {1-2j
\over \pi} ( e^{-\phi} + |\gamma - x|^2 e^\phi )^{-2j}.
}
The labels $x,\bar x$ are introduced to keep track of
the $SL(2,C)$ quantum numbers \ref\ggv{I. M. Gelfand,
M. I. Graev, and N. Ya Vilenkin, {\it Generalized
functions Vol 5,} Academic Press, 1966.}.\foot{In the
string theory interpretation discussed in section 3,
$(x,\bar{x})$ is identified as the location of the
operator in the dual CFT on $S^2$ on the boundary of $H_3$
\deboer .}
The $SL(2,C)$ currents act on it as
\eqn\slaction{
 J^a(z) \Phi_j( x, \bar{x}; w,\bar{w})
  \sim {D^a \over z-w} \Phi_j(x, \bar{x};w,\bar{w}),
~~a=\pm,3,}
where $D^a$  are differential operators
with respect to $x$ defined by
\eqn\whatdx{
  D^+ = {\partial \over \partial x},~~
   D^3 =  x {\partial \over \partial x} + j,~~
      D^- = x^2 {\partial \over \partial x} + 2jx.}
By using this and the Sugawara construction of the
energy momentum tensor,
\eqn\sugawara{
 T(z) = {1 \over k-2}\left(  J^+(z) J^-(z)  -  J^3(z) J^3(z)  \right),}
we find the precise
expression for the conformal weights of these operators as
\eqn\confweight{
 \Delta(j) = - {j(j-1) \over k-2 } = {s^2 + {1\over 4} \over k-2}.
}
Operators with $j=\half + is $ have positive conformal weight,
as we expect for normalizable operators in a well-defined theory
with Euclidean target space. It was shown in \gawedzki\ that
states corresponding to these operators and their current
algebra descendants make the complete Hilbert space of the
$SL(2,C)/SU(2)$ model.

The vacuum state of the $SL(2,C)/SU(2)$ model is
not normalizable. This again is not unfamiliar; the vacuum state
for the free scalar field on ${\bf R}$ is also non-normalizable since
its norm is proportional to ${\rm vol}({\bf R})=\infty$.
In this case, we do not
consider the vacuum in isolation. The vacuum state always appears
 with an operator,  such as in $e^{ipX(0)} |0\rangle$.
Similarly, on $H_3$, the vacuum state $|0\rangle$ is not normalizable,
but we can consider a state given by operators of the form \operators\
with $j = \half + is$ acting on it.\foot{In the flat space case,
the vacuum $|0 \rangle$ can be regarded as $p\rightarrow 0$ limit
of $e^{ipX(0)} |0 \rangle$, and therefore it is a part of the
continuum normalizable states. Such an interpretation is not
possible in the case of $H_3$ since there is a gap
of ${1 \over 4(k-2)}$ between the conformal
weight \confweight\ of the normalizable states
and that of the vacuum.}

The two and three point functions of operators like \operators\ were
computed in \refs{\teschnerone,\zam,\teschnersecond}.
The two point function has the form,
\eqn\teschnertwo{\eqalign{&
\langle \Phi_j(x_1,\bar x_1;z_1,\bar z_1 )
 \Phi_{j'}(x_2,\bar x_2;z_2,\bar z_2 ) \rangle\cr
& =
{ 1\over |z_{12}|^{4\Delta(j)}}
\left[ \delta^2(x_1-x_2) \delta(j+j'-1)
+{B(j)\over |x_{12}|^{ 4j }} \delta(j-j')\right] ,}
}
The
coefficient $B(j)$ is given by
\eqn\defofbtes{
B(j) =  {k-2 \over \pi} { \nu^{1-2j} \over \gamma\left(
{ 2 j -1 \over k-2 }\right)} ,}
where
\eqn\defofsmallgamma{
\gamma(x) \equiv {\Gamma(x) \over
\Gamma(1-x) }.}
The choice of the constant $\nu$ will not affect
the discussion in the rest of this paper.
In \teschnersecond , it is set to
be
\eqn\teschnerchoice{
\nu = \pi {\Gamma\left(1 - {1\over k-2} \right)
\over \Gamma\left( 1 + {1 \over k-2} \right)},}
by requiring a certain consistency between the
two and three point functions.

The three point function is expressed as
\eqn\teschnerthree{\eqalign{
\langle \Phi_{j_1}&(x_1,\bar x_1;z_1,\bar z_1 )
 \Phi_{j_2}(x_2,\bar x_2; z_2,\bar z_2)
\Phi_{j_3}(x_3,\bar x_3;z_3,\bar z_3 )
\rangle = \cr
&=C(j_1,j_2,j_3)
{1
 \over | z_{12}|^{2(\Delta_1+ \Delta_2 - \Delta_3) }
|z_{23}|^{2(\Delta_2+ \Delta_3 - \Delta_1) }
|z_{31}|^{2(\Delta_3+ \Delta_1 - \Delta_2)}} \times\cr
&~~~~~~~{1 \over
|x_{12}|^{2(j_1+j_2-j_3)}|x_{23}|^{2(j_2+j_3-j_1)}
|x_{31}|^{2(j_3+j_1-j_2)} }  .}
}
The $z$ and $x$ dependence is determined by $SL(2,C)$ invariance of the
worldsheet and the target space.
The coefficient $C(j_1,j_2,j_3)$ is given by
\eqn\threecoeff{
 C(j_1,j_2,j_3) = -{
G(1-j_1-j_2-j_3) G(j_3-j_1-j_2) G(j_2-j_3-j_1)
G(j_1-j_2-j_3) \over   2\pi^2
 \nu^{j_1+j_2+j_3-1}
\gamma\left( {k-1\over k-2}\right)G(-1) G(1-2j_1) G(1-2j_2)G(1-2j_3)},}
where
\eqn\whatg{
 G(j) = (k-2)^{{j(k-1-j) \over 2(k-2)}} \Gamma_2(
-j ~|~1,~ k-2)
 \Gamma_2(k-1+j ~|~1,~ k-2).}
and $\Gamma_2(x| 1, \omega)$ is the Barnes double Gamma function
defined by\foot{The sums over $n,m$ in the right-hand side
are defined by  analytic regularization.
Namely, the sums are defined for ${\rm Re} (\epsilon) > 2$, where
they are convergent, and the result is
analytically continued to $\epsilon \rightarrow 0$.}
\eqn\barnes{\eqalign{&
  \log \left(\Gamma_2(x|1,\omega)\right)\cr
&= \lim_{\epsilon \rightarrow 0}
 {\partial \over \partial \epsilon}\left[
 \sum_{n,m=0}^\infty
(x+n+m\omega)^{-\epsilon}
-
 \sum_{n,m = 0 \atop  (n,m)\neq(0,0)}
(n+m\omega)^{-\epsilon}\right].}}
This shows that $\Gamma_2$ has poles at
$x=-n-m\omega$ with $n,m=0,1,2,\cdots$.
The function $G(j)$ defined by \whatg\ then
has poles at
\eqn\gpolesfirst{
 j = n + m (k-2),~~-(n+1) - (m+1)(k-2),~~~(n,m=
0,1,2,\cdots).}
These will play an important role in the following
discussion.

Another important fact about $G(j)$ is that it
obeys the functional relations,
\eqn\grelation{
\eqalign{ &G(j+1)  = \gamma\left(-{j+1 \over k-2}\right)
G(j) \cr
&G(j-k+2) = {1\over (k-2)^{2j+1} } \gamma(j+1)G(j).}}
For example, one can use the first of these relations to show that
\eqn\deltafunction{\eqalign{&
\lim_{\epsilon\rightarrow 0}
{G(j_1-j_2+\epsilon)G(j_2-j_1+\epsilon)
\over G(-1) G(1-2\epsilon)}\cr
&=(k-2) \gamma\left({k-1\over k-2}\right) \lim_{\epsilon \rightarrow
0}
{2\epsilon \over (j_1-j_2)^2 - \epsilon^2}\cr
& = -2\pi (k-2) \gamma\left({k-1\over k-2}\right) \delta(s_1-s_2),}}
when $j_1={1\over 2} + is_1$ and
$j_2={1\over 2} + is_2$. From this, it follows that,
\eqn\twoandthree{
C(j_1,j_2,0)
 = B(j_1) \delta(j_1-j_2),}
verifying that the three point function including the
identity operator $\Phi_{j=0}$ is in fact equal to the two point
function. Similarly, by using the second of \grelation ,
we can show,
\eqn\anotherdelta{
 C\left(j_1,j_2,\khalf\right)
 = (k~{\rm dependent~coefficient}) \times \delta(j_1+j_2-k/2 ).}
Unlike the case of \twoandthree , the
proportionality factor depends only on $k$
and not on $j_1, j_2$. This identity is used in later sections
when we evaluate correlation functions involving spectral flowed
states.

These two and three point functions are perfectly well behaved and finite
for normalizable operators with $j= \half + is $.
Similarly one  expects that
the four point function of such states to be given by summing over
intermediate normalizable states \refs{\teschnersecond,\teschnernew}.\foot{
Recently it was shown in \teschnerrecent\ that the four point function
of the $SL(2,C)/SU(2)$ model has the same form as that of the five point
function of the Liouville model where the cross ratio of
four $x_i$'s in the $SL(2,C)/SU(2)$ model is related to the location
fo the fifth vertex operator
in the Liouville model. This in particular shows that
the four point function obeys the crossing symmetry, the monodromy
invariance, and so on, assuming that Liouville correlation functions
also satisfy these properties. The monodromy invariance of the four
point function is proven explicitly in section 4.2.}
 The four point
function will be discussed in detail in section 4.
These properties are familiar and happen in all conformal field theories.
The non-compactness of the  target space does not pose a problem;
we  deal with it as in the case of a free non-compact scalar field.

\subsec{Analytic continuation and singularities}

Life would be relatively simple if all we were interested in
were operators like \operators\ with
$j = \half + is$.

The complications in our case show up because the operators we
are going to be interested in are non-normalizable operators
\refs{\shenkerseiberg,\seiberg}.
This is also familiar in  standard flat space computations
in string theory. There, we are interested
in vertex operators which go as $e^{p_0 X_{Euclid}^0}$,
where $p_0$ is the energy carried by the operator and is
real, and $X_{Euclid}^0$ is the scalar field representing
the Euclidean time coordinate.
It is sometimes said that we compute amplitudes in Euclidean signature
space (with pure imaginary $p_0$) and
then we analytically continue the results
in $p_0$. This analytic continuation is possible
if  correlation functions with non-normalizable operators
of the form $ e^{p_0 X_{Euclid}^0}$ make sense in the
model with Euclidean target space. There might be singularities
for complex values of $p_0$, but we should be able to go around them
to arrive at real values of $p_0$.
Original correlators with pure imaginary $p_0$ are well-defined
in the Euclidean theory and never infinite since these operators
correspond to normalizable states of the theory.
When we analytically continue to real (or complex) $p_0$, there
can be singularities  where the amplitudes diverge.
In flat space string theory,
these singularities arise when we integrate
over the positions of the operators on the worldsheet.
The integrated four point function
can become singular as a function of the momenta.
 The interpretation of these singularities is
of course well-known in  flat target spacetime; they
corresponds to poles in the $S$ matrix and
they are due to the propagation of an intermediate on-shell state.
 The lesson from the flat space case
is that we should be able to interpret
any singularity that appears in the physical computation
of string amplitudes.
Part of the definition of the physical theory is the choice of operators
we consider. In the flat space case, $p_0$ has to be
pure imaginary in order for the vertex operator $e^{p_0 X^0_{Euclid}}$
to be normalizable. These are the operators  that are most natural
($i.e.$, normalizable)
from the point of view of the Euclidean worldsheet theory.
On the other hand, for applications to
string theory, we need to consider the case when
$p_0$ is real as these are the ones that correspond to
physical states in target space.

In our case we can
define non-normalizable operators by taking $j$ away from the line
$j= \half + is$. In the string theory application, we will
be interested in the case when $j$ is real.
One can  define correlation functions of these
operators  by analytically continuing  the
well-defined expressions that were found for  $j= \half + is$.
In fact, the expressions for complex $j$ were derived in
\teschnerone\  by using
special properties of operators at particular real  values of  $j$,
so analyticity in $j$ was an input to the calculation.
A feature of this analytic continuation is that correlation functions
that were perfectly finite and well behaved can develop singularities
for particular values of $j$. In the following subsections we will
explain the origin of these singularities in the $SL(2,C)/SU(2)$ model.
We will also explain that there are other non-normalizable operators that
are necessary for the string theory application which are {\it not}
obtained by analytic continuation in $j$ of \operators .
In section 5, we will discuss how to compute correlation functions
of these operators.


\subsec{Singularities in two point functions}

The first thing we need to understand is
how the operators \operators\ with
real $j$ are defined. It seems that all we need to do is to insert
the vertex operators \operators\ in the path integral.
As usual we need to remove short distance singularities in the
worldsheet theory when we insert these operators. This is the
standard renormalization procedure we need to use to define vertex
operators. In this case, however,
 we also need to be careful with singularities
on the worldsheet theory that arise due to the fact that the
sigma model is non-compact.
The vertex operator $\Phi_j(z,x)$ defined by \operators\
has the property that, depending on whether $\gamma(z)=x$
or $\neq x$, it behaves as $\Phi_j \sim e^{2j\phi}$ or
$\sim e^{-2j\phi}$ for large $\phi$.
For ${\rm Re}(j) <1/2$, we see that, once we take into
account the measure factor $e^{2 \phi}$, the two point
function
will have a divergence. This divergence comes from the region
where $\gamma \not =x$ and $\phi \rightarrow \infty$,
 and therefore it is not localized near
$x$ in target space; it is spread all over the $x$ space.
On the other hand, if ${\rm Re}(j)>1/2$ this divergence is
localized at $\gamma =x$. This distinction between these
two cases will be very important for
the string theory application discussed in the next section.
{} From the worldsheet point of
view, operators of the form ${\rm Re}(j) \not =1/2$ are not normalizable.
Analytic continuation is defining  these  operators in some way.
We also get a divergence in the two point function coming from the
delta function $\delta (j-j')$ in \teschnertwo.
 This comes from the volume of the
subgroup of target space global
$SL(2,C)$ transformations that leave two points fixed
(the two points, $x_1$ and $x_2$, where the operators are inserted).

The analytically continued expression \defofbtes\
has  other divergences. It
has poles at
\eqn\twopoles{
 j = {n\over 2} (k-2) + {1\over 2},~~~~
n=1,2,\cdots.}
Let us understand these poles when $k$ is large.
 Before we continue,
let us note that we know the exact expression \teschnertwo, and
there is no need to re-evaluate it approximately. The purpose
of this exercise is to understand the origin of these
singularities. This will help us interpret them in
the context of string theory later. It
may also be useful in analyzing
similar singularities in situations where we do not know
exact answers.


Let us start with the $n=1$ case.
Since $j\sim k$ and the semi-classical limit corresponds to
$k \rightarrow \infty$, these poles can be thought of as arising
from non-perturbative effects
on the worldsheet. The non-perturbative effect we have in mind is due to
a worldsheet instanton.
The target space has a boundary that is an $S^2$, and our worldsheet
instanton approaches to it while wrapping on this $S^2$ once.
These are sometime called ``long strings'' \ref\gks{
A. Giveon, D. Kutasov, and N. Seiberg,
``Comments on string theory on $AdS_3$,''
Adv.\ Theor.\ Math.\ Phys.\ 2 (1998) 733;
{\tt hep-th/9806194}.}, which are related to
the long strings in the spectrum of
the $SL(2,R)$ WZW model. To
evaluate effects of the instanton, it is useful to use  global
coordinates in $H_3$.
As we discussed earlier, the worldsheet action \largerhoaction\
grows exponentially large toward the boundary $\rho \rightarrow
\infty$ unless the worldsheet obeys the holomorphicity condition
 \holcondition.
For a holomorphic worldsheet,
the action grows linearly as $S \sim 2 k\rho$ for large $\rho$.
The effect is of the order $e^{-2 k\rho}$, which is
indeed nonperturbative if we identify $k \sim {1\over g^2}$
where $g$ is the coupling constant on the worldsheet.
These worldsheet instanton effects are similar to the ones
which appear in the computation of the Yukawa coupling of
the type II string compactification, where the
instantons wrap topologically
non-trivial 2-cycles in a Calabi-Yau 3-fold.
In our case, however, the $S^2$ is contractible in $H_3$.
In fact, the instanton action $\sim 2 k\rho$ is not
a topological invariant, but it depends on the size
$\rho$ of the worldsheet.
Thus the instanton configuration is not topologically stable,
and it is continuously connected to the vacuum.\foot{
In several respects, these instantons are similar to instantons in
ordinary Yang Mills theory in four dimensions. In this latter case
their action depends logarithmically on the size of the instanton
(analogous to $e^{-\rho_0}$ in our case) and if we are in a given
theta vacuum, the instanton can dissolve into the vacuum. }
Without additional effects, the factor $e^{-2 k\rho}$
tends to suppress large instantons.

\ifig\nontrivial{If ${\rm Re}(j) > {k-1 \over 2}$, the
worldsheet for the two point function grows uniformly on $S^2$
toward the boundary.}
{\epsfxsize2in\epsfbox{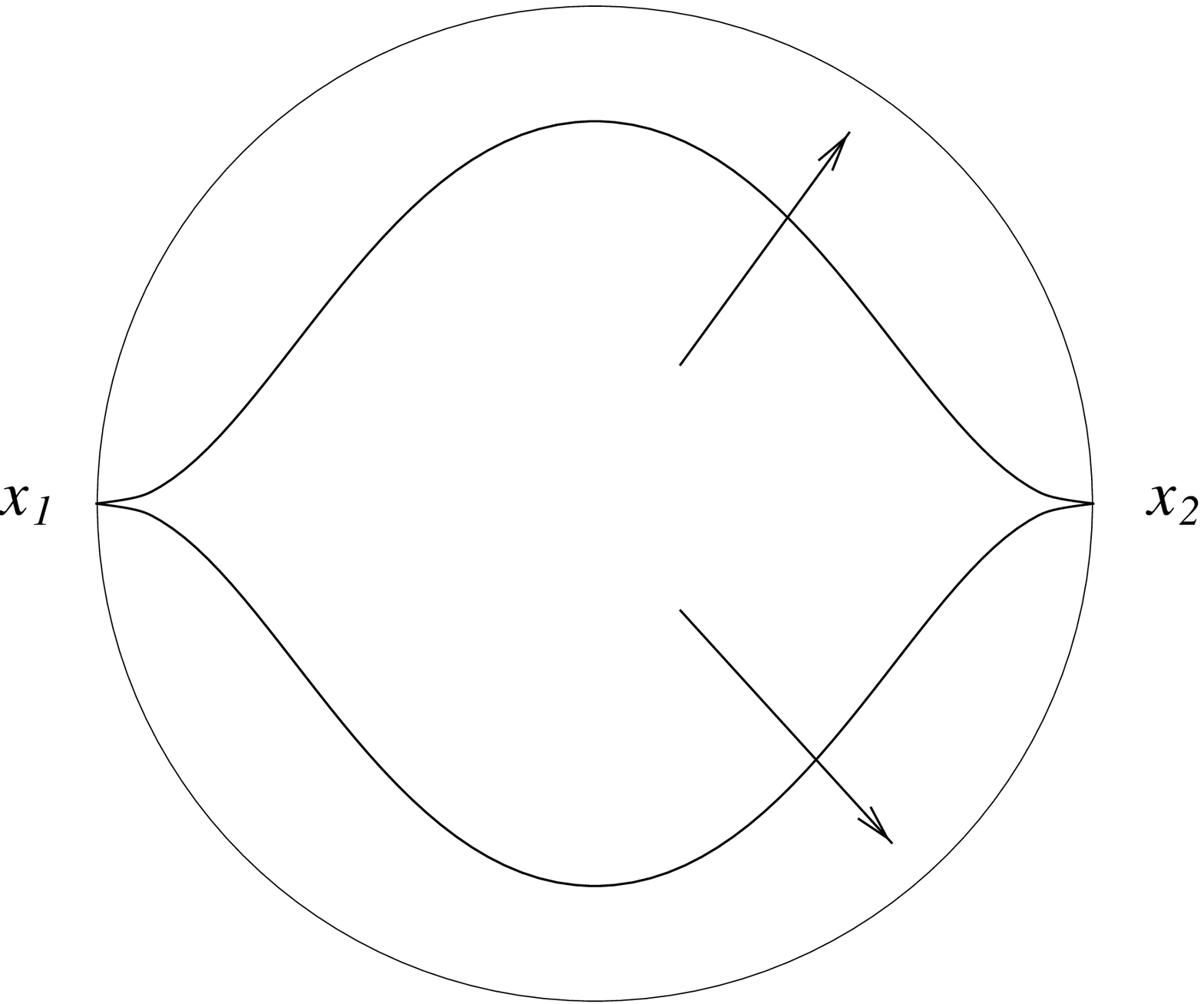}}

This observation can be used to explain the poles
in the two point function at $2j \sim k$ in the
following way. As we noted, depending on whether
$\gamma(z) = x$ or $\gamma(z) \neq x$,
the vertex operator behaves as $\Phi_j(z,x) \sim e^{2j\phi}$
or $\sim e^{-2j\phi}$ for
large $\phi$.
On the worldsheet $S^2$ with the two vertex
operators inserted, one can always
find a holomorphic map such that $\gamma(z_i)=x_i$
($i=1,2$). In fact there is a one-complex parameter
family of instantons, generated by dilatation
and rotation which keep fixed the two points, and
the integral over the family is responsible for the
delta function $\delta(j-j')$
in the two point function. On such instantons,
the vertex operator is evaluated as
$\Phi_j(z_i,x_i) = e^{2j\phi}$ in  Poincare coordinates.
In the global coordinates, it behaves
as $\Phi_j \sim e^{2j\rho}$ for large $\rho$. Therefore,
in the two point function,
the integral over the zero mode $\rho_0$ of the instanton
size is of the form
\eqn\zeromode{ \int d\rho_0 e^{-2 k \rho_0 } e^{ 2j\rho_0} e^{2j\rho_0},
}
where the first factor is the instanton action, and the last two
factors come from the vertex operator insertions.
We see that the integral \zeromode\ converges at large $\rho_0$
only for $j<\khalf $
(the exact answer \defofbtes\
is finite only for $j<{k-1\over 2}$).\foot{
In principle, we expect the computation in \zeromode\ to give
us only the  leading order in $k$ behavior.
By being a bit more careful
about the integral over quadratic fluctuations, we can see that
the amplitude can be
better approximated as
$\int d\rho_0 e^{2 \rho_0} e^{-2 (k-2) \rho_0 } e^{ 2(j-1)\rho_0}
 e^{2(j-1)\rho_0}$, where the first factor comes from the measure of the
$\rho_0$ integral, the shift in $k$ comes from the determinants and
the shift in $j$ comes from the integral over $\gamma,\bar \gamma$.
This gives the exact bound $j < {k-1 \over 2}$. }
Thus the instanton effect explains the origin of the singularity
as due to the non-compact direction in field
space which can be explored with finite cost in the action.
Since this divergence
is coming from the large $\rho$ region, it does not matter that the
instanton is not topologically stable in the full space of the
worldsheet fields. What is important is that the large $\rho$ region gives
a dominant contribution to the functional integral.
We can therefore say that this divergence is an IR effect in the target
space. It is interesting that the divergence is not localized
on the worldsheet and therefore cannot be considered as an UV
effect there. The standard lore about the
correspondence between IR effects in the target space
and UV effects on the worldsheet does not hold in this case.

Thus we have shown that there is a divergence for
${\rm Re}(j) \geq  {k-1\over 2}$ due to  large
worldsheet instantons.
In the analytic regularization, the divergence is converted into
a pole at $j =  {k-1\over 2}$. Of course
the formula \teschnertwo\ is precisely the result of such
analytic continuation.
These poles were also discussed in \gklast\ in
the context of the $SL(2,R)/U(1)$ coset
model using a dual
description \zam .
Similarly, by considering an instanton which wraps $n$-times
the $S^2$, we can explain the pole at $2j \sim nk$ in
the two point function.


\subsec{Singularities in the three point function}

Three point function \teschnerthree\
has various poles which come from the poles in $G(j)$ \gpolesfirst .
One finds that $C(j_1,j_2,j_3)$ has poles at
\eqn\cpoles{\eqalign{
j =& n + m(k-2), ~~~~-(n+1)-(m+1)(k-2),~~~~
(n,m=0,1,2,\cdots), \cr
&  {\rm where}~~~
 j = 1-j_1-j_2-j_3, ~~j_3-j_1-j_2,~~j_2-j_3-j_1,~~
{\rm or}~~j_1-j_2-j_3.}}
Our first task is to understand the origin of these singularities
from the point of view of the $SL(2,C)/SU(2)$ sigma model
on the worldsheet.

Let us first consider the poles at
\eqn\morepoles{ j_3-j_1-j_2 = n,~~(n=0,1,2,\cdots).}
Here we use the standard large $k$ approximation treating $\phi$ and
$\gamma , \bar \gamma$ as constant on the worldsheet (this is the point
particle approximation).
The vertex operator \operators\ goes like $e^{ 2 j \phi}$
at $\gamma = x$ and
it decays like $e^{-2j \phi}$ for $\gamma \not = x$.
When $j_3 > j_1+j_2$, a divergence in the
three point amplitude arises from the integral region
where $\gamma = x_3$ (and therefore $\gamma \neq x_1, x_2$)
so that $\Phi_{j_3}(x_3) \sim e^{2j_3\phi}$
and $\Phi_{j_1}, \Phi_{j_2} \sim e^{-2j_1\phi}, e^{-2j_2\phi}$.
The integral over $\phi$ then takes the form
\eqn\div{\int d\phi
e^{2( j_3 - j_1 - j_2)\phi},
}
where the measure factor $e^{2\phi}$ is cancelled by the
integral over $\gamma, \bar{\gamma}$.
The amplitude is divergent for $j_3\geq j_1+j_2$,
and  analytic regularization gives a  pole at
$j_3 = j_1 + j_2$. This explains the pole with $n=0$
in \morepoles.
To reproduce the other poles with $n=1,2,\cdots$,
we just have to expand $\Phi_{j_3}(z_3,x_3)$
in powers of $|\gamma(z_3)-x_3|^2$ and repeat the above
exercise. Thus we have interpreted the poles \morepoles\
in the exact expression \threecoeff\ from the point of
view of the worldsheet theory.
There are also poles when $(j_2-j_3-j_1)$
and $(j_1-j_2-j_3)$ are non-negative integers
and they are explained in a similar way.
In section 3, we will discuss how these divergences
are dealt with in string theory.
We will see that these are very analogous to poles in the $S$ matrix
in the flat space computation.

The other poles
in \cpoles\ can be explained by the worldsheet instanton
effects.
Since one can always find a holomorphic map
from the worldsheet to the target space such that
$\gamma(z_i) = x_i$ ($i=1,2,3$), the worldsheet instanton
can grow large whenever ${\rm Re}(j_1+j_2+j_3)$ exceeds $\sim k$.
This explains the first pole in \cpoles\
with $(n,m)=(0,1)$. As in the case of the two point function,
this divergence is non-local in target space.
 The remaining poles in \cpoles\
can be interpreted in similar ways.


\subsec{Singularities in four point functions}

Let us now move on to the four point function.
By worldsheet conformal invariance and target space isometries, it
depends non-trivially only on the cross ratios of $z_i$'s and $x_i$'s
($i=1,\cdots,4$),
\eqn\crossratios{ z = { (z_1-z_2)(z_4-z_3)
\over (z_1-z_3)(z_4-z_2)},~~~~~~~~  x = { (x_1-x_2)(x_4-x_3)
\over (x_1-x_3)(x_4-x_2) }.}
For special values of  $j_i$, the labels of the four operators,
 the dependence of the four point
function of $z$ and $x$ can be determined by  differential
equations. These values of $j_i$  are outside the range which
leads to physical operators in the string theory.

For  generic values of $j_i$,
one very useful piece of information is
that it obeys the Knizhnik-Zamolodchikov (KZ) equation, which follows
from the Sugawara construction of the stress tensor \sugawara.
The idea is to compute $\langle T(w) \Phi_{j_1}(z_1, x_1)
\cdots \Phi_{j_4}(z_4,x_4)\rangle$ in two different ways.
One is to convert $T(z)$ into derivatives with respect to
$z_i$'s using the conformal Ward identity. Another is to
use \sugawara\ to express $T(z)$ in terms of the currents
$J^a$ and to turn them into differential operators on
$x$ by the $SL(2,C)$ Ward identities \slaction.
Combining these two expressions together
and going over to the cross ratios \crossratios,
one finds \teschnersecond\
that the four point function ${\cal F}_{SL(2)}=
\langle \Phi_{j_1} \cdots \Phi_{j_4} \rangle$ obeys
\eqn\kzequation{
    {\partial\over \partial z} {\cal F}_{SL(2)}
=  {1 \over k-2} \left({P \over z}
+ {Q \over z-1} \right) {\cal F}_{SL(2)},}
where $P$ and $Q$ are differential operators with
respect to $x$ defined by
\eqn\pandq{
 \eqalign{ P  = & x^2 (x-1) {\partial^2 \over \partial x^2}
 + \left( (-\kappa + 1) x^2 - 2j_1 - 2j_2x(1-x) \right) {\partial
\over \partial x}
- \cr
& - 2\kappa j_2 x - 2j_1 j_2, \cr
Q =& -(1-x)^2 x {\partial^2\over \partial x^2} +
\left( (\kappa - 1) (1-x)^2 + 2j_3 (1-x)
   + 2j_2 x(1-x) \right) {\partial \over \partial x} - \cr
& - 2\kappa j_2 (1-x) - 2j_2 j_3,}}
with
\eqn\whatkappa{
 \kappa =  j_4 -j_1 - j_2 - j_3  .}

Because of the factor $z^{-1}$ and $(z-1)^{-1}$ in the right-hand
side of the KZ equation \kzequation , the amplitude
${\cal F}_{SL(2)}(z,x)$ has singularities at $z=0, 1$ and $\infty$.
Such singularities are familiar in conformal field theory
and appear when locations of two operators coincide on the
worldsheet. This leads to the operator product expansion,
which will be discussed extensively in section 4.

Quite unexpectedly,
the equation also implies a singularity at $z=x$.
This is because the coefficients
on front of $\partial^2/\partial x^2$ in $P$ and $Q$ cancel with each
other at $z=x$. Substituting the ansatz ${\cal F}_{SL(2)}
\sim (z-x)^\delta$ into \kzequation\ and solving the
equation to the leading order in $(z-x)$, the exponent
$\delta$ is determined as
\eqn\whatexponent{\delta = 0 ~~{\rm or}~~~k-j_1-j_2-j_3-j_4.}
The solution with $\delta=0$ is regular at $z=x$.
However, as we will see in section 4,  monodromy
invariance of the amplitude
${\cal F}_{SL(2)}$ around
$z=0,1,\infty$ as well as around $z=x$ requires
that we include the other solution with $\delta
=k-j_1-j_2-j_3-j_4$. Therefore
${\cal F}_{SL(2)}$ has to have a singularity of the form
\eqn\sing{{\cal F}_{SL(2)} \sim |z-x|^{2(k-j_1-j_2-j_3-j_4)}.}
Here we combined holomorphic and anti-holomorphic
parts so that the amplitude is monodromy invariant around
$z=x$.

The presence of the singularity at $z=x$
is very surprising from the point of
view of the worldsheet theory since this is a point in the middle
of moduli space. In a standard conformal field theory,
amplitudes become singular only at boundaries of moduli spaces.
A very closely related divergence
appears in the
one loop diagram \second .\foot{In \second, we considered
the finite temperature situation where we periodically
identify the target space Euclidean time, and computed
a partition function on a worldsheet torus. We found that,
in addition to the divergence at the  boundary
of the worldsheet moduli space
$\tau \rightarrow i \infty$, there are singularities
when $\tau$ is related to the periodicity of the
target space Euclidean time. These singularities are
interpreted as due to worldsheet instantons from the
worldsheet torus to the finite temperature target space
($i.e$ the Euclidean black hole in $AdS_3$).}
The interpretation of this singularity is again associated with
instanton effects. In the case of the four point function,
worldsheet instantons can grow
large if and only if $z =x$ since there has to be a holomorphic
map from the worldsheet to the boundary $S^2$ of the target
space such that $\gamma(z_i) = x_i$ ($i=1,\cdots,4$).
Such a map exists only when the worldsheet modulus $z$
coincides with the target space modulus $x$. The instanton
approximation also explains the value of $\delta$ in the following way.
If $z$ is not equal to $x$ but close to it,
there is a harmonic map $(\theta,\phi)$ for which
\eqn\nearlyhol{
\int (\partial \theta -i \sin\theta \partial \phi)
(\bar{\partial} \theta + i \sin \bar{\theta} \partial \phi)
\sim |z-x|^2.}
 We can then insert this into \largerhoaction\
to estimate the action for large $\rho$ as,
\eqn\approxiaction{
  S \sim 2k \rho_0 + \alpha e^{2\rho_0} |z-x|^2,}
for some positive constant $\alpha$. Here we only
show the dependence on the zero mode $\rho_0$ of $\rho$.
The functional integral for the four point function is then
approximated as
\eqn\action{\int d\rho_0 e^{2\rho_0}
 e^{ - 2 (k-2) \rho_0  + \alpha |x-z|^2 e^{2 \rho_0} } e^{ 2 \sum_i (j_i-1)
\rho_0} \sim |z-x|^{2(k-j_1-j_2-j_3-j_4)}.}
reproducing the singularity \sing .
This is  related to the remark
in \sw\ that the dynamics of long strings is approximated
by the Liouville theory --- here $|x-z|^2$ plays the role of  the
cosmological constant.
By a simple extension of this argument, we expect that
$n$ point amplitudes have
singularities when the worldsheet moduli coincide with the
target space moduli.  For $n > 4$, there can also be
singularities when a subset of the worldsheet moduli
coincides with a subset of the target space moduli.
In this case, only the corresponding part of the worldsheet
grows large.

\subsec{Correlation functions of spectral flowed states}

So far, we have discussed some general properties of
(analytically continued)
correlation functions of the operators \operators\ in
the $SL(2,C)/SU(2)$ model, and we have explained
the origin of various singularities in the correlation
functions. It turns out that there are other non-normalizable
 operators we will need to consider for the string theory application.

%

The operators $\Phi_j$ and their descendents by
the $SL(2,C)$ current algebra are not
the only operators we will be interested in.
The current generators $J^a(z)$ act on $\Phi_j$ as
\slaction , which means that
$\Phi_j$ and their analytic continuations
also  obey the conditions
\eqn\standard{J^\pm_n |\Phi\rangle = 0,~~
J^3_n |\Phi\rangle =0, ~~~(n=1,2,\cdots).}
These lead to the conventional representations of the
current algebra. In WZW models based on compact Lie groups,
these are all the operators we need
to consider; other operators are just current algebra descendents of
these. In the $SL(2,R)$ WZW model, there are other states
one needs to take into account. These are states in spectral flowed
representations of the types described in
\hwsdisc\ and \hwscont .
Correspondingly, there are
non-normalizable operators in the $SL(2,C)/SU(2)$ model
that are different from the ones obtained by analytic continuation
of $\Phi_j$.
In fact, by taking worldsheet OPE of operators of the form
\operators , which obey \standard ,
we can produce operators which are not in the conventional
representations obeying \slaction .
For example, we shall see in detail in section 5.2
 that we can construct an operator which generates
 spectral flow from the operator in
\operators\ with $j= k/2 $; the
spectral flowed representations
are generated by the worldsheet OPEs with this operator.

In the remainder of this section  we will argue from a semiclassical
point of view that these are natural operators to consider.
In particular we will build operators that are non-normalizable, but
such that their ``non-normalizability'' is concentrated at a point
$x$ on target space.

To formulate the problem, let us consider a vertex operator
$\Psi_j(z_0,x_0)$ defined so that it imposes the boundary condition,
\eqn\wsbc{
 \eqalign{ \phi(z) & \sim -{j\over k} \log|z-z_0|^2 ,\cr
           \gamma(z) & \sim x_0 + o(|z-z_0|^{2j/k}).}}
The reason that the subleading term in the second line
of \wsbc\ has to be smaller than $|z-z_0|^{2j/k}$
will become clear below. We will also show that,
when ${1\over 2} < {\rm Re}(j) < {k-1\over 2}$,
the operator $\Psi_j$ coincides to the operator $\Phi_j$.
What happens when $j$ is outside of this range?
Let us express $j$ as $j = \tilde{j}
+ {k \over 2} w$ with ${1\over 2} < {\rm Re}(\tilde{j})
< {k-1\over 2}$.
The semi-classical analysis that  follows
shows that the operator $\Psi_j$ defined by
\wsbc\ is identified as $\Phi^w_{\tilde{j}}$,
which is defined by acting the $w$ amount of the
spectral flow on $\Phi_{\tilde j}$.
In the semi-classical approximation, the spin $\tilde{j}$
will actually be found to be in the range $0 <
{\rm Re}(\tilde{j}) < {k\over 2}$. In the
exact computation, this becomes ${1\over 2} < {\rm Re}(\tilde{j})
< {k-1\over 2}$.

To explain this, let us consider the two point function
of the vertex operators
$\Psi_j$ at $(z,x) = (0,0)$ and $(\infty, \infty)$. We consider the
case when $j$ is real. It was shown in
\deboer\
that a general solution to the classical equation
of motion for \lagrangian\ is given by
\eqn\generalsol{
\eqalign{ \phi & = \rho(z) + \bar{\rho}(\bar{z}) +
\log(1 + b(z)\bar{b}(\bar{z})) , \cr
 \gamma & = a(z) + {e^{-2\rho(z)} \bar{b}(\bar{z})
\over 1 + b(z)\bar{b}(\bar{z})}, \cr
\bar{\gamma} & = \bar{a}(\bar{z}) + {e^{-2\bar{\rho}(\bar{z})} b(z)
\over 1 + b(z)\bar{b}(\bar{z})}, }}
for some holomorphic functions $\rho, a, b$ of $z$.
The simplest solution obeying the boundary conditions
\wsbc\ is
\eqn\simplest{ \eqalign{\phi & = - {j \over k} \log|z|^2, \cr
   \gamma & = 0.}}
This solutions corresponds to $\rho = -{j\over k} \log z$
and $a=b=0$ in \generalsol.
This clearly satisfies the boundary conditions at $z=0$.
To see that it also obeys the boundary conditions at
$z=\infty$, we use
the inversion of Poincare coordinates as,
\eqn\inversion{
\eqalign{e^{\phi'}&= e^{-\phi}\left(1 + e^{2\phi} |\gamma|^2
\right), \cr
 \gamma' & = - {e^{2\phi}\bar{\gamma}\over
1 + e^{2\phi} |\gamma|^2},\cr
\bar{\gamma}' & = - {e^{2\phi}\gamma\over
1 + e^{2\phi} |\gamma|^2}.}}
Note that, at $\phi \rightarrow \infty$, this corresponds
to the inversion $\gamma' = - 1/\gamma$ of the complex coordinates
on $S^2$. We then find
\eqn\simplesttwo{\eqalign{ \phi' & = - {j \over k} \log|z'|^2, \cr
                  \gamma' & = 0,}}
where $z'$ is the worldsheet coordinate appropriate near
$z=\infty$,
\eqn\wsinversion{ z' = - {1 \over z}.}
Thus the solution \simplest\ obeys the boundary conditions
both at $z=0$ and $\infty$.
This solution describes a cylindrical worldsheet of zero radius,
connecting $x=0$ and $\infty$.

Now let us examine what type of perturbations are allowed to
this solution. The simplest ones are of the form,
\eqn\perturb{\eqalign{
\phi & = -{j\over k} \log|z|^2,\cr
\gamma & = \epsilon z^n,}}
for small $\epsilon$.
We claim that this deformation corresponds
to the action of the current algebra generator $J_n^+$
on the solution \simplest. To see this, we note that
the point $g$ in the coset $SL(2,C)/SU(2)$ is parameterized
by the coordinates $(\phi, \gamma, \bar{\gamma})$ as
\eqn\groupelement{
g = \left( \matrix{ e^{-\phi} + \gamma \bar{\gamma}e^\phi &
e^\phi \gamma \cr
e^\phi \bar{\gamma} & e^\phi\cr} \right) ,}
and the action of $J_n^+$ is given by
\eqn\jnaction{
 J_n^+: ~g \rightarrow g + \left(\matrix{0 & \epsilon z^n \cr
0 & 0 \cr}\right) g
 .}
One can easily verify that \jnaction\
indeed maps \simplest\ to \perturb.

One should ask whether this
perturbation is normalizable or not.
The norm of worldsheet fluctuations
is defined using the target space metric as\foot{
Here the worldsheet metric is set to
$|z|^{-2} dzd\bar z$, which is
appropriate when the worldsheet is an infinite cylinder,
since we will use this computation to idenfity the state
corresponding to the vertex operator $\Psi_j$.}
\eqn\norm{
||(\delta\phi, \delta \gamma, \delta \bar{\gamma})||^2
 = \int {d^2 z \over |z|^2}
    \left( \delta\phi ^2 + e^{2\phi} \delta\gamma
 \delta\bar{\gamma}\right).}
Therefore the perturbation \perturb\ is normalizable
(at small $z$) if
\eqn\normalizable{
 n = w+1, w+2, w+3, \cdots,}
and non-normalizable if
\eqn\nonnorm{
 n = w, w-1, w-2, \cdots.}
Normalizable perturbations should be integrated out
when we perform the functional integral over
the worldsheet and therefore do not change
the boundary conditions. This explains why
we require that the subleading term in the
second line of \wsbc\ has to be smaller than
$|z-z_0|^{2j/k}$ since any perturbation equal to or
greater than is non-normalizable.
Non-normalizable perturbations change boundary
conditions and correspond to inserting different
operators on the worldsheet. Since these perturbations
correspond to the action of $J_n^+$ on the worldsheet
as in \jnaction, one can say that the vertex operator
$\Psi_j$ is annihilated by $J_n^+$ which generates
normalizable perturbations, $i.e.$,
\eqn\hwcond{
  J_n^+ \Psi_j = 0, ~~~ n = w+1, w+2, w+3, \cdots.}

One can repeat this analysis for the action of $J_n^-$.
This gives a perturbed solution of the form,
\eqn\anotherperturb{\eqalign{
  \phi& = -{j\over k} \log|z|^2, \cr
  \gamma & = \epsilon |z|^{4j\over k} \bar{z}^n.}}
A similar analysis shows that this perturbation is
normalizable\foot{Here we assume $2j/k$ is not
an integer. See the discussion below.} for
\eqn\anothernorm{
 n = -w, -w+1, -w+2, \cdots,}
and is non-normalizable for
\eqn\anothernon{
 n=-w-1, -w, -w+1, \cdots.}
This means $\Psi_j$ is annihilated by $J_n^-$ as
\eqn\anotherhw{
  J_n^- \Psi_j = 0, ~~~ n = -w, -w+1, -w+2, \cdots.}

Combining \hwcond\ and \anotherhw , we find that
$\Psi_j$ corresponds to the highest weight state
of a discrete representation with with $w$ amount
of spectral flow. By evaluating $J^3$ for
the solution \simplest, one finds that it carries
the $J^3$ charge $j$. According to the rule of the
spectral flow \introflow, this means that the Casimir
of the representation before the spectral flow
is given by $-\tilde{j}(\tilde{j}-1)$ where
$\tilde{j} = j - {k \over 2} w$.

Something special must happen when $2j/k$ is an integer
since the amount $w$ of spectral flow jumps there.
What happens is that the solution \perturb\
with $n=w$ coincides with the solution \anotherperturb\
with $n=-w$ and both are non-normalizable. This means
that we have a new type of state, not annihilated by
$J_{-w}^-$ and $J_w^+$. It is in the continuous representation
with $w$ amount of spectral flow. The fact that the
two solutions coincide means that there is a new solution.
In fact, when $2j/k = w$, there is a new solution,
\eqn\newsol{\eqalign{
  \phi & = -{w\over 2} \log|z|^2, \cr
   \gamma& = \epsilon z^w \log|z|^2 .}}
One can think of $\epsilon$ as the radial momentum carried by the long
string. This is a Euclidean version of the phenomenon
discussed in section 3 of \first\ in the context of string
theory in the Lorentzian $AdS_3$.

Here we have explained how to define the vertex operators
$U_{j}(z,x)$ for the spectral flowed representations. In
section 5, we will give exact  expressions for correlation functions
of these operators.

\newsec{ Spacetime interpretation of the singularities in two and three
point functions }

In the previous section, we have discussed properties of non-normalizable
operators in the $SL(2,C)/SU(2)$ model in general. In this section, we
will discuss which subset of those operators we will consider as
physical operators.
The physical theory we have in mind is string theory on
$H_3 \times {\cal M}$ where ${\cal M}$ is a compact target space
represented by some standard
unitary CFT. We will interpret
singularities in the amplitudes discussed in the previous section
from the point of view of this string theory. According to the
$AdS$/CFT conjecture, the string theory is dual to a boundary conformal field
theory on $S^2$, which we denote from now on as ``BCFT'' \magoo .
The observables of BCFT are local normalizable
operators on the boundary of the target space.
In  string perturbation theory, they are represented on
the worldsheet by products of non-normalizable operators
in the $SL(2,C)/SU(2)$ theory times normalizable operators
in the unitary CFT for ${\cal M}$.\foot{ More precisely, these are
what ``single particle'' operators correspond to \magoo .}
The same is true in flat space computations where  normalizable plane
waves in the target space theory are represented by
non-normalizable operators of the form $e^{p_L^0 X^0_E}$
times normalizable
operators in the internal CFT
in the Euclidean worldsheet theory.
(In this discussion we have neglected the tachyon which could be
both normalizable in the Euclidean worldsheet theory
and physical in the string theory; it is projected out
in superstring.)
Notice that in the $AdS_3$ case the Euclidean worldsheet computations
are directly related to the Euclidean BCFT computations. We will
concentrate on the interpretation of the string theory as a Euclidean
field theory. The rotation to Lorentzian target space then should
be the standard rotation of the BCFT to Lorentzian signature.

\subsec{ Two point functions}

Our first task will be to pick a set of non-normalizable operators
in the $SL(2,C)/SU(2)$ model which we will use to
construct physical observables. The BCFT is a unitary CFT
and it makes sense to analytically continue the target space
to $AdS_3$ with Lorentzian signature metric. By the standard
state-operator correspondence, a normalizable operator of the
BCFT corresponds to a normalizable state in the BCFT in
the Lorentzian signature space. In the regime where perturbative
string theory is applicable, these states correspond to
single particle states and multi particle states of
string theory on  Lorentzian $AdS_3 \times {\cal M}$.
The worldsheet theory of the string on the Lorentzian $AdS_3$
is the $SL(2,R)$ WZW model. The spectrum of the WZW model
was proposed in \first\ based on a semi-classical
analysis, and the proposal was verified by an exact computation
of one-loop free energy in \second. The spectrum of
the WZW model is decomposed into a sum of irreducible
representations of the $SL(2,R) \times SL(2,R)$ current
algebra. As shown in \hilbertspace,
it contains the discrete representations ${\cal D}_j^0 \otimes
{\cal D}_j^0$ with ${1\over 2} < j < {k-1\over 2}$ and their spectral
flow images corresponding to short strings, and
the continuous representations ${\cal C}_{j,\alpha}^0
\otimes {\cal C}_{j,\alpha}^0$ with
$j = {1\over 2} + i s$ for real $s$, and their
spectral flow images corresponding to long
strings.

Going back to the $SL(2,C)/SU(2)$ model,
these states correspond to the operators with
\eqn\cont{
j = \half + is
}
or
\eqn\range{ \half < j < {k-1\over 2},
}
and all their spectral flow images. Though operators with
$j=\half +is$ are normalizable in the worldsheet theory,
 their spectral flow images are not.
After imposing the physical state conditions, the only
states with $j=\half + is$ and $w=0$ are tachyons. Neglecting
the tachyons, we see that all the operators of interest are
non-normalizable on the worldsheet theory.

Though we just argued for the conditions \cont\ and \range\
 on the basis of the Lorentzian
theory, we can make a similar argument purely in the Euclidean
theory. The operators on the worldsheet that can correspond to
good spacetime BCFT operators are those non-normalizable operators
for which the divergences are localized at the point $x$ which
we want to interpret as the point where the BCFT operator is
inserted. In other words, the ``non-normalizability'' of the
worldsheet vertex operator should be concentrated around $\gamma \sim x$
in target space. Indeed we saw in section 2 that if $j$ is outside
the range \range , there are divergences on the worldsheet theory
that are not localized on the boundary $S^2$. For $j<\half$
these can  be interpreted in the usual point particle limit, while
for $j>{k - 1 \over 2}$ the divergences came from worldsheet instantons.
Let us clarify the target space implication of the latter.
Instead of the analytic regularization, one may choose to compute
the two point function by using an explicit target space
cut-off regularization
by limiting the functional integral to be over $\rho < \rho_0$
for some large value of $\rho_0$. From the discussion in section
2.2, we expect that, if $j$ is in the range \range ,
the worldsheet never grows large for generic $\gamma$ and all cutoff
dependence is localized near $\gamma \sim x_i$.
On the other hand,
if $j$ exceeds the upper bound, the amplitude
depends on $\rho_0$ since the
worldsheet can grow larger than $\rho_0$. So the large $\rho_0$
dominates the functional integral and the two point function
is divergent.
The divergence is  not localized in target space around the
points $x_i$, but it is spread all over target space, as
shown in \nontrivial .
Thus the two point function of the operator $\Phi_j$
in the Euclidean theory makes sense as a local operator in $x, \bar x$
only in the region \range . One can nevertheless
define the worldsheet operators $\Phi_j$ outside the range \range  , via
analytic continuation. In this definition, one
 is implicitly subtracting
counter-terms that are not localized in $x$. From the point of view
of the worldsheet theory, there seems to be nothing wrong with this.
In fact,  operators outside
\range\ are very useful for
computing correlation functions on the worldsheet
\refs{\teschnerone,\zam,\teschnersecond}.
However, worldsheet operators outside \range\ cannot be identified
with local operators in the BCFT.
In fact, our analysis in section 2.5 shows that, if
one tries to exceed the upper bound in the Euclidean
worldsheet theory, one is naturally lead  to  operators
in spectral flowed representations.

The coefficient $B(j)$ in the worldsheet two point function
\teschnertwo\ given by  \defofbtes\
is well-defined and positive
for $j$ belonging to the range \range . In the string theory computation, we need to
divide the amplitude by the volume of the conformal group $V_{conf}$
which keeps the two points fixed. It cancels the divergence coming
from evaluating the delta function $\delta(j-j')$ in \teschnertwo\ at $j=j'$,
leaving a finite answer, as explained  in \ks .\foot{
The target space two point function receives contribution
from the internal CFT. Since this part is diagonal in the conformal
weight, the physical state condition for the short string
implies that we need to set $j=j'$ to have non-zero two point
function in the target space.}
The cancellation of the two divergent factors
requires some care since it
may leave some finite $j$ dependent factor.
In section 5, we will given a heuristic argument
to say that the target
space two point function comes with an extra factor of $(2j-1)$
as
\eqn\targetwo{
  \eqalign{& \langle \Phi_j(x_1) \Phi_j(x_2) \rangle_{target}\cr
&=
{1\over V_{conf}}
 \langle \Phi_j(x_1; z_1=0) \Phi_{j}(x_2; z_2=0)
)\rangle_{worldsheet}\cr
&={(2j-1) B(j) \over |x_{12}|^{4j}}.}}
A more rigorous derivation of the extra factor $(2j-1)$
is given in Appendix A, where we show that this is required
by the consistency with the target space Ward identities.
The target space two point function \targetwo\
is also well-behaved in the physical range \range .

We can also compute target space two point functions for
any spectral flowed states; this will be done explicitly in
section 5.
We will find that they are all regular and have
positive definite two point functions in the region \range .
The extra factor $(2j-1)$ mentioned
in the above paragraph is generalized to
$|2j-1+(k-2)w|$ when $w\neq 0$.

As shown in \first,
the spectral flowed
continuous states ($j={1\over 2} + is$)
correspond to operators in the BCFT which have
continuous dimensions.
We conclude from this that the BCFT has a non-compact
target space (at least it is non-compact in the leading order
in string perturbation theory).
The nature of this non-compactness was discussed in \sw\ in the
case of $AdS_3 \times S^3 \times M_4$ where $M_4 = K3 $ or $T^4$.
In these cases, BCFT is the supersymmetric sigma model whose
target space is the moduli space of the Yang-Mills instantons
on $M_4$. The non-compact directions are related to the limits
where instantons become small. The relation between
existence of the continuous spectrum in CFT and the
non-compact directions in its target space is familiar
in the case  of a free non-compact scalar. We would like
to stress that there is nothing particularly non-local about
the sigma model with continuous spectrum. The operators corresponding
to these states are local on the space where the BCFT is defined.
This is for the same reason that
an operator like $e^{ikX}$ is local on the worldsheet
of the free scalar field $X(z,\bar{z})$.
In our case, these operators are the spectral flowed versions
of $j=\half + is$. Their target space two point function
 will be computed
in section 5 and are given by
\eqn\correcontf{\eqalign{
&\langle \hat \Phi^{j w}_{J \bar J}(x_1)  \hat
\Phi^{j' w}_{J{\bar J}}(x_2) \rangle_{target}
\cr
&\sim \left[ \delta(s+s')  + \delta( s- s')
{ \pi B( j )\over \gamma(2j)}
{\Gamma(j -  \khalf w + J)
\over \Gamma( 1- j-  \khalf w + J  )}
{ \Gamma(j +  \khalf w - \bar J )\over
\Gamma( 1 - j +  \khalf w - \bar J) } \right]
{1 \over x_{12}^{ 2 J} {\bar x}_{12}^{2 \bar J} }.}
}
Here $j={1\over 2} + is, j'={1\over 2} + is'$,
the spacetime conformal weight of the operator $J$ is
given by
\eqn\jvalue{
J =  {k \over 4}w + { 1 \over w} \left(  { s^2 + {1\over 4}  \over k-2}
+ h -1 \right),
}
and $h$ is the conformal weight of the vertex operator
for the internal CFT,
whose two point function we assumed to be unit normalized in \correcontf .
Equation \jvalue\ comes from the $L_0 =1$ condition.
Unlike the case of short strings,
the two point function of long strings does not receive
the extra factor of $|2j-1+(k-2)w|$ when we transform
the worldsheet computation into the target space computation.
Note that the term multiplying the second delta function in \correcontf\
is a pure phase as,
\eqn\purephase{
\eqalign{
e^{i\delta(s)} & \equiv
{ \pi B( j ) \over \gamma(2j)}
{\Gamma(j -  \khalf w + J)\over
\Gamma( 1- j-  \khalf w + J  )}{
 \Gamma(j +  \khalf w - \bar J )
\over \Gamma( 1 - j +  \khalf w - \bar J) } \cr
& = - \nu^{-2is} {\Gamma\left(-{2is\over k-2}\right)
\over \Gamma\left(+{2is \over k-2}\right)}
{\Gamma\left(-2is\right) \over
\Gamma\left(+2is\right)}
{\Gamma\left({1\over 2} + is - {k \over 2} w + J\right)
\over \Gamma\left({1 \over 2} -is -{k\over 2} w + J \right)}
{\Gamma\left({1 \over 1} + is + {k \over 2}w - \bar J \right)
\over \Gamma\left({1 \over 2} -is + {k \over 2}w
+ \bar j \right)} .}}
 This  is the
phase shift that occurs when a long string comes from the boundary
and back, which in terms of the BCFT is a small instanton becoming large
and small again.

In summary, the singularities in the two point function are
outside of the range \range\ of our choice of operators.
Now we can ask whether this choice removes all singularities in all
$n$ point functions. The answer is {\it no}. We will see however that the
singularities can be interpreted physically and we will give a
prescription for how to deal with them. In other words, all
singularities that appear are interpretable in the BCFT.

\subsec{Three and four  point functions}

The three point function has poles at $j_3 = j_1 + j_2 + n$ and
their permutations in $j_1, j_2, j_3$.
These poles are standard and easy to understand. They appear in
all $AdS_{d+1}$/CFT$_d$  examples \lref\freedman{
D.~Z.~Freedman, S.~D.~Mathur, A.~Matusis and L.~Rastelli,
``Correlation functions in the CFT$_d$/AdS$_{d+1}$ correspondence,''
Nucl.\ Phys.\ B 546 (1999) 96;
{\tt hep-th/9804058};
``Comments on 4 point functions in the CFT/$AdS$
correspondence,'' Phys. Lett. B452 (1999) 61;
{\tt hep-th/9808006}.} \lref\liu{
H. Liu, ``Scattering in anti-de Sitter space and
operator product expansion,'' Phys.
Rev. D60 (1999) 106005; {\tt hep-th/9811152}.}
\refs{\freedman, \liu}.
These poles are due to mixing with two particle states. The string
perturbation expansion in $AdS$ corresponds to a $1/N$ expansion in the
boundary theory. To leading order in $1/N$ the operators are single
particles and multiparticle states in $AdS$. When we compute $1/N$ corrections
these operators can mix. The mixing is  generically small, of
order $1/N$,  but
if two operators have the same conformal weight at leading order in $1/N$,
then the mixing can be of order one, since we are doing degenerate
perturbation theory. If $j_3 = j_1 + j_2 + n$, then we have
two operators with the same conformal weight,  namely $O_{j_3}$
and $: \partial^n_{12}O_{j_1}
O_{j_2}:$ where the $O_{j_1}$ are single
particle operators and the derivatives act on both
operators
in such a way that the result is a primary operator
under $SL(2,R) \times SL(2,R)$ symmetry at large $N$.
These two operators can mix in the subleading order in $1/N$,
and the divergence in the three point function is cancelled
if we take into account this mixing effect.

\ifig\figone{ Here we see the change in behavior of the semiclassical
geodesics when we go from the case of $j_3 < j_1+j_2$ in (a) to
the case $j_3 > j_1 +j_2 $ in (b).
}
{\epsfxsize2.8in\epsfbox{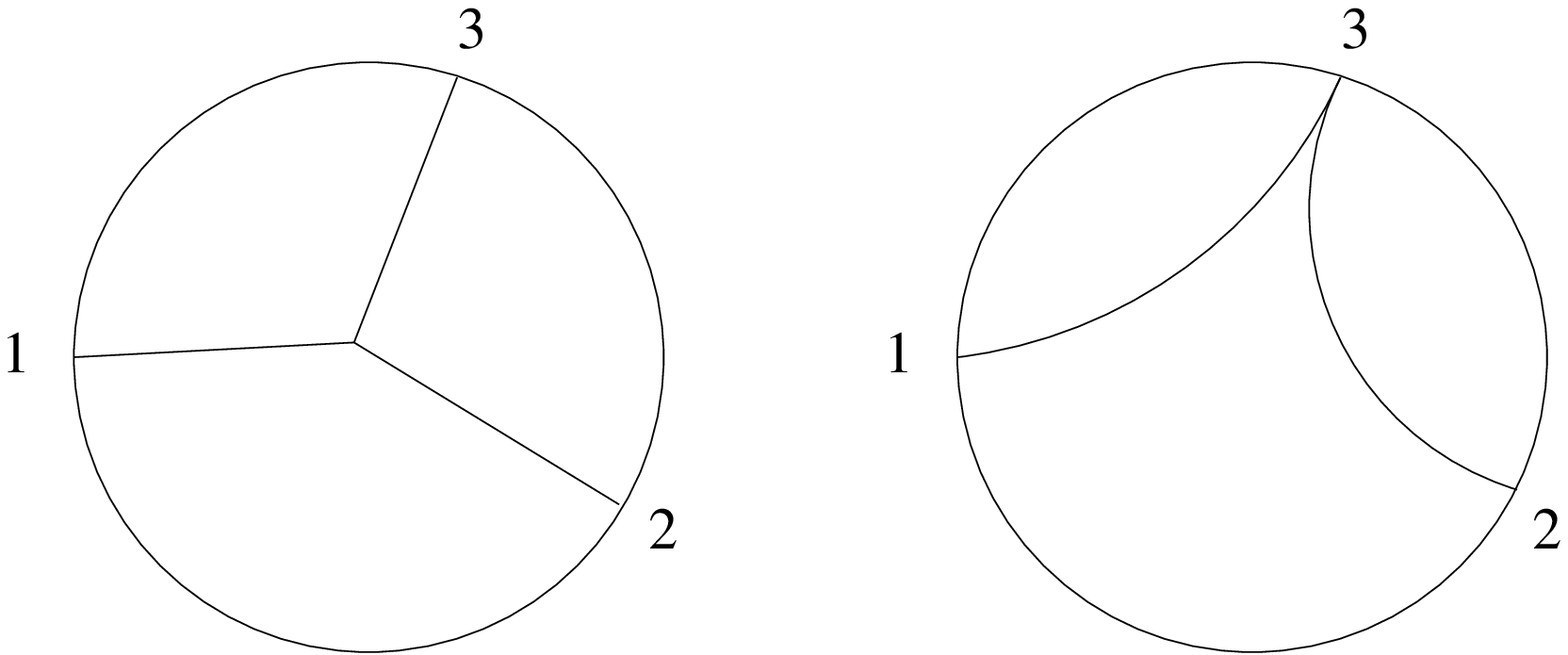}}

It is instructive to
 look at the semiclassical description of this divergence.
Suppose $j_i$ are large, then
correlation functions can be computed by considering a particle of masses
proportional to $j_i$ with  trajectories
that intersect the boundary at the points where the operator are
inserted \wittensurfaces .
If $j_3 < j_1 + j_2$ (and the same holds for other permutations of 123)
the dominant contribution is given in \figone a. On the other hand
if $j_3> j_1+j_2$ we cannot find a configuration where the interaction
point is in the interior, the interaction point moves to the boundary as
shown in \figone b. In the semiclassical approximation
$n>0$ becomes a continuous variable. If  we quantize the
fields we see that $n$ is an integer.
This divergence is eliminated by  a redefinition of the operator
$O_{j_3}$ which mixes the single particle operator with the two particle
operator.
 That a local redefinition of the operator can cancel the
divergence is related to the fact that the divergence is coming from
the region close to the point on the boundary where $O_{j_3}$ is inserted.

The three point function has also a divergence at $\sum_i j_i = k$.
This divergence appears even if all $j_i$'s are within the
range \range.
{}From the point of view of the worldsheet theory,
this divergence is due to instanton corrections as
we saw in section 2. This means that the divergence appears because
the worldsheet can be very close the the boundary of $AdS$ with no cost
in action, see figure 3.

\ifig\big{ Change in behavior of the classical worldsheet when
$\sum j_i < k$ in (a) to the case where $\sum j_i > k$ in (b).
In (b) the worldsheet is driven to the boundary of $AdS$.
}
{\epsfxsize2.5in\epsfbox{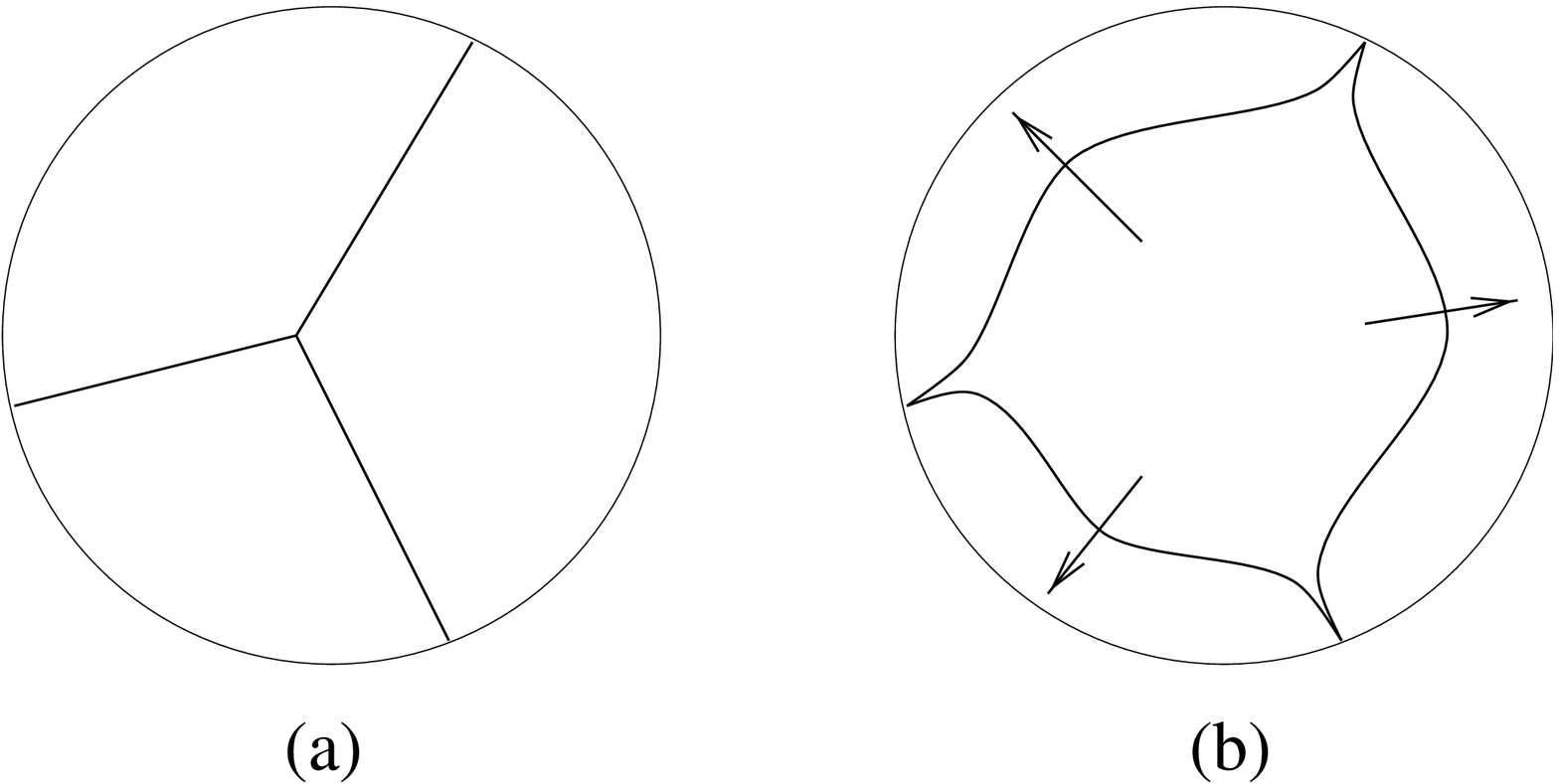}}

One  might  think that this is a  non-local
effect in the BCFT.  In order to remove it it seems
that we need counter-terms which are spread all over the $S^2$ where
the BCFT is defined. We would like to
propose a different interpretation.
The BCFT is local and  this divergence is simply
due to the non-compactness of
the BCFT target space. In other words, we do not remove the divergence.
The origin of this divergence, which we will explain
below, suggests that only three point functions with
$\sum j_i < k$ make sense in the BCFT.

In order to clarify this point, let us consider a quantum mechanical
example which has a phenomenon very analogous to what we are dealing with.
Suppose that we have the quantum mechanics of a particle in a potential
well, where the potential asymptotes to zero at infinity and it is
negative at the origin, so that the system has a normalizable ground
state wave-function $\psi(x) $ which for large $x$ decays as
$\psi(x) \sim e^{- \kappa x/2}$. In this system we can consider
operators of the form $O_\lambda = e^{ \lambda x}$. The expectation
value on the ground state of the product of two
 of these operators is well-defined as long as $\lambda < \kappa/2 $.
If we insert several operators and we try to compute
$\langle \psi |O_{\lambda_1}(t_1) \cdots O_{\lambda_n}(t_n) |\psi \rangle
$, we will find that we can only do the computation if
$\sum \lambda_i <  \kappa $.
In other words, there seems to be a non-local constraint (in time)
on the operators whose correlators we can compute. The theory is
perfectly local, and the divergence is just an IR effect in the target
space coming from the non-compactness of the target space.
It is a well-known fact that there are operators in quantum
mechanics have a domain and a range, and some operators can take
a state out of the Hilbert space.\foot{
As a trivial example, consider a harmonic oscillator and imagine
the Hamiltonian acting on the state $|\psi\rangle
= \sum { 1 \over n} |n \rangle $.}
In this quantum mechanical model, there are other operators, of
the form $e^{ikx}$ for example, which are perfectly well-defined
for any real value of $k$.

Our BCFT is very similar to this quantum mechanical example.
It has a normalizable ground state, and the vacuum
expectation value of discrete states
with $\sum j_i > k$ is not defined.
There are other operators, the ones in the spectral flowed
 continuous representations,
which we can consider. These operators are analogous to
$e^{ikx}$ in the quantum mechanical model.
Correlation functions of these are well-defined without any
additional constraint.
Notice that the target space BCFT has a normalizable ground state,
despite having a non-compact target space since there is
a gap between the ground state energy and the threshold
where the continuum starts due to the non compactness.

Based on these observations, we claim that correlation functions of
discrete states are only well-defined if
$\sum j_i < k$. The expression \teschnerthree\ can be defined for
$\sum j_i > k$ by analytic continuation,
but it does not make physical sense as it does not
represent a well-defined computation in the BCFT. In
order to define it we need to add counter-terms that are spread over
$S^2$ in target space.

For the four point function, the singularity at $z=x$ \sing\ implies,
after integrating over $z$, that there is a divergence in the
four point function if $\sum_i j_i  = k +1 $.\foot{
In an $n$ point function we expect a divergence when $\sum j_i =
k + n-3$.}
So a four point function makes sense only for $\sum j_i < k+1$.
It might be possible to
extend the four point function to $\sum j_i > k+1$
by analytic continuation, but it does not have any immediate
physical interpretation.

Note that we are not saying that there is a bound on the spacetime
conformal weight of the operators we add. By using spectral flowed
operators, we can compute correlation functions of operators whose
conformal weights are as high as we like.
These spectral flowed operators were defined
precisely to avoid the divergences associated to long strings.

In order to stress once again that these divergences have nothing to
do with non-local behavior of the BCFT, let us consider an example
with $N=4 $ SYM in $d=4$ where this feature appears. Consider
  $N=4 $ SYM on $T^2 \times S^1\times$(time)
 with anti-periodic boundary conditions
for the fermions on $S^1$ and periodic on $T^2$. The supergravity
solution describing the ground state of this
theory was described in \wittenthermal , it is the near extremal
black three brane doubly Wick rotated. It is a non-singular geometry
with topology $T^2 \times D^2$, where $D^2$ is a disk whose boundary is
the $S^1$ (we concentrate on the geometry of the radial direction
and the three spatial dimensions of the brane). This theory has
finite energy excitations which correspond to placing  a D3 brane
 at some radial position and winding on $T^2 \times S^1$. These
are analogous to the long strings described above. They  lead to
divergences in computations of certain correlation functions,
in a very similar fashion to how long strings lead to divergences
in the $AdS_3$ case. These divergences come from the fact that
there is a Coulomb branch  that we can explore
with finite cost in energy.

Finally let us note that, both in the $AdS_3$ case and in the $N=4$ SYM
example we have given above, we can remove the non-compact direction
in field space by deforming the Lagrangian of the theory. In the
$AdS_3$ case we can add some RR fields fields, which in the BCFT has
the effect of making the target space compact.
In the $N=4$ example we can add mass terms for all scalar fields.

In $AdS_3$ with RR backgrounds the continuum states
 become discrete and we can compute the correlation functions
of any number of operators.
If we take the limit of RR fields going to zero, we will find
that  states with high conformal weight with
 $j> {k-1 \over 2} $ will lead to operators in the
$SL(2)/SU(2)$ model which are spectral flowed.
Similarly we expect that if we compute a three point function for
three discrete states with $\sum j_i <k$ the result will go over
smoothly to \teschnerthree\ as we take the RR fields to zero.
On the other hand there is no reason why the correlation function
of states with $\sum j_i >k$ should go over smoothly to
\teschnerthree\ when we remove the RR fields; in fact, we expect
that the correlation function diverges in the limit.

\newsec{Four point function }

In this section, we compute four point functions in
 target space by performing the integration over the
moduli space of the string worldsheet. A four point
amplitude depends non-trivially on the cross ratio $x$
of the four points on the boundary of $AdS_3$ where
the operators ${\cal O}_1, \cdots, {\cal O}_4$ are inserted.
In other words, we can use conformal
invariance to fix the operators as
\eqn\fourp{
 {\cal F}_{target}(x,\bar x)=
\langle {\cal O}_1(0){\cal O}_2(x) {\cal O}_3(1)
 {\cal O}_4(\infty) \rangle .
}
 Our
main objective is to derive the operator product
expansion by evaluating the small $x$ expansion of
${\cal F}_{target}$.
%
%
If the amplitude ${\cal F}_{target}(x,\bar x)$ in the BCFT
obeys the factorization condition,
we should be able to expand it for $|x| < 1$ in powers of
$x$ as
\eqn\expan{
 {\cal F}_{target}(x,\bar x) = \sum_{J,\bar J}
  x^{ J - J_1 - J_2} {\bar x}^{\bar J - \bar J_1 - \bar J_2}
 {\cal C}_{target}(J,\bar J),
}
where $(J, \bar J)$ are the target space conformal weights
and ${\cal C}_{target}(J , \bar J)$
is given in terms of three and two point functions as
\eqn\formofc{
 {\cal C}_{target}(J,\bar J) = \langle {\cal O}_1(0) {\cal O}_2(1)
{\cal O}_{J,\bar J}(\infty)\rangle
{1 \over
\langle {\cal O}_{J, \bar J}(\infty) {\cal O}_{J \bar J}(0) \rangle }
\langle {\cal O}_{J, \bar J}(0) {\cal O}_3(1) {\cal O}_4(\infty)\rangle
}
and $\{ {\cal O}_{J, \bar J} \}$ is a complete set of operators
in BCFT.

Before we start the detailed computation, let us summarize our
result. We will focus on the case when the operators
${\cal O}_1, \cdots, {\cal O}_4$ correspond to short strings
with $w=0$, $i.e.$, they correspond to states
in discrete representations ${\cal D}_j^0
\otimes {\cal D}_j^0$ of the current algebra $\widehat{SL}(2,R)
\times \widehat{SL}(2,R)$.
We find that, if their conformal weights $j_1,\cdots,j_4$
obey the inequalities,
\eqn\physcond{
j_1 + j_2 < {k+1\over 2} ~,~~~~~~~j_3 + j_4 < {k+1 \over 2},
}
the string amplitude \fourp\ can indeed be expanded in
powers of $x$ as
\formofc , and the intermediate states ${\cal O}_{J,\bar J}$
are either short strings with $w=0$ and in the range \range ,
 long strings with $w=1$,
or two particle states of short strings.
All other
physical states do not appear. In section 5, we
will show that this is  because the three point functions in
\formofc\ vanish for the other cases.
If \physcond\ is not obeyed, then there are terms
in the $x$ expansion that cannot be interpreted as coming from the
exchange of physical states. We explain at the end of this section
that this is due to  the non-compactness of the target space of BCFT,
and it is the physically correct behavior. For CFT's with
compact target spaces, the operator product expansion \formofc\
should always be valid. In our
case, we expect it to hold only if \physcond\ is obeyed.
Now we proceed to explain these statements in more detail.

\subsec{The four point function in the $SL(2,C)/SU(2)$ coset model}

Each spacetime operator is associated to a worldsheet
vertex operator
$ {\cal O}_i(x,\bar x)  \to  \int d^2 z \Phi_i( x, \bar x; z,\bar z;)$.
If we want to calculate the spacetime four point function ${\cal F}_{target}$,
we should calculate the four point function ${\cal F}_{worldsheet}$
of the corresponding
worldsheet vertex operators and integrate it over their positions.
Using worldsheet conformal invariance, we can fix the worldsheet
position of three of them, and the worldsheet correlator
depends only on the cross ratio $z$.
So we need to compute
\eqn\comput{
 {\cal F}_{target}(x,\bar x) =
\int d^2 z~ {\cal F}_{worldsheet}(z, \bar z; x , \bar x )
}
There are two factors that contribute to the worldsheet
correlation function as,
 \eqn\nnn{
{\cal F}_{worldsheet}(z,\bar z ; x , \bar x ) =
{\cal F}_{SL(2)}( z,\bar z ; x , \bar x) {\cal F}_{internal}(z,\bar
z),}
where ${\cal F}_{SL(2)}$ is the correlation function of
the $SL(2,C)/SU(2)$ coset model and ${\cal F}_{internal}$
is that of the internal CFT.\foot{In general ${\cal
F}_{worldsheet}$ could be a sum of such
products.}

A closed form expression of ${\cal F}_{SL(2)}$ is not known for generic
values of $j_1, \cdots , j_4$
 for the external states. We will use an expression
for it given in \teschnersecond, which involves an integral
over a continuous family of solutions to the KZ equation \kzequation.
Let us review the derivation.
The KZ equation \kzequation\ has an infinite number of solutions
reflecting the
fact that the Hilbert space of the $SL(2,C)/SU(2)$ model is
decomposed into infinitely many representations
of $\widehat{SL}(2,C)$. It turns out that there is
a unique combination of these solutions that satisfies
the factorization properties on the worldsheet, $i.e.$,
the $z$ expansion of the amplitude should be expressed
as a sum over normalizable states
when all four external
operators, labeled by $j_1,\cdots j_4$, are also normalizable
(or close enough to normalizable).  It was shown in
\gawedzki\
that the Hilbert space of the $SL(2,C)/SU(2)$ coset theory
is a sum of the representations with $j={1\over 2} + is$
($s$: real, $>0$) with the conformal weight $\Delta(j)$. Therefore
it is reasonable to expect that the four point function is
a sum of products of the conformal block ${\cal F}_j(z,x)$ of the form
\eqn\smallz{
{\cal F}_j(z,x) = z^{\Delta(j) - \Delta(j_1) - \Delta(j_2) } x^{j-j_1-j_2}
 \sum_{n=0}^\infty f_n(x) z^n.
}
Substituting this into the KZ equation, one finds
that $f_{0}(x)$ has to obey the hypergeometric equation in $x$
with two linearly independent solutions
\eqn\whatf{\eqalign{&
F(j-j_1+j_2, j+j_3-j_4, 2j; x) \cr & {\rm or}~~
   x^{1-2j} F(1-j-j_1+j_2, 1-j+j_3-j_4, 2-2j; x).}}
As we will
discuss below, we need  both solutions
to construct a monodromy invariant four point function.
Taking into account
the factor $x^{j-j_1-j_2}$ in \smallz, one sees that
the two solutions in \whatf\ are related to each other by
the reflection $j \rightarrow 1-j$, or $s \rightarrow -s$
if we write $j={1\over 2} + is$. Therefore, instead of
requiring $s > 0$ and use both solutions, we can allow
$s$ to be any real number and always pick the first
solution in \whatf.

It was shown by Teschner that,
for generic values of $j$, all other $f_n(x)$
($n=1,2,\cdots$) are determined iteratively by the KZ equation
once we fix $f_0(x)$ as the initial condition at $z \rightarrow 0$.
They take the form
\eqn\whatfn{
  f_n(x) = \sum_{m=-n}^\infty c_{nm} x^m.}
Therefore, by demanding that $f_0(x)$
is given by the first solution in \whatf, we can
uniquely determine
${\cal F}_j$ as a solution to the KZ equation.
Note that, unlike $j_1,\cdots , j_4$, the parameter $j$ does not
appear in the KZ equation
\kzequation, but it is used as a label of the solution of the KZ
equation whose small $z$ behavior is as in \smallz .

The full four point function ${\cal F}_{SL(2)}(z,x)$
is then given by the worldsheet factorization ansatz
\teschnersecond\ as
\eqn\expres{
{\cal F}_{SL(2)}(z,\bar z; x , \bar x ) =
\int_{{1\over 2} + iR}
 dj ~{\cal C}(j)
| \cf_{j}(z,\bar z; x,\bar x)|^2,}
where the normalization factor ${\cal C}(j)$ is given by
\eqn\whatc{
{\cal C}(j) =  C(j_1,j_2,j) { 1 \over B(j)} C(j,j_3,j_4)
}
where $C(j_1,j_2,j_3)$ and $B(j)$ are defined in \threecoeff\
and \defofbtes .
The integral is over $j={1\over 2} + is$ with $s \in R$.
As we mentioned, the $j$ integral covers both solutions \whatf\
because of the reflection symmetry $j \rightarrow 1-j$ of
the integration region. As shown in \teschnersecond , including both
solutions is necessary
in order for the four point function to be monodromy invariant
around $x=1$ and $\infty$.
In appendix B we argue that the integral
over $j$ in \expres\ is convergent.

\ifig\polesone{ The solid line indicates
the integration contour for \expres\ in the
$j$ complex plane. We highlighted the location of some poles
in $C(j)$. Here all external $j_i$ are of the form $j_i = \half + is_i$.
 There are similar poles with $j_1,j_2 \to j_3,j_4$, there
are also some other poles that will not be important for our purposes.
}{\epsfxsize3.5in\epsfbox{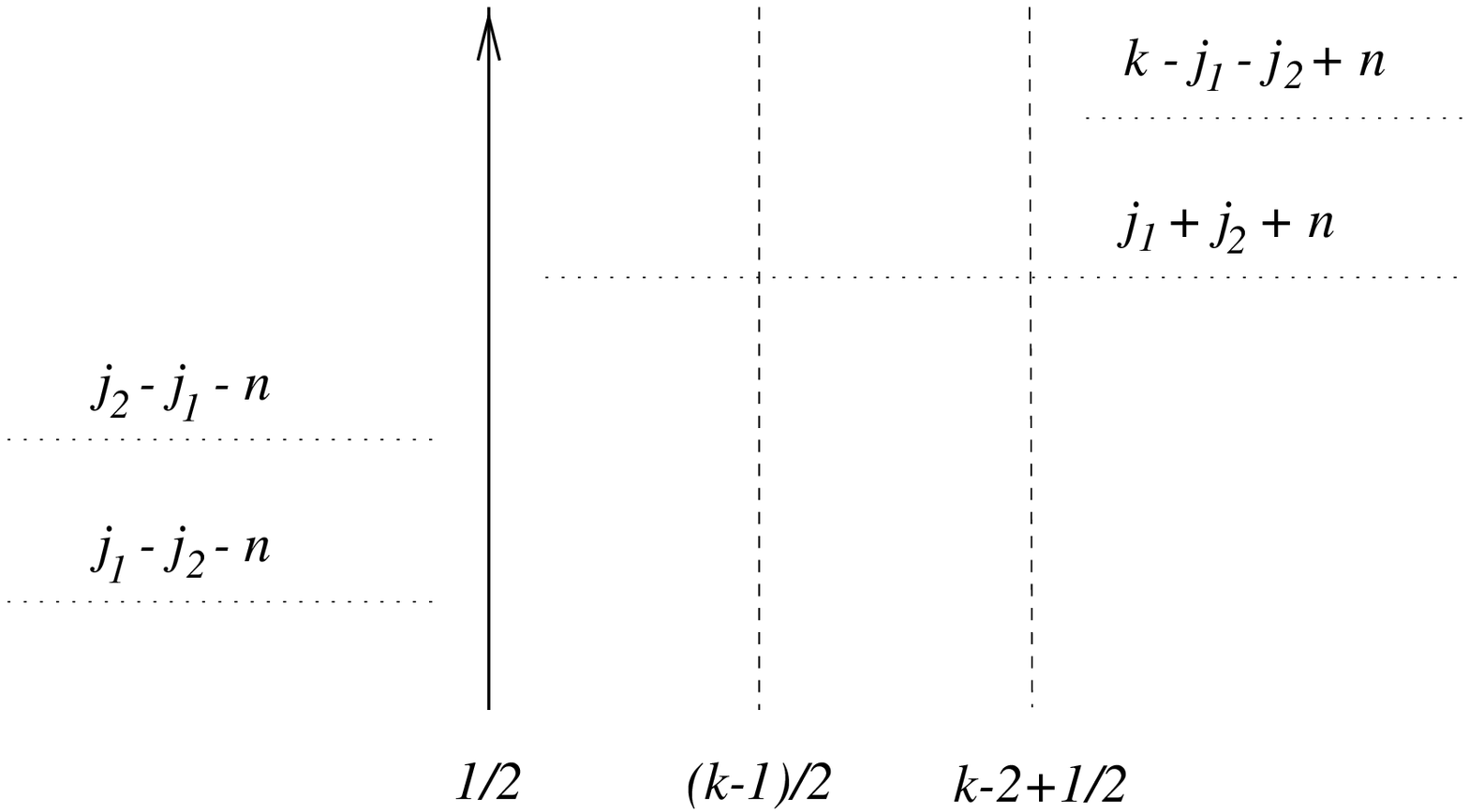}}

\ifig\polestwo{
The solid line indicates the integration contour after we analytically
continue \expres\ in the external $j_i$. Some  poles of the form
$|j_1-j_2| -n$  have crossed the integration contour so we should
include their residues. There are similar poles with $j_3,j_4$.
We separated the poles along the imaginary direction for
clarity, although they are all along the real axis when $j_i$ are real.}
{\epsfxsize3.5in\epsfbox{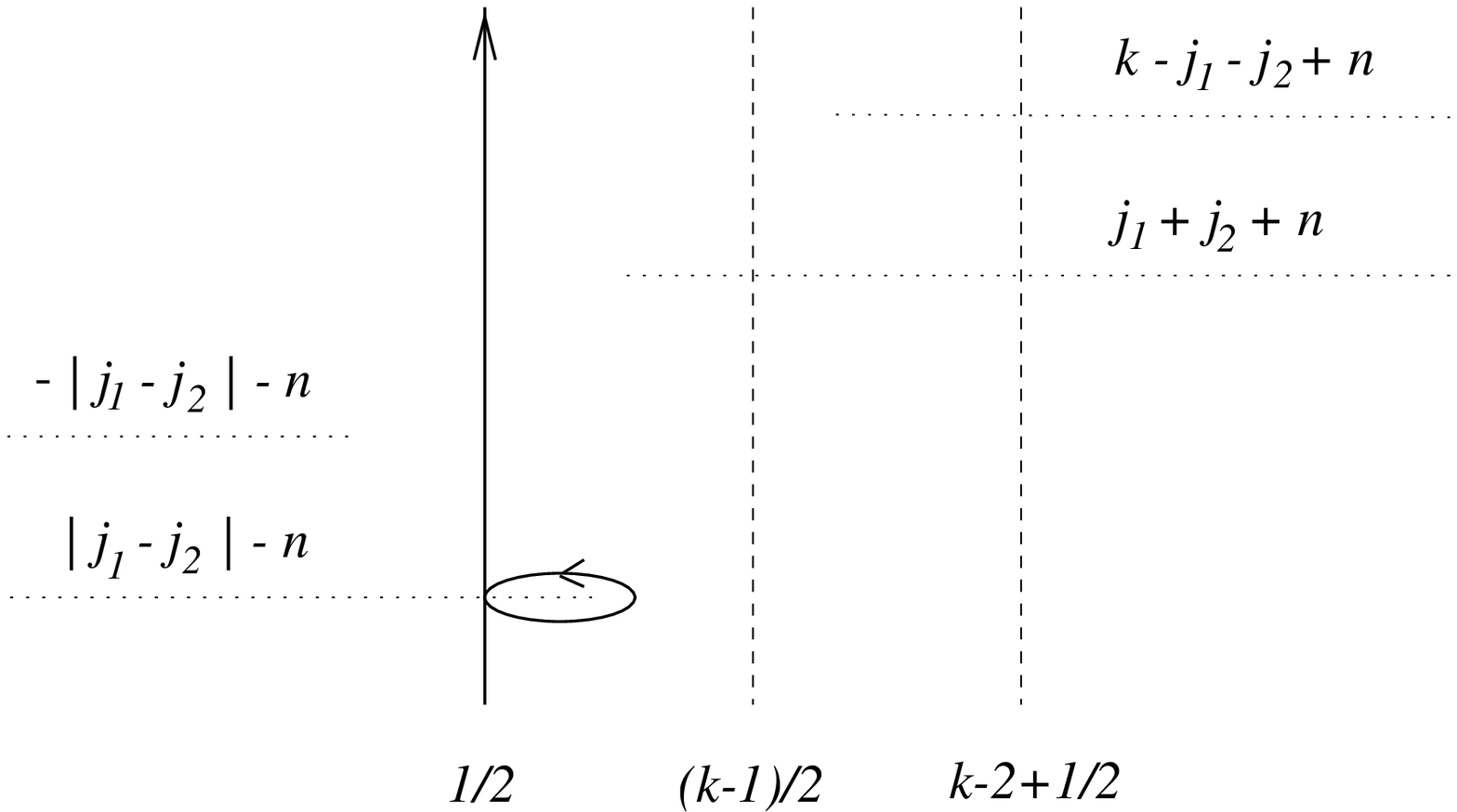}}

The expression \expres\ is valid
if all external labels $j_1, \cdots, j_4$
are close to the line $j = {1\over 2} + is$.
The expression for other values
of $j_1, \cdots j_4$ is defined by  analytic continuation.
When we do this, some poles in the integrand
 cross the integration contour.
The four point function is then
\expres\ plus the contribution of all poles
that have crossed the integration contour.
We need to know the pole structure of ${\cal C}(j)$ and
${\cal F}_j(z,x)$.
As we discussed in  earlier sections,
the  three point function $C(j_1,j_2,j)$  in
\whatc\ has poles at
\eqn\qpoles{
\eqalign{& j =  1-j_1-j_2- j_p, ~~~j_1+j_2 +j_p,~~~~ \pm(j_1-j_2) - j_p
\cr
&{\rm where}~~~j_p =  n + m(k-2), ~~-(n+1)-(m+1)(k-2),~~~~
(n,m\geq 0)
}}
To compute the correlation function of short strings
with $w=0$, we need to analytically
continue $j_1, \cdots, j_4$
from the line $j_i= {1\over 2} + is$ to the interval
${1\over 2} < j_i < {k-1 \over 2}$ on the real axis.
The poles that cross
the contour of the $j$ integral in \expres\ are of the form
\eqn\polecrossing{
 j = |j_1-j_2| -n,~~~~n=0,1,2, \cdots,}
with $j>{1\over 2}$.
There are similar poles in $C(j,j_3,j_4)$ at
\eqn\polecrossingtwo{
 j = |j_3-j_4| -n,~~~~n=0, 1 , 2 , \cdots.}
There are no poles in $B(j)^{-1}$
and ${\cal F}_j$ that cross the contour when we do
the analytic continuation. Therefore, after the analytic
continuation in $j_1, \cdots, j_4$, the correlation
function ${\cal F}_{SL(2)}$ is defined by the integral
\expres\ plus the contribution from the poles at \polecrossing\
and \polecrossingtwo . Stated in another way,
the contour of the $j$ integral is deformed from the line
$j={1\over 2} + is$ to avoid these poles. See \polestwo .

\ifig\polesthree{We shifted the integration contour to
$j = \khalf -\half + is$.
We picked up contributions from Poles$_1$ and Poles$_2$.
This figure represents the case when
$j_1+j_2 < \khalf$ and $j_3+j_4 > \khalf$.
}{\epsfxsize3.5in\epsfbox{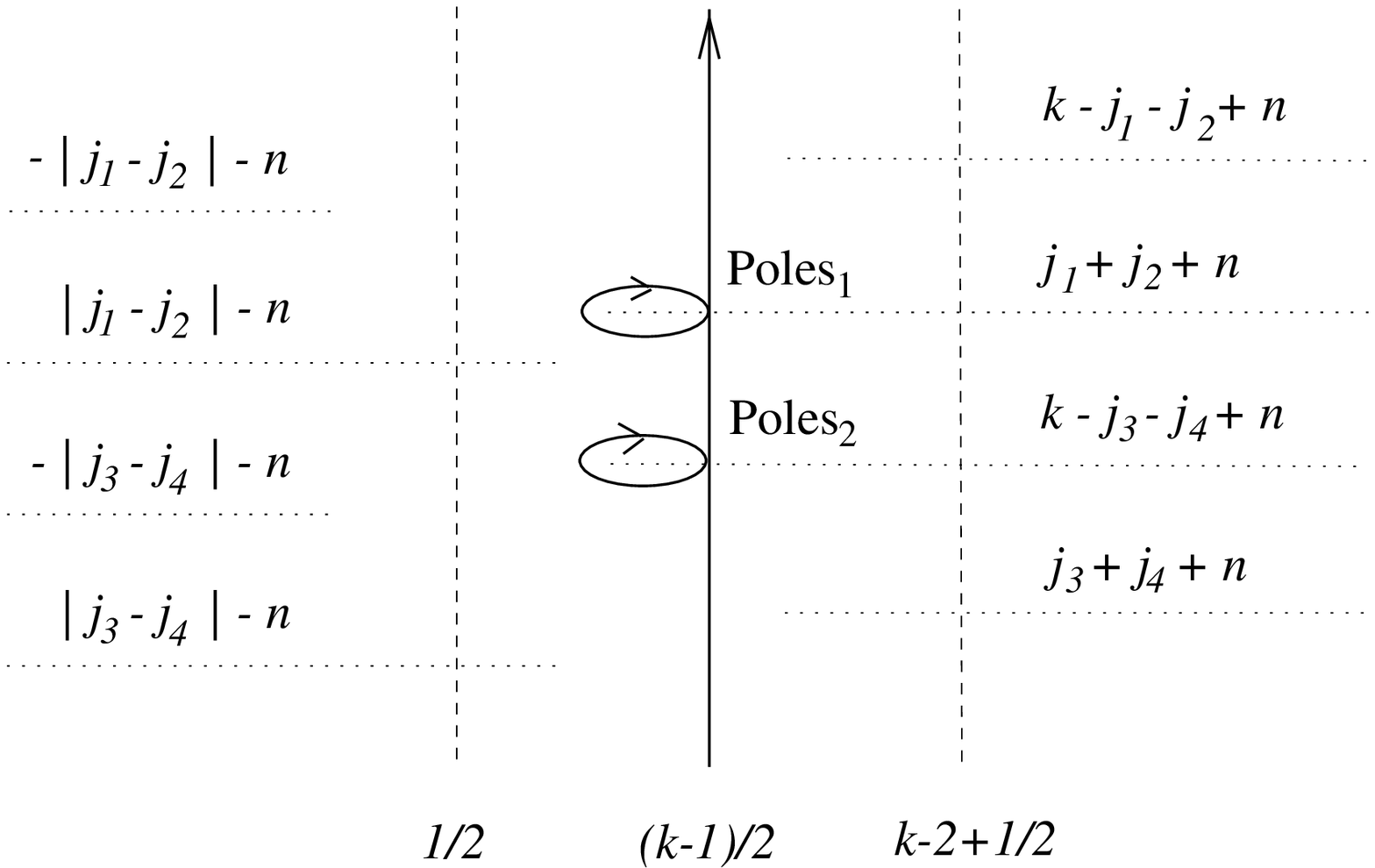}}

This completes the specification of ${\cal F}_{SL(2)}(z,x)$.
The next task is to multiply the factor ${\cal F}_{internal}(z,\bar{z})$
coming from the internal CFT and integrate the resulting
expression over the $z$ plane as in \comput .
We will find it useful to deform the contour of the $j$ integral.
We will deform the contour of the $j$ integration in \expres\
 within the region
\eqn\deformregion{
{1 \over 2} \leq Re~j \leq {k-1\over 2}.}
In this process, we will pick up poles in
${\cal C}(j)$
and ${\cal F}_j$, so it is useful to list them here.
Among the poles \qpoles\ in $C(j,j_1,j_2)$,
the relevant ones in the region \deformregion\
are of the form
\eqn\poles{
 \eqalign{{\rm Poles}_1:&~~j=j_1+j_2 + n, \cr
 {\rm Poles}_2:&~~j=k-j_1-j_2+n \cr
&~~~~n=0,1,2,\cdots .}}
Here we are assuming that $j_1, j_2$ are in the physical
range ${1\over 2} < j_1, j_2 < {k-1 \over 2}$. Note that
\deformregion\ imposes a constraint on allowed values of
 $n$ in \poles .
The poles \polecrossing\ are also in the
region \deformregion , but the contour of the
$j$ integral is defined to avoid these poles, as we discussed
in the previous paragraph, see \polesthree .
There are similar poles in $C(j,j_3,j_4)$ given
by exchanging $j_1,j_2 \rightarrow j_3,j_4$.
%
{}From \defofbtes\ we can see that $1/B(j)$
has no poles in the region \deformregion .

One may also ask if there is
pole coming from the conformal block ${\cal F}_j$. It turns out that there
is no such pole in the region \deformregion .
This has been shown in \teschnersecond\ using properties of
the Kac-Kazhdan determinant.
To see this explicitly, it is useful to rearrange the expansion
\smallz\ as
\eqn\regionone{
\cf_j(z,x) = x^{\Delta(j)-\Delta(j_1)-\Delta(j_2) +j-j_1-j_2 }
             ~ u^{\Delta(j)-\Delta(j_1)-\Delta(j_2)}
 \sum_{m=0}^\infty g_m(u) x^m,}
where $u = z/x$. This expansion will also be used in
the next subsection to evaluate the $z$ integral in the
region where $|z| < 1$. If we substitute this expansion
in the KZ equation, we find that the first term
$g_0(u)$ in the expansion should obey the hypergeometric
equation in $u$. The solution which agrees with the
initial condition \smallz\ for small $z$ is
\eqn\gzero{ g_0(u) = F(j_1+j_2-j,
j_3+j_4-j,k-2j;u).}
By looking at the standard formula for the Taylor expansion
of the hypergeometric function, one can check explicitly
that $g_0(u)$ has no poles in the region \deformregion.
Given  that $g_0(u)$ has no poles, we can prove inductively
that the same is true for all $g_m(u)$, $m \geq 1$. The
proof of this statement is given in Appendix B.

In the following subsections, we consider the case $|x| < 1$
and expand the expression \expres\ in powers
of $x$. We will then integrate it over $z$.
We will not impose a restriction on $z$ since we
must  integrate over $z$ to obtain the physical string amplitude.
We will divide the range of $z$ into two  regions:
$$ \eqalign{ &{\rm Region ~I} ~:~~  |z| < 1 \cr
             &{\rm Region ~II}~:~~  |z| > 1.}$$
Since ${\cal F}_{SL(2)}(z,x)$ has the singularity \sing\
at $z=x$, one may consider dividing the region I further into
two regions where $0<|z|<|x|$ and $|x|<|z|<1$, but it turns
out to be unnecessary to do so as we shall see below.

\subsec{Integral over the region I}

To integrate the four point function over the region I,
it is useful to define the variable $u = z/x$
and use the expansion \regionone.
We will mostly concentrate on the first term $g_0(u)$
of the expansion. As we mentioned,
the KZ equation implies that the first term $g_0(u)$
obeys the  hypergeometric
equation whose solutions are $F_j(u)$ and $F_{k-1-j}(u)$,
where $F_j$ is defined by,
\eqn\hypsol{\eqalign{ F_j(u) \equiv
& F(a,b,c;u)
\cr
a =&j_1+j_2-j ~,~~~b =j_3+j_4-j ~,~~~~~ c =k-2j.
}}
At $u=1$ these solutions
behave as  $ c_1 + c_2 (u-1)^{k-\sum j_i}$,
where the coefficients $c_1,c_2$ are both non-zero
for generic values of $j_1, \cdots, j_4$.
It is therefore clear that the first solution \hypsol\
on its own is not monodromy invariant
at $u=1$. For a given $j$, there is
a unique monodromy invariant combination given by\foot{
Note that $j$ is not complex conjugated in this expression.
In other words $ |a(j) x^{f(j)}|^2 \equiv  a^2(j) |x|^{2 f(j)}$.}
\eqn\monodinv{\eqalign{
 {G}_{j,0}(u,x)
= &  \left|x^{\Delta(j)-\Delta(j_1)-\Delta(j_2) +j-j_1-j_2 }
  u^{\Delta(j)-\Delta(j_1)-\Delta(j_2)}\right|^2 \times
\cr &~~~
\Big( |F_j(u)|^2 + \lambda |u^{1-c} F_{k-1-j}(u)|^2
\Big),}}
where
\eqn\whatlambda{
\lambda
 = - {\gamma(c)^2 \gamma(a-c+1) \gamma(b-c+1)
\over (1-c)^2 \gamma(a)\gamma(b)},
}
and  $\gamma(x)$ is given in \defofsmallgamma .
The subindex $0$ is there to remind us that we are examining
the first term in the $x$ expansion
in \regionone .
It is useful to note that we can write it as
\eqn\anothermonod{
{\cal C}(j) G_{j,0}(u) =
 {\cal C}(j) |{\cal F}_{j,0}(u,x)|^2 +
{\cal C}(k-1-j)|{\cal F}_{k-1-j,0}(u,x)|^2 ,}
where
\eqn\whatfzero{
{\cal F}_{j,0}(u,x) \equiv
x^{\Delta(j)-\Delta(j_1)-\Delta(j_2) +j-j_1-j_2 }
  u^{\Delta(j)-\Delta(j_1)-\Delta(j_2)} F_j(u),}
is the first term in the $x$ expansion of ${\cal F}_j$
in \regionone .
We can show \anothermonod\ by using the identities,
\eqn\useful{\eqalign{
 &C(k-1-j)  = \lambda {\cal C}(j), \cr
 &\Delta(k-1-j)+(k-1-j) =
 \Delta(j) + j, \cr
 &\Delta(k-1-j) = \Delta(j) + 1-c.}}

The problem with the monodromy invariant combination
\monodinv\ is that it does not satisfy the small $z$ expansion condition
\smallz\
because of the factor $u^{1-c}$
in the second term in the parenthesis.
On the other hand, the solution \hypsol\ satisfies the expansion
\smallz\
but is not monodromy invariant around $z=x$.
This puzzle  is resolved by performing the $j$ integral.
We can show that, after the $j$ integral, the amplitude \expres\
is monodromy invariant.
To see this, we need to deform the contour from $j={1\over 2} + is$
to $j={1\over 2} + is + {k-2\over 2}$, see \polesthree .
  The new contour is
such that, if it includes the point  $j$, it also includes
 the point $ k-1-j$.
Therefore we write the integral of the solution \hypsol\
as ${1\over 2}$ of the integral of the
monodromy invariant combination $G_{j,0}(u,x)$.
As we deform the contour, we pick up some residue contributions from
the poles at  \poles . It turns out that
each of those contributions is  monodromy
invariant by itself.
This can be seen by noting that, for the values of $j$  in \poles ,
the coefficient $\lambda$ in \monodinv\ vanishes.
More specifically, we find that the contributions from Poles$_1$ in \poles\
are non-singular at $u=1$,
while those of Poles$_2$ in \poles\ contain only the singular
solution at $u=1$, and therefore both are monodromy invariant
by themselves.
We can now express \expres\ in the manifestly
monodromy invariant form as
\eqn\monodromy{
 \eqalign{& \int_{{1\over 2} + iR}  dj {\cal C}(j) {\cal F}_j(z,x)= \cr
    & = \int_{{k-1\over 2}+ iR} {\cal C}(j) {\cal F}_j(z,x) +
 ({\rm contribution~from~Poles}_1~{\rm and~Poles}_2) \cr
 & = {1\over 2} \int_{{k-1\over 2} + iR} dj ~{\cal C}(j)~\Big[ G_{j,0}(u,x)
 + \cdots \Big] + ({\rm contribution~from~Poles}_1~{\rm and~Poles}_2) ,}}
where the dots represent  higher order terms
in the $x$ expansion. It is convenient to combine
the integrand into the monodromy invariant form
$G_{j,0}(u,x)$ given by \monodinv\ because, in the following,
 we will perform the $z$ integral
before the $j$ integral. (We will
be careful about justifying the exchange of the
$j$ integral and the $z$ integral by regularizing
the $z$ integral.)
 In conclusion, we have shown that
after integrating over $j$,  Teschner's expression \expres\
for the four point function is monodromy
invariant around $z=x$.

The contribution from Poles$_1$ is of the form
$x^{j-j_1-j_2} f(z,\bar{z})$ with $j=j_1+j_2+n$.
Since the integral of $f(z,\bar{z})$ times ${\cal F}_{internal}(z,\bar z)$
is independent of $x$, we conclude that the conformal
weight of the intermediate states is $J = j=j_1 + j_2 + n$.
These conformal weights  can be identified with the conformal
weights of two particle contributions. In other words, when we
compute the spacetime operator product expansion, the intermediate
operators could be two particle operators.
There can be  other contributions with these quantum numbers in the
intermediate channel which come from two disconnected
sphere diagrams in string perturbation theory.
The $z$ integral of this contribution contains
divergences at small $z$. They are canceled by another contribution
which will be discussed later.

If \physcond\ is satisfied, \monodromy\ does  not receive
any contributions from Poles$_2$ in \poles .

Before we perform the integral over the $z$ plane,
we need to multiply ${\cal F}_{SL(2)}(z,x)$
by a four point function ${\cal F}_{internal}(z, \bar z)$
of the internal CFT. In
region I, $i.e.$, $|z|<1$, we can expand ${\cal F}_{internal}$
as
\eqn\wsint{
{\cal F}_{internal}(z,\bar z) =
 \sum_{h,\bar h} z^{h-h_1-h_2} {\bar z}^{\bar h- h_1-h_2}
{\cal C}_{internal}(h,\bar h) ,}
where the coefficient is given by
\eqn\whatcint{
{\cal C}_{internal}(h,\bar h) = C_{internal}(h_1,h_2,h)
{1 \over B_{internal} (h \bar h) } C_{internal}(h,h_3,h_4),}
and $B$ and $C$  are given by the two and three
point functions of the internal CFT.

Now we are ready to integrate ${\cal F}_{worldsheet}
={\cal F}_{SL(2)}\times {\cal F}_{internal}$
over $z$ in  region I, namely
 over the region $|u| \leq |x|^{-1}$.
One problem is that this integral might diverge at $u=0$.
This would not be a problem if we were actually integrating
${\cal F}_{worldsheet}$ since
we can remove the divergence by analytic continuation,
which is the standard procedure in string theory
computation. The problem arises if we try to do
the $z$ integral before the $j$ integral in \expres\ since these
two integrals may not commute if there are divergences.
In fact, it is necessary to keep track of these possible
divergences and to be careful about the exchange of
the $z$ and $j$ integrals in order to recover the correct
pole structure. The two integrals commute if we regularize
the $z$ integral by introducing a cutoff $\epsilon$
and integrate over $\epsilon \leq |u| \leq |x|^{-1}$.
We will keep track of the $\epsilon$ dependence and
send $\epsilon \rightarrow 0$ after we perform the $j$ integral.
In practice, what we do is first integrate
over the whole $u$ plane and define the integral by analytic
continuation. We then subtract the contributions from
$|u| < \epsilon$ and $|x|^{-1} < |u|$.
If we use the same analytic continuation technique to evaluate
the integrals over these three regions,
the result after the subtraction of the two contributions
gives the regularized
integral over $\epsilon < |u| < |x|^{-1}$.

\bigskip
\noindent
(1) $\underline{{\rm Integral~over~the~whole~}u{\rm -plane}}$
\bigskip

Let us start with the integral over the whole $u$ plane.
\eqn\integral{\eqalign{
R_1 & \equiv \int dz^2 {\cal F}_{SL(2)} {\cal F}_{internal} \cr
&= \sum_{h \bar h} \int dj {\cal C}(j)
{\cal C}_{internal}(h,\bar h)  \sum_{m,\bar m=0}^\infty
 x^m{\bar x}^{\bar m}
 I^{h,\bar  h}_{j;m,\bar m}(x).}}
The first term in the $x$ expansion is given by
\eqn\initegraltwo{\eqalign{
I_{j;0, 0}^{h, \bar{h}}(x) = &  x^{\Delta(j) + h -1  +j-j_1-j_2 }
\bar{x}^{\Delta(j) + \bar{h} -1  +j-j_1-j_2 } \times \cr
&~~
{1\over 2} \int  d^2u u^{d -1}\bar{u}^{\bar{d}-1}
\Big( |F_j|^2 + \lambda |u^{1-c} F_{k-1-j}|^2\Big),}}
where
\eqn\whath{
d=  \Delta(j) + h -1,~~~~~  \bar{d} = \Delta(j) + \bar{h} -1.}
This integral can be done using the formula (C.1) in Appendix C.
We  find
\eqn\whatrone{
\eqalign{
 R_1 =
{\cal C}_{internal}(h , \bar h)
& \int_{{k-1\over 2} + i R} dj ~{\cal C}(j) ~x^{d +j- j_1 - j_2}
\bar{x}^{\bar{d}+j-j_1-j_2} \times \cr
&~~~~ {\pi \over 2}
{\Gamma(d)\Gamma(a-\bar{d}) \Gamma(b-\bar{d})
\Gamma(1-c+d)
\over \Gamma(1-\bar{d})\Gamma(1-a+d)\Gamma(1-b+d)
\Gamma(c-\bar{d})}
{\gamma(c)\over \gamma(a)\gamma(b)} + \cdots.}}
where the dots indicates terms with higher integer powers of $x, \bar x$.
By looking at the powers of $x, \bar x $, we can read off the
 conformal weight of the intermediate states as
\eqn\longstring{
J = d  +j =\Delta(j) + j+ h-1 = {k \over 4} + {s^2 + {1 \over 4} \over k-2}
 + h-1,}
where $j = {k-1 \over 2}  + is$ and a similar expression for $\bar J$ obtained
by replacing $h \to \bar h$ in \longstring .
We conclude that \whatrone\ represents the
contribution of  long strings with winding number $w=1$ in
the intermediate channel. In section 5, we will show
that the coefficient  in \whatrone\
 is precisely  what we expect from  \expan\  and \formofc .

The subleading terms $I_{j;m,\bar m}^{h,\bar h}$ with $(m,\bar m)
\not = (0,0) $
in the $x$ expansion \integral,
represented by the the dots in \whatrone ,
are identified as coming from the global $SL(2,R)\times SL(2,R)$
descendents of the long strings considered above.
Indeed their $J^3_0$ and $\bar J^3_0$
eigenvalues are
\eqn\moreenergy{J^3_0 =  J + m ~,~~~~~~~\bar J^3_0 = \bar J + \bar m }
with $J$ as in \longstring . In principle, there could be new
contributions from conformal primary fields with these quantum
numbers, but they seem hard  to disentangle
from the descendent contributions.

\bigskip
\noindent
(2) $\underline{{\rm Integral~over~}|u| < \epsilon}$
\bigskip

{}From the integral \integral\  that
we just computed, we need to subtract contributions
from $|u| < \epsilon$ and from $|x|^{-1} < |u|$. Here we will evaluate
the integral over $|u| < \epsilon$. As in the case of $R_1$ \integral ,
 let
 us focus on the leading term in the $x$ expansion in \regionone .
The integral we need to evaluate is
\eqn\intadd{
- \sum_{h \bar h}  \int_{{k-1\over 2} + iR} dj~ {\cal C}(j)~
C_{int}(h, \bar h)x^{d+j-j_1-j_2}
\bar{x}^{\bar{d}+j-j_1-j_2}
\int_{|u|< \epsilon}d^2u u^{d-1} \bar{u}^{\bar{d}-1} | F(a,b,c;u)|^2 .}
Here we used the reflection symmetry $j \rightarrow k-1-j$
of the contour at ${k-1\over 2} + is$ ($s:$ real) to combine the
two terms in \monodinv\ into one.
We can carry out the $u$ integral by expanding $F(a,b,c;u)$ in
powers of $u$,
\eqn\expandF{
 \eqalign{&
\int_{|u|< \epsilon} u^{d-1}\bar{u}^{\bar{d}-1} | F(a,b,c;u)|^2 \cr
&= \sum_{n,\bar{n}=0}^\infty
 {\pi \over d+n} \epsilon^{2(d+n)}
\times \cr
&~~~~~~~ \delta_{n+h, \bar{n}+\bar{h}}
{\Gamma(a+n)\Gamma(b+n) \Gamma(a+\bar{n})\Gamma(b+\bar{n})
\Gamma(c)^2 \over \Gamma(a)^2\Gamma(b)^2\Gamma(c+n)
\Gamma(c+\bar{n})}.}}
Note that the condition
 $h+n = \bar{h}+\bar{n}$ is imposed by the angular integral over $u$.
In order to take the limit $\epsilon \rightarrow 0$,
we move the contour to $j = {1\over 2} + is$ with $s:$ real.
There the exponent $d+n$ of
$\epsilon$ is positive (if we ignore the tachyon) since
\eqn\notachyon{
 d + n = {s^2 + {1\over 4} \over k-2} + h + n - 1.}
Thus the contribution from the contour integral
along  $j={1\over 2} + is$ vanishes in the limit $\epsilon \rightarrow
0$. This does not mean that the original integral \intadd\ vanishes
in the limit $\epsilon \rightarrow 0$. As we are going
see, the integral picks up pole residues as we move the contour
from $j = {k-1\over 2} + is$ to $j = {1\over 2} + is$.

There are four types of poles that contribute when we deform
the contour of the $j$ integral in \intadd\ from
$j={k-1\over 2} + is$ to ${1\over 2} + is$.
The first type of poles comes from the zeros of
$d + n $ in \expandF. At the pole, we have
\eqn\shortstring{
 d + n = -{j(j-1) \over k-2} + h + n -1=0.}
The $x$ dependence of the
pole contribution is $x^{j-n-j_1-j_2}$ so that the spacetime
conformal weight of the corresponding operator is $J=j-n$.
We can identify this state as coming from a particular
current algebra descendent of a $w=0$ short string representation
of the form
\eqn\discsta{
 (J^-_{-1})^n (\bar J^-_{-1})^{\bar n} | j ,j\rangle
}
 in the
$SL(2, R)$ WZW
model times an operator of dimension $h,\bar h$ in the
the internal CFT. In fact \shortstring\ is the
$L_0=1$ condition for such an intermediate state. The $L_0=\bar L_0$
condition follows from the condition
 $h+n = \bar h + \bar n$ in  \expandF .
In section 5, we will check that the coefficient in \intadd \expandF\
evaluated at the pole \shortstring\ exactly agrees with
what we expect from the operator product expansion
 \expan , \formofc .
 The states in \discsta\ are global $SL(2,R) \times SL(2,R)$ primaries,
although those with $n \geq 1$ are descendents of the current
algebra. Higher order terms in the $x$ expansion \integral\
 produce terms which have the quantum number of descendents
of the states in \discsta\ under the global
$SL(2,R)\times SL(2,R)$. Note that due to the fact that we only
shifted the contour within the range \deformregion ,
the values of $j$ of these discrete state contributions to the
OPE are naturally bounded by \deformregion .
This reproduces the constraint on the spectrum of the
short string found in \refs{\first , \second}.

The second type of poles are at
$j=j_1+j_2+n$, ($n=0,1,2,\cdots$).
These cancel the $\epsilon$ dependence of
the contribution from Poles$_1$ that emerged
when we originally moved the contour from
$j = {1\over 2} + is$ to ${k-1\over 2} + is$.
Thus the net result is that we can compute
the $z$ integral for the contribution from
Poles$_1$ by the standard analytic continuation
method. The resulting contribution can
be interpreted as a contribution to the OPE from
two particle operators.

Similarly the third type of poles are at
$j =k -j_1 -j_2 + n$ ($n=0,1,2,\cdots$) cancel the
$\epsilon$ dependence of the contributions from Poles$_2$.
These poles do not appear if \physcond\ is obeyed.

Finally the fourth type of poles
are at $j= |j_1-j_2| - n$. In the original
contour of \expres, we avoided these poles
since they crossed the contour when we
performed the analytic continuation in $j_1, \cdots, j_4$.
We now pick up contributions from these
poles since we have to move the contour all
the way to the line at $j={1\over 2} + is$.
The contributions from these poles have
explicit $\epsilon$ dependence. We believe that these should
be explicitly subtracted.

All that we said regarding $j_1,j_2$ should be repeated for
$j_3,j_4$.

To summarize, the integral over $|u| < \epsilon$
reproduces the exchange of short string states
with $w=0$ and mixing with two particle states. These
are the only contributions to the integral as long
as \physcond\ is satisfied.

\bigskip
\noindent
(3) $\underline{{\rm Integral~over~}|u| > |x|^{-1}}$

Finally let us evaluate the integral over $|u| > |x|^{-1}$
and subtract it from $R_1$.
It is convenient to use the expansion
of \monodinv\ for large $u$. It is   given by
\eqn\largeu{\eqalign{
G_{j,0} =& \left| x^{\Delta(j)-\Delta(j_1)-\Delta(j_2) +j-j_1-j_2 }
u^{\Delta(j)-\Delta(j_1)-\Delta(j_2)}\right|^2 \cr
&\left\{ \left[ {\cal C}(j){ \G(j_1 + j_2 - j_3 - j_4)^2 \G(k-2 j)^2 \over
\G(j_3 + j_4 - j)^2 \G(k-j -j_1 -j_2)^2} + ( j \to k-1-j) \right]
 \times\right.\cr
&~~ \left|
\left({z\over x}\right)^{j - j_1 -j_2}
 F\left(j_1+j_2 - j, j_1 + j_2 -k + j + 1, j_1 + j_2 -j_3-j_4 + 1;
{x \over z}\right)\right|^2
+
\cr
 &\left. ~~~~+ \Big( (j_1,j_2) \leftrightarrow ( j_3 , j_4) \Big)
\right\}. }}
Note that this is the large $u$ expansion of the leading term
 \monodinv\ in the $x$ expansion in region I  \regionone .
 The large $u$ expansion of the full solution KZ solution is different
and
will be discussed later when we study the
integral in the region II.
In \integral ,  we integrated this leading term
over the whole plane. Thus we need to subtract the integral over
$|u|>|x|^{-1}$ using the same integrand to obtain
an approximate expression for the integral
of the full solution of the KZ equation over $|u|<|x|^{-1}$.
 Using \largeu , we find that the integral gives terms of the form,
\eqn\largeuint{\eqalign{
& x^{\Delta(j) + h -1  +j-j_1-j_2 } {\bar x}^{\Delta(j) + \bar h -1 + j
-j_1-j_2}
\times \cr
&~~~~~ \int_{|u|>|x|^{-1}}
{du^2 } u^{d-1}\bar{u}^{\bar{d}-1}
\left(a_{n \bar n} u^n {\bar u}^{\bar n}
|u|^{2(j -j_1-j_2)} +
 b_{n\bar n} u^n {\bar u}^{\bar n} |u|^{2(j -j_3-j_4) }  \right)
 \cr
&\sim  \tilde a_{n \bar n}  x^{n} {\bar x}^{\bar n}
  + \tilde b_{n \bar n}  x^{j_3+j_4 - j_1 - j_2 + n} {\bar x}^{
j_3 + j_4 -j_1 -j_2 + \bar n},
}}
for some $a_{n,\bar n}$, $b_{n, \bar n}$, $\tilde a_{n, \bar n}$
and $\tilde b_{n, \bar n}$.
{}From the exponents of $x$,
we see that these terms all have the form of two particle contributions.
It seems possible that we could shift the contour of integration in $j$
to a region where it becomes convergent. This shift might produce
extra contributions, but they all have these powers of $x$ and therefore
 will be of the form of two particle exchanges.

This completes the evaluation of the $z$ integral in region I.

\subsec{Integral over region II}

It remains now to do the integral over the region II. In this region, we
can expand any solution of the KZ equation as
\eqn\regtwoexp{
F(z,x) \sim x^\alpha \sum_{m= 0}^\infty \tilde{g}_m(z) x^m.
}
Substituting this into the KZ equation, we find that
$\alpha = 0$ or $\alpha = j_3+j_4 - j_1-j_2$. This means that
the full contribution from this region is  interpreted as two
particle contributions. In this region, we also have to expand the
internal part in a different way. But in any case the $x$ dependence
is just that of the two particle contributions.

Thus we have completed the computation of the integral over the $z$ plane
with the results summarized at the beginning of the section.
The intermediate states in the small $x$ expansion are identified
and are found to be consistent with the operator product expansion
in BCFT interpreted in the standard way as in \expan ,
provided \physcond\ is satisfied. Note that as long as \physcond\ is
satisfied  the three point functions that appear in the factorization
on intermediate discrete states automatically obey the
constraint $ \sum j_i < k$. This is consistent with our previous
statement that only those three point functions make sense in
the theory.

\subsec{ When the OPE does not factorize}

Let us now discuss what happens when \physcond\ is not satisfied.
In this case, besides the terms we discussed above, we get
contributions from the residues of Poles$_2$ in \poles .
If we were to read off naively
the dimension $J$ of an intermediate operator from the power of $x$
appearing in these contributions,
we would find $J = k-j_1-j_2 + n $ (or a similar expression with
$j_3,j_4$). For generic values of $k, j_1, \cdots, j_4$, there is no
physical
operator with this value of $J$.  Therefore  these contributions do
not have an interpretation as exchange of intermediate physical
states as in \expan .
Their presence signals a breakdown in the operator product expansion.

One may naively interpret this as saying
 that we need to include more physical states
in the theory. We claim this is not the correct interpretation.
Instead we propose that, in this case, the operator product
expansion is not
well-defined in the target space theory. This is due to the non-compactness
of the target space of BCFT. To clarify this issue, it is useful
to go back to the simple quantum mechanics example we gave in
section 3.2,
$i.e.$, that of a quantum particle moving in a one-dimensional
space with coordinate $x$ under
a potential that is zero for $|x|\gg 1$
such that the wave-function of the ground state
decays as $\langle x |0\rangle =
\psi(x) \sim e^{-{\kappa \over 2} x}$
 for large $x$. In these circumstances, we
consider the operators $ {\cal O}_i(t) =  e^{\lambda_i x(t)}$
and try to evaluate their
 correlator $\langle 0|{\cal O}_4(t_4){\cal O}_3(t_3)
 {\cal O}_2(t_2){\cal O}_1(t_1)
|0\rangle $. This correlation function is well-defined if
$\sum \lambda_i < \kappa$. Now we can try to perform the OPE when
$t_1 \to t_2$ and $t_3 \to t_4$.
Naively one may expect to find normalizable (and also
continuum normalizable) states running in the intermediate channel.
It is easy to see
that this will be the case only if $\lambda_1 + \lambda_2
<{\kappa  \over 2}$
 and $\lambda_3 + \lambda_4 <{\kappa \over 2} $.
 These conditions are analogous
to \physcond . If these conditions are not obeyed, the intermediate
state is not in the Hilbert space of the theory. In other words,
the product ${\cal O}_1 {\cal O}_2$ maps the state $|0\rangle$
outside the Hilbert space. This is effect is not a UV divergence;
rather it is
an IR divergence in the target space of the quantum mechanical
system.

These contributions from Poles$_2$ that we are discussing are important
for reproducing the general properties of the amplitude that we explained
in section 3. The four point function should  have a pole at
$\sum j_i - k =1$. This pole is absent from all the terms in the amplitude
that can be written as \formofc . But it is present in the term coming
from Poles$_2$, as it can be checked explicitly by performing the integral
over $z$ for the Poles$_2$ contribution. Note that
 \physcond\ cannot be obeyed
if we are at the pole at $\sum j_i - k =1$, so we definitely have
Poles$_2$ contributions in this region.

Note that we have assumed that all the $j$'s involved in the
computation of the OPE are generic enough so that there are
no coincident poles. Coincident poles can produce terms involving
$\log x$ . These were studied in \refs{\freedman,\liu}, and they have
the same interpretation here as they had in their case.

\newsec{Two and three point functions with spectral flowed states}

In the last section, we have shown that the four point function of
short strings with $w=0$ is factorized
into a sum of products of three point functions. We found that
the intermediate states
are long strings with $w=1$, short strings with $w=0$,
and two particle states. These intermediate states are
identified by evaluating the $x$-expansion of the amplitude
and by comparing exponents of $x$ with the  spectrum of physical
states of the short and long strings. One of the
purposes of this and the next sections
is to prove that the coefficients
in the $x$-expansion are what we expect from the factorization
of BCFT.  To this end, we need to compute two and three point
functions involving spectral flowed states.
We  will also
explain the origin of the constraint on the winding number
violation. In appendix D, we will use the representation theory
of the $SL(2,R)$ current algebra to show that two short strings
with $w=0$ can only be mixed with short strings with
$w=0, 1$ or long strings with $w=1$. This almost accounts
for the winding number violation rule we saw in the factorization
of the four point function, but leaves out the question of
why short strings with $w=1$ do not appear in the intermediate
channel. In this section,
we will show that, if we normalize the vertex operators so
that their target space two point functions are finite,
the three point function of two short
strings with $w=0$ and one short string with $w=1$ vanishes
identically, thereby explaining the additional constraint on
the winding number violation. We will also discuss  other
aspects of these correlation functions.

In \first, it is shown how to construct vertex operators for
the spectral flowed representations. This can be done most
easily in the $m$ basis, where the generators $(J_0^3, \bar{J}_0^3)$
of the global $SL(2,C)$ isometry are diagonalized. On the
other hand, in \teschnertwo\ and \teschnerthree,
we used the $x$ basis to express
the two and three point functions.
Therefore, to compute correlation functions involving
spectral flowed states, we first have to convert \teschnertwo\
and \teschnerthree\ into the $m$ basis, perform the spectral
flow operation as described in \first, and then transform the
result back in the $x$ basis.

One thing we need to be careful about in this procedure
is that the spectral flow changes the way
the global $SL(2,C)$ isometry acts on
states since the currents are transformed as
\eqn\sfagain{
J_0^\pm = \tilde{J}_{\mp w}^\pm,~~~
J_0^3 = \tilde{J}_0^3 + {k \over 2} w .}
For example, consider a representation of the
current algebra whose worldsheet energy $\tilde L_0$
is bounded from below. (${\cal D}_j^{w=0}$
and ${\cal C}_{j,\alpha}^{w=0}$ are an example of such
representations, but here we do not assume that
the lowest energy states of the representation
makes a unitary representation of the global $SL(2,R)$.)
We then pick
one of the lowest energy ($\tilde L_0$) states
$|\psi \rangle$, satisfying\foot{Here $j$ is what we called $\tilde j$
in \first .}
\eqn\annihilated{
 \eqalign{& \tilde{J}_n^{\pm , 3} | \psi \rangle = 0~,~~~~~~~~n = 1, 2, 3,
\cdots\cr
 & \tilde{J}_0^3 |\psi \rangle = m |\psi \rangle \cr
 & \left[ - \tilde J_0^3 \tilde J_0^3 + {1 \over 2}
\left(\tilde J_0^+ \tilde J_0^-
 + \tilde J_0^- \tilde J_0^+ \right) \right]
 | \psi \rangle =
 - j(j-1) |\psi\rangle.}}
If $m = \pm (j+n)$ for a non-zero integer $n$,
the state $|\psi \rangle$ belongs to the discrete representation
$d_j^\pm$ with respect to the  $SL(2,R)$ algebra  generated by
$\tilde J^a_{0}$.
Otherwise it is in the
continuous representation $c_{j,\alpha}$,
where $m = \alpha $ (mod integer).\foot{
We are using the symbols $d_j^\pm$ and $c_{j,\alpha}$
to label representations of an $SL(2,R)$ algebra,
to  distinguish them with the representations
of the full  current algebra.}
If $w$ is positive,
the same state $|\psi \rangle$ is seen
in the spectral flowed frame \sfagain\ as obeying
\eqn\newframe{ J_0^- |\psi \rangle = 0~,~~J_0^3
 |\psi \rangle = \left( m + {k \over 2} w \right)
| \psi \rangle.}

\ifig\flowfigure{Under the spectral flow,
a global $SL(2,C)$ descendant $|\psi\rangle$ of spin $J_0^3=m$
among the lowest energy states in ${\cal D}_j^{w=0}$ or
${\cal C}_{j,\alpha}^{w=0}$ turns into the lowest weight
state of the discrete representation $d^+_J$ with
$J=m + {k \over 2}w$. The figure shows the flow of
${\cal D}_j^{w=0}$. The resulting operator is denoted by
$\Phi_{J,\bar J}^{w,j}(x,z)$.}
{\epsfxsize4in\epsfbox{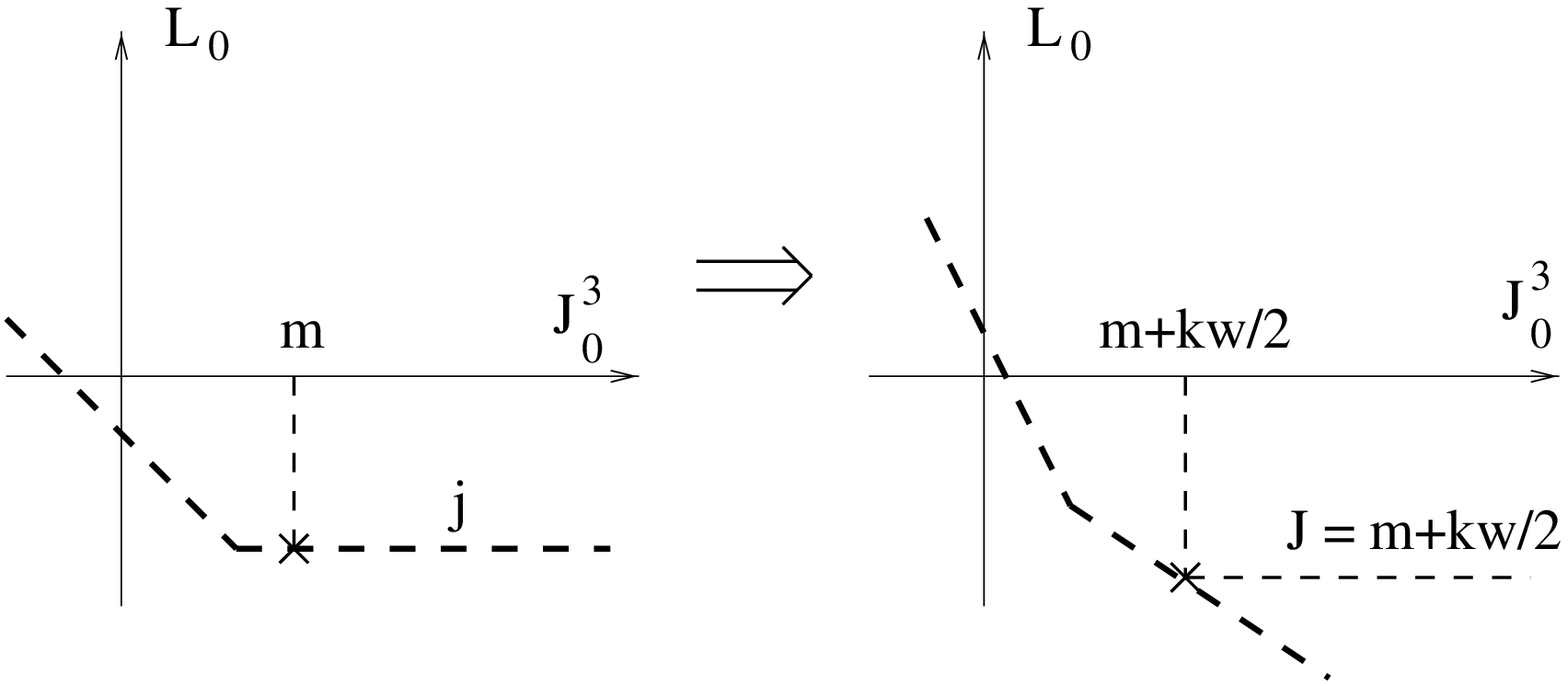}}
With respect to the
global $SL(2,R)$ algebra generated by $J^a_0$ the state $|\psi\rangle$
is the lowest weight state of a discrete
representation $d_{J}^+$ with $J=m + {k \over 2} w$,
independently of whether $|\psi\rangle$ was in
$d_j^\pm$ or $c_{j,\alpha}$ of the $SL(2,R)$ algebra generated
by $\tilde J^a_0$.
Similarly,  spectral flow with $w < 0$ turns
$|\psi\rangle$ into the highest weight state of
$d_J^-$ with $J = -m + {k \over 2} | w|$.
In our physical application we identify the $SL(2,R)$ algebra
generated by $J^a_0$ with the spacetime isometries of the background,
and the global $SL(2)$ symmetries of the BCFT.
In  what follows we will indicate by $J$ and $M$ the global
$SL(2)$ spin and $J_0^3$ eigenvalue respectively.

The transformation between the $x$ basis and the $m$ basis is
carried out as follows.
Consider an operator $\Phi_{J, \bar{J}}(x,\bar{x})$
in the $x$-basis, with the spacetime conformal weights
$(J,\bar J)$.
In general, the difference $(J-\bar J)$ has to be
an integer in order for their correlation functions
to be single-valued in the $x$ space,
and we will consider such cases only. The integral transform
\eqn\basis{
\Phi_{J, M ; \bar J ,\bar M} = \int {d^2 x \over |x|^2 }  x^{J-M}
{\bar x}^{\bar J- \bar M} \Phi_{J ,\bar J} (x, \bar x ),
}
turns the operator into the $M$ basis
 where
$M$ and $\bar M$ are eigenvalues of $J_0^3$ and $\bar J_0^3$,
respectively.\foot{
We reserve the small case letters
$m, \bar m$ to denote eigenvalues of $J_0^3, \bar J_0^3$
in the $w=0$ sector, $i.e.$, for states before we perform
the spectral flow.} Note that
$(\bar M, \bar M)$ are not necessarily
complex conjugate of $(J,M)$. Since
$(J-\bar J)$ is an integer, the integral vanishes unless
$(M-\bar M)$ is also an integer and we will assume this
in the following.

In practice,
the $x$ integral in \basis\ is carried out after computing
correlation functions and using analytic continuation in
the variables, $J, M, \cdots$.
When $J$ is real, we have to keep in mind
that the $x$ integral
gives poles at  $ M= J+n $ and $ \bar M = \bar J + \bar n$,
with non-negative integers $n , \bar n$. We will see this
explicitly in the two and three point function computations
in the following.\foot{
Although there are two conditions on $M$ and $\bar M$,
the pole is only in one variable
of the form, $(M + \bar M -J - \bar J -n - \bar n)^{-1}$.
The second condition is imposed by the angular integral in the
$x$-space.} These are precisely the values at which
the operator $\Phi_{J,M;\bar J, \bar M}$ belongs to
a discrete representation $d_J^+ \otimes d_{\bar J}^+$ of
the global $SL(2,R) \times SL(2,R)$ symmetry.
In such cases, we have
to keep track of this additional divergent factor.
There are also similar poles when $M=-J-n$, $\bar M = -\bar J - \bar n$,
with non-negative integers $n, \bar n$ and they
form $d_J^- \otimes d_{\bar J}^-$.
We will call the poles with positive $M$ as
``incoming states'' and the poles with negative $M$ as
``outgoing states''. In this way, we see that the single
operator $\Phi_{J,\bar J}$ in
the $x$ basis gives rise to both $d_J^+$ and $d_J^-$,
depending on the value of $M$ we choose in evaluating
the integral transform \basis.

Correlation functions of spectral flowed operators are
then evaluated as follows. We start with $n$ point
correlation functions in the $w=0$ sector, which are known for
for $n=2,3,$ and $4$.
We perform the integral transform \basis\ to turn them
into expressions in the $m$ basis. We then act the spectral
flow operator to find expressions for $w\neq 0$ (as described
in detail in the following subsections). Finally we use
\basis\ to transform the expressions back into the $x$ basis.

Alternatively one can perform the spectral flow operation
directly in the $x$ basis. In the case of $w=1$,
the spectral flowed operator
$\widehat{\Phi}_{J, \bar J}^{w=1, j}(x,z)$ is
constructed from $\Phi_j(x,z)$ in the $w=0$ sector as
\eqn\flowinxbasis{
 \widehat{\Phi}_{J, \bar J}^{w=1, j}(x,z)
\equiv \lim_{\epsilon \rightarrow 0}
 \epsilon^m \bar{\epsilon}^{\bar m}
 \int d^2 y y^{j-m-1} \bar y^{j-\bar m-1}
\Phi_j(x+y, z+\epsilon) \Phi_{k/2}(x,z).}
Here we put $~\widehat{}~$ on the spectral flowed operator
since its normalization is different from the one naturally
defined by going through the $m$ basis as described in the
above paragraph. In Appendix E, we will prove that \flowinxbasis\
in fact defines the spectral flowed operator by showing that
it has the correct operator product expansions with the
currents $J^{3,\pm}$. We will then use \flowinxbasis\ to
compute their two and three point functions.

In this section, we will use the spectral flowed operator defined
through the $m$ basis. This approach has an advantage of being
able to treat all values of $w$ simultaneously.

\subsec{ Two point functions }

Let us start with a two point function in $x$ space
for generic values of $J$, $\bar J$.  The two point functions
in the following typically take the form,
\eqn\start{
\langle \Phi_{J,\bar J}(x_1) \Phi_{J, \bar J}(x_2)\rangle=
{ D(J,\bar J) \over
x_{12}^{2 J} {\bar x}_{12}^{2\bar  J}}
}
where we have suppressed  a possible  $z$ dependence.
Performing the integral using the formula (C.5)
in Appendix C, we
find
\eqn\startm{
\langle \Phi_{J,M; \bar J,\bar M}
\Phi_{J,M';\bar J' \bar M'} \rangle =
\delta^2(M+M')
  { \pi \Gamma(1-2\bar J)  \Gamma(J+M) \Gamma(\bar J - \bar M )
\over \Gamma (2J) \Gamma( 1- J +  M)
 \Gamma( 1 - \bar J - \bar M) } D(J,\bar J)
}
The delta function $\delta^2 (M)$  the standard
delta function for the sum $(M+\bar M)$ and the Kronecker delta
for the difference $(M-\bar M)$, which is an integer.
Using the formula $\Gamma(x)\Gamma(1-x) = \pi / \sin\pi x$
and  the fact that $(J-\bar J)$ and $(M-\bar M)$ are both integers,
one can check that the expression \startm\ is symmetric under exchange
of $(J, M)$ and $(\bar J, \bar M)$.

Conversely, if we are given
the expression \startm , we can turn it back into
of the form \start\ in the $x$ basis. To do this,
it is not necessary to know the
expression for all possible values of $M ,\bar M$.
For example, the expression \startm\ has a pole
at $M=J$ and $\bar M = \bar J$, and the residue
is equal to $D(J,\bar J)$ times a simple factor.
Thus it is sufficient to know the pole residue there in
order to recover the $x$ space expression \start.
Similarly we can reconstruct \start\ from
the residue of the pole at $M=-J$ and $\bar M = - \bar J$.
In the following, we will encounter such situations.

We now consider the two point function of $w=0$ states given
by \teschnertwo\ and convert it into a two point function
with $w\not =0$ states.  As we mentioned, we first turn the expression
 into the $m$ basis,
perform the spectral flow, and then transform this back into
the $x$ basis.
In transforming the second term \teschnertwo\ into the $m$ basis,
 we can use
\startm\ with $D(j,j) = \delta(j-j') B(j)$;
the $x$ integral of the first term is easy to do directly.
In the $m$ basis, it is straightforward to apply the spectral
flow. As explained in \first , the only change in the two point
function is that the power of
$z$ is modified in an $m$ dependent fashion reflecting
the change in the worldsheet conformal weight,
\eqn\weightmodify{
  \Delta(j) \rightarrow \Delta(j) -wm - {k \over 4} w^2,}
 without any modification to the coefficient.
We should also remember that the assignment
of the global $SL(2,C)$ charges is changed
according to the discussion after \sfagain.
To perform
the spectral flow explicitly, we bosonize
the $J^3$ current as $J^3 = i\sqrt{{k \over 2}} \partial \varphi$ and
write an operator with $J^3$ charge $m$ as
\eqn\parafermion{
\Phi_{j,m} \sim e^{i m \sqrt{2 \over k} \varphi} \psi_{j,m}.}
The operator $\psi_{j,m}$ carries  no $J^3$ charge and is analogous
to the parafermion field in the $SU(2)$ WZW model.
We then make the replacement,
\eqn\spectflowact{
  e^{ i m {\sqrt{ 2 \over k}} \varphi } \to
e^{ i (m + w {k \over 2}) {\sqrt{ 2\over k }}\varphi},
}
and similarly for $\bar m$.
As explained in \first ,  the operator we find in this way
has  $ J = M = m +  \khalf w $, $\bar J = \bar M = \bar m +  \khalf
w$, namely it is the lowest weight state in the representation
$d_J^+ \otimes d_{\bar J}^+$ of the global $SL(2,C)$ isometry.
Including the modified $z$ dependence that comes from applying the
spectral flow operator, we obtain the two point function \first ,
\eqn\twoptfl{\eqalign{
& \langle \Phi^{w, j}_{J, M; \bar J , \bar M}(z_1)
 \Phi^{-w , j'}_{J , M'; \bar J ,  \bar M '} (z_2) \rangle =\cr
& = { 1  \over
 z_{12}^{ 2(\Delta(j) - w M +  {k \over 4}w^2 ) }
{\bar z_{12}}^{ 2(\Delta(j) - w \bar M +  {k \over 4}w^2  ) } }
\delta^2(M+M') \times \cr
 &~~~~~~~ \left[ \delta (j+j'-1) + \delta(j-j')
 { \pi B(j) \over \gamma(2j)}
{\Gamma(j+m) \over \Gamma(1-j+m)}
{\Gamma(j - \bar m )
\over  \Gamma( 1 - j - \bar m) }  \right], \cr
& {\rm where} ~~~~~
J = M = m +  \khalf w~,~~~~~~~ \bar J = \bar M = \bar m +  \khalf w.
}}
Note that $j$ is the spacetime conformal weight of the original
$w=0$ operator and it should be distinguished from
$J, \bar J$ for the operator we get after spectral flow.
The amount of  spectral flow of the second operator is $-w$;
this is necessary in order to preserve the total $J_0^3$ charge.
If $w, m>0$, we can interpret the first operator as an incoming
state and the second as an outgoing state.

We would like to convert \twoptfl\ back to  the $x$ basis.
According to our previous discussion, this can be done
by evaluating the pole residue at $J=M$ and $\bar J = \bar M$.
Unlike a generic  two point function such as \start\ and \startm ,
the expression \twoptfl\ is finite at this location.\foot{There
is an important exception when the
$w=0$ operator is in a discrete representation, in which
case $\bar m = j+\bar n$ and there is a pole. We will come back
to this point later.} The pole that we are missing here
comes from the divergent integral of
the form $\int d^2z / |z|^2$. We recognize that it has
the same form as the volume $V_{conf}$
of the conformal group of $S^2$ with
two point fixed,
\eqn\confgr{
  \int {d^2 z \over |z|^2} = {\int {d^2z d^2 w d^2 u \over
 |z-w|^2 |w-u|^2 |u-x|^2 }\over
 \int {d^2z d^2 w \over |z-w|^4}} = V_{conf}.}
 Since evaluating the pole residue of \twoptfl\
at $J=M, \bar J = \bar M$ is same as evaluating it at the pole
and dividing it by $V_{conf}$ (with an appropriate regularization
of the $z$-integral), we can interpret \twoptfl\
as resulting from a two point function in the $x$ basis of
the form,
\eqn\flowedx{\eqalign{&
\langle \Phi^{w, j}_{J, \bar J }(x_1,z_1)
\Phi^{w, j'}_{J \bar J}(x_2,z_2) \rangle =
\cr
&
= {1 \over V_{conf}}
\left[ \delta(j+j'-1) +  \delta( j- j')
 { \pi B(j) \over \gamma(2j)}
{\Gamma(j+m) \over \Gamma(1-j+m)}
{\Gamma(j - \bar m )
\over  \Gamma( 1 - j - \bar m) }\right]
\times \cr
&~~ ~~~~~~~~~
{1 \over x_{12}^{ 2 J} {\bar x}_{12}^{2 \bar J} }
{1 \over z_{12}^{ 2(\Delta(j) - w M +   {k\over 4} w^2) }
{\bar z_{12}}^{ 2(\Delta(j) - w \bar M +  {k\over 4}w^2) } }
}}
The factor $V^{-1}_{conf}$ will eventually be cancelled in
the string theory computation that follows.
In going from \twoptfl\ to \flowedx, we have switched the sign
of $w$ in the second operator. This is due to the fact that
an outgoing state with negative $w$ is the same as an incoming state
with positive $w$. In other words, in the $x$ basis we can label
the operators with $w\geq 0$.

Some readers may be disturbed by
the appearance of the infinite factor $V_{conf}$
in our computation. We
can avoid the use of $V_{conf}$ altogether
if we work directly in the
$x$ basis using \flowinxbasis . This will be explained in Appendix E.
For $w=1$, both approaches give the same result. For $w > 1$,
computations in the $x$ basis become cumbersome.
For this reason, we will continue to work in the $m$ basis
in this section so that we can find expressions for all $w$
at once.

So far, we have taken $j $ to be arbitrary.
Let us now set  $ j = {1 \over 2}  + is $, so that we
have a  continuous representation at $w=0$. In this case,
the spectral flowed expression \flowedx\ gives
the two point function of the vertex operator
for the long string with $w=1$.
In order to compute the spacetime two point function,
we need to take into account the contribution from the
internal CFT. We choose the internal
conformal weight $(h, \bar h)$ such that the long string
obeys the physical state condition,
\eqn\longstringphysical{ \Delta(j) - w M +  {k \over 4} w^2
 + h =1,~~\Delta(j) - w \bar M + {k \over 4} w^2
 + \bar h = 1.}
Assuming that the operator in the internal CFT
is unit normalized, its
effect is to  multiply the factor $z^{-2h}\bar z^{-2\bar h}$ to \flowedx.
We then need to integrate over $z$ and divide it by the volume of the
conformal group on the sphere. This produces another factor
of $V_{conf}^{-1}$. By changing the normalization of the operator as
$\hat \Phi = V_{conf} ~ \Phi$, the two point function
in the target space is given by
\eqn\correcont{
\eqalign{ & \langle \hat \Phi^{w, j}_{J, \bar J}(x_1)
\hat{\Phi}^{w, j'}_{J,{\bar J}}(x_2) \rangle_{target}\cr
& =
{1\over V_{conf}}
V_{conf}^2 ~ \langle \Phi_{J,\bar J}^{w,j}(x_1,z_1=0)
 \Phi_{J,\bar J}^{w,j}(x_2,z_2=1) \rangle_{worldsheet}
\cr& \sim
 \left[\delta(s+s') +  \delta( s- s')
{ \pi B(j)\over \gamma(2j)}
{\Gamma(j - \khalf w+ J) \over
 \Gamma( 1- j- \khalf w+ J  )}
{\Gamma( j + \khalf w - \bar J )
\over \Gamma( 1 - j + \khalf w - \bar J) }
\right]
{1 \over x_{12}^{ 2 J} {\bar x}_{12}^{2 \bar J} },}}
where
\eqn\valj{\eqalign{
j &= \half + is
\cr
J &=  {k \over 4} w+
{1 \over w } \left({ {1 \over 4} + s^2 \over k-2} + h -1 \right) ,
}}
and a similar expression for $\bar J$ in terms of $\bar h$.
As far as the two point function is concerned, we can of course
normalize the operator $\Phi$ as we like. All we are saying here
is that this normalization removes the divergent factor $V_{cont}$
and keeps the target space
two point function finite. In the next subsection,
we will see that the rescaling $\hat{\Phi} = V_{conf}~ \Phi$ also
gives finite results for the three point functions that appear
in the factorization of the four point function.

We would like to make a couple of comments about the
two point function of the long strings \correcont.
Unlike the case of the short string, the on-shell condition does
not require $j=j'$. However the two point function has
the delta functions $\delta(s+s')$ and $\delta(s-s')$,
giving constrains on the labels $s, s'$. For the
operator before the spectral flow,
the term proportional to  $\delta(s+s')$ in \teschnertwo\
 is multiplied
by $\sim \delta^2(x_{12})$, $i.e.$, it is a contact term
in BCFT.
After the spectral flow  \correcont, the corresponding
term contributes to the long range correlation of the
two operators
$\sim x_{12}^{-2J} \bar{x}_{12}^{-2\bar J}$ in the same
way as the term proportional to $\delta(s-s')$.
Thus, when
we discuss the factorization of the four point function,
we need to take into account both the first and the second
terms in \correcont .
Another remark we would like to make is
 that the factor multiplying the $\delta(s-s')$
in the second term in
\correcont\ is a pure phase $e^{i \delta(s)}$, see \purephase .
We can interpret it as
 the phase shift for a scattering experiment where
we let a long string come from
the infinity of $AdS_3$, shrink to the origin, and go back to the
infinity again \first.
In fact, the operators labeled by $s$ and $-s$ are not independent, and
they are related by the reflection coefficient $\Phi^{ \half + is } \sim
e^{i \delta(s)} \Phi^{\half - is }$ as shown in \teschnerone .

Now let us turn to discrete representations. We start with a
global $SL(2,C)$ descendent with $ m= j + q$ and $ \bar m = j + \bar q $
where $q, \bar q$ are non-negative integers.
After the flow, we obtain a state with
$J = M =  j + q + \khalf w $, $\bar J = \bar M = j + \bar q + \khalf w$.
In this case,
we get  a pole from one of the  Gamma functions in \flowedx ,
and it cancels the factor $V_{conf}^{-1}$.
Thus the expression in the $x$ space is
finite. As in the case of long string, turning this into string
theory two point function generates an additional factor of
$V_{conf}^{-1}$, but this is also cancelled by
$\delta( j -  j')$ in \flowedx\ evaluated at $j = j'$.\foot{
For a short string, the physical spectrum of $j$ is discrete
and we need to evaluate the delta function right at $j=j'$
rather than leaving  the delta-functions, $\delta(s+s')$ and
$\delta(s-s')$, as in the case of long strings in
 \correcont .}

With all the factors $V_{conf}$ cancelled out, we
have a finite correlation function in the target space.
There is one subtlety here since there is a possibility
that a $j$ dependent factor appears when we cancel
$\delta(j-j')$ at $j=j'$ with the volume of the conformal group
$V_{conf}$.  We claim that, in fact, a finite factor of the
form $|2j -1 + (k-2)w|$ remains after the cancellation.
One heuristic  way to see this is the following.
(A more rigorous derivation
of this factor in the case of $w=0$ is given in Appendix A.)
If we regularize the computation by taking $j$ to
be slightly away from on-shell $L_0(j) - 1 =0$ and
introduce a cutoff $\epsilon$ in the $z$ integral,
the volume $V_{conf}$ of the conformal group would be
$\delta_\epsilon(L_0(j) -1)$, where $\delta_\epsilon$ is
a Gaussian with a short tail
which becomes the delta function in the limit
 $\epsilon \rightarrow 0$.
This is the factor that cancels the $\delta(j-j')$ term
in the worldsheet two point function.
Thus we expect that the cancellation of the two divergences
leaves the finite factor given by
\eqn\ratiojm{\left| {\partial L_0(j) \over \partial j}\right| \sim
|2j-1 + (k-2)w |,}
up to a $k$ dependent coefficient.
Taking this into account, the two point function of the short
string with winding number $w\neq 0$ is   of the form
\eqn\floweddis{
\eqalign{&
\langle \Phi_{J \bar J}^{w,j}(x_1)  \Phi^{w,j}_{J \bar J}(x_2)
\rangle_{target}\cr
&= {1 \over V_{conf}} \langle \Phi_{J,\bar J}^{w,j}(x_1,z_1=0)
\Phi_{J,\bar J}^{w,j}(x_2, z_2=0) \rangle_{worldsheet}\cr
& \sim
|2j-1 +(k-2)w|
 { \Gamma(2  j + q)  \Gamma(2  j + \bar q)
\over \Gamma(2j)^2 q! \bar q !  }
{ B( j) \over x_{12}^{2 J} {\bar x}_{12}^{2 \bar  J}},
}}
where $q=J-j-\khalf w$, $\bar q = \bar J - j - \khalf$.
Unlike
the case of the long string, we do not have to rescale the
operator $\Phi_{J,\bar J}$.
We note that the coefficient in \floweddis\ is positive as long as
$j$ is in the physical range $\half < j < {k-1 \over 2} $.
This
of course is consistent with the positivity of the physical
Hilbert space of the string in $AdS_3$.
When $w=0$, the two point function is given by
\eqn\twoptcorr{
\langle \Phi_j^{w=0}(x_1) \Phi_j^{w=0}(x_2) \rangle_{target} \sim
(2j-1){ B(j) \over |x_{12}|^{4j} }.
}
Later in this section, we will show that
this additional factor of $(2j-1)$ is
precisely what one need in order to reproduce the
factorization of the four point function onto
the short string with $w=0$.
In general, we have to be careful about
a possible $j$ dependent factor that could appear when
we go from the worldsheet expression to the
target space expression, and \twoptcorr\ is an example
of this.

For a short string,
another useful computation one can do is to evaluate
the two point functions of operators $\hat{\Phi}_{J, \bar J}^{j; q,q}$
corresponding to the state of the form,
\eqn\statef{
(J_{-1}^-)^p (\bar J_{-1}^-)^{\bar p} | j; m=\bar m = j \rangle .
}
where $J = j -p$ and $\bar J = j- \bar p$ are the spacetime conformal
weights under global $SL(2,C)$. Although they are
descendants of the current algebra, they are the lowest weight
states of the global $SL(2,C)$. These states appear in the
intermediate channel of the factorization of the four point
amplitude discussed later in this section, so it is useful
to compute their two point functions here.
 They are computed in the following way.
Let us view these states as given by performing one-unit of
spectral flow the lowest
energy states in ${\cal D}_{\khalf -j}^{-0} \to
{\cal D}_{\khalf -j}^{- ~ w=1}= {\cal D}_{j}^{+0}$.
We start with the state $ |j'; m = -j'-p, {\bar m} = -j'-{\bar p}
\rangle$
with $j' = \khalf  -j$.  Under one unit of spectral flow,
this state is mapped into a state of the form \statef .
 So we first compute the correlation
function of the state labeled by $j'$ in the $m$ basis, perform
spectral flow using the formulas \twoptfl , and finally  we go to the
$x$ basis as in \floweddis . We find
\eqn\pretwoneg{
\langle \Phi^{jp \bar p}_{J \bar J} (x_1)  \Phi^{jp \bar p}_{J \bar J
}(x_2)
\rangle
\sim (2j-1)
{ \Gamma(k - 2 j +  p)  \Gamma(k - 2 j + \bar p)
\over   \Gamma(k - 2 j)^2 p ! \bar p !}
{ B\left(\khalf - j\right)
\over x_{12}^{ 2(j-p)} {\bar x}_{12}^{2(j - \bar p) },
}}
where again we have assumed that the amplitude is multiplied by a unit
normalized primary field in the
internal CFT operator so that the total worldsheet conformal weight
of the vertex operator is one, and we integrated the
resulting two point function over the worldsheet.
We have taken into account the factor $(2j-1)$ discussed at
\twoptcorr. Notice that, up to a  $k$ dependent factor,
$B(\khalf -j)$ is equal to $B(j)^{-1} $ as one can see from \defofbtes .
If we set $p=\bar p =0$ in \pretwoneg , we
recover the original result \twoptcorr\
but with a different normalization;
instead of $B(j)$, we have $B^{-1}(j)$.
What this shows is that the natural normalization of the operator in ${\cal
D}_j^{w=0}$ and that of the operator in ${\cal D}_{\khalf - j}^-$
are different. It is therefore more convenient to
define the operator corresponding to the state \statef\ as
\eqn\newnorm{
{\hat  \Phi}^{jp \bar p}_{J \bar J} (x) \sim
B(j) \Phi^{jp \bar p}_{J \bar J} (x) .
}
In this way, for $p =\bar p =0$, we recover
the $w=0$ $SL(2)$ current algebra primaries with the standard
normalization \defofbtes . Their two point function is then
given by
\eqn\twoneg{
\langle \hat{\Phi}^{jp \bar p}_{J \bar J} (x_1)
\hat{\Phi}^{jp \bar p}_{J \bar J
}(x_2)
\rangle
\sim (2j-1)
{ \Gamma(k - 2 j +  p)  \Gamma(k - 2 j + \bar p)
\over  \Gamma(k - 2 j)^2  p ! \bar p !}
{ B\left(j\right)
\over x_{12}^{ 2(j-p)} {\bar x}_{12}^{2(j - \bar p) }.
}}
We will use this formula in section 5.5 where we examine
effects of intermediate short strings with $w=0$
in the four point function.

\subsec{Three point functions in $m$ basis}

Let us now turn to three point functions.
In the case of the two point functions, the winding number
$w$ is preserved in the $m$ basis \twoptfl . This simply
reflects the fact that the worldsheet Hamiltonian $L_0 + \bar{L}_0$
can be diagonalized by states carrying fixed amounts of $w$.
However the winding number
can be violated by string interactions. In this subsection,
we will compute the four point function with
three vertex operators  and  one  spectral flow operator.
This computation has been done in \zam , and we
reproduce it here. In \giribet\ this  was done using the free
field theory approach.
 In the next subsection, we will use
this result to derive
the three point functions with winding number violations.

The spectral flow operator
changes the winding number of another operator by one unit.
According to \first, we can view it
as the lowest weight
state in $d_j^+ \otimes d_j^+$ with $j=k/2$. This operator is outside the
allowed range \range\ for physical operators in the target space theory.
We will not use this operator by itself for  an operator in the
target space theory, but it is used in an intermediate step
to construct physical operators with non-zero winding numbers.
A very important property of  the spectral flow operator is that
it  has a null
descendant of the form,
\eqn\nulldec{
J_{-1}^- |j={k\over 2}; m={k \over 2} \rangle =0 .}
 We can then compute a four point function where one
of the operators is $|j=k/2, m=k/2 \rangle$ since
it obeys the differential equation which follows from
the existence of the null state \nulldec . The
equation turns out to have a unique solution up to
an overall normalization, and we can use it to derive
a three point function with winding number violation.
This computation also serves as a simple example where
we can find an explicit expression
for ${\cal F}_{SL(2)}$ in \nnn\
(though in the non-generic case) and, it gives us
some intuition about how four point functions
look like in general. In particular,
we will find that the solution indeed has a singularity
at $z=x$ with the exponent for $(z-x)$ expected from the
general argument given in section 2.4.

We want to   compute the four point function,
\eqn\fpt{\eqalign{
 \langle \Phi_{j_1}&(x_1,z_1) \Phi_{\khalf}(x_2,z_2) \Phi_{j_3} (x_3,z_3)
\Phi_{j_4} (x_4,z_4)\rangle =
\cr
& =|z_{43}|^{2(\Delta_2 + \Delta_1 - \Delta_4 - \Delta_3)}
|z_{42}|^{-4 \Delta_2}
| z_{41}|^{2(\Delta_3+ \Delta_2 - \Delta_4 - \Delta_1)}
|z_{31}|^{2(\Delta_4 - \Delta_1 - \Delta_2 - \Delta_3)}\times
\cr
&~~~~
|x_{43}|^{2(j_2 + j_1 - j_4 - j_3)}
|x_{42}|^{-4 j_2}
|x_{41}|^{2(j_3+ j_2 - j_4 - j_1)}
|x_{31}|^{2(j_4 - j_1 - j_2 - j_3)}\times \cr
&~~~~ \tilde{C}(j_1,j_3,j_4)
|{\cal F}(z,x)|^2,
}}
where the coefficient
$ \tilde C(j_1,j_3,j_4)$ will be determined later.
We have written the dependence on
the cross ratios $z = { z_{21} z_{43} \over z_{31 } z_{42} }$
and $x={x_{21} x_{43} \over x_{31} x_{42}}$ of the worldsheets
and the target space coordinates in the form of
a square of some homomorphic function ${\cal F}$ in \fpt ,
anticipating that there is only one state in the intermediate channel.
This fact will be derived by explicitly solving the differential
equation below.
The null state condition \nulldec\
for the operator at $z_2$ implies the equation
\eqn\nulle{
\left\{ \left[ {x \over z} - {x-1 \over z-1} \right]
   x(x-1) \partial_x
-\kappa \left[ {x^2 \over z} - {(x-1)^2 \over z-1}
 \right] - {2j_1 x \over z} - {2j_3 (x-1) \over z-1}
 \right\} {\cal F}(z,x) = 0.}
Here $j_2 =  k/2 $ and $\kappa = j_4-j_1-j_2-j_3$.
On the other hand, since
\eqn\kzsimplify{
  L_{-1} | j = {k \over 2}  \rangle
 = -  J_{-1}^3 | j = {k \over 2}  \rangle,}
the KZ equation takes the form,
\eqn\kz{
 \partial_z {\cal F} =
 -  \left\{ {x(x-1) \over z(z-1)} \partial_x + \kappa
\left[ {x\over z} - {x-1\over z-1} \right]
 + {j_1 \over z} + {j_3 \over z-1} \right\} {\cal F}.
}
Using  \nulle\  we can eliminate $\partial_x$ from \kz,
and we obtain
\eqn\anotherkz{
 \partial_z {\cal F}
=  \left\{
 {j_1 \over z} +{j_3\over z-1}
 - { (j_1+j_3+j_4-k/2)
 \over z-x}\right\}
{\cal F}.}
This equation can be easily integrated and we can  insert the
resulting general
solution in \nulle\ to determine the $x$ and $z$ dependence
of ${\cal F}$ completely. We
find
\eqn\solutiontwo{
 {\cal F} =  z^{j_1} (z-1)^{j_3}
     (z-x)^{-j_1-j_3-j_4+ {k\over 2} }
 x^{2j_3+\kappa} (x-1)^{2j_1+\kappa}.
}
The solution is unique up to an overall normalization,
and the four point function is
indeed given by the absolute value squared of this function
as anticipated in \fpt .
Note also that there is a singularity at $z=x$ with precisely
the expected form.

We also need to determine the coefficient $\tilde{C}(j_1,j_3,j_4)$
in \fpt . We use the same method as
in \refs{\zam , \teschnerone} .
The standard operator product expansion formula
gives $C(j_1,\khalf,j)
B(j)^{-1} C(j,j_3,j_4)$, where $j$ is for the intermediate
state. As we mentioned earlier in \anotherdelta,
the factor $C(j_1, \khalf, j)$ is equal
to the delta function
$ \delta(j_1+j - \khalf)$ modulo a $k$ dependent
($j_1$ independent) coefficient.
This is consistent with the fact that, in
\solutiontwo , only the state with
$j= \khalf  - j_1  $ is propagating in the intermediate channel
for $z\to 0$. Thus
the coefficient $\tilde{C}$ is determined as
\eqn\fourpt{
 \tilde C(j_1,j_3,j_4) \sim B\left(\khalf-j_1\right)^{-1}
C\left(\khalf-j_1,j_3,j_4\right)
\sim B(j_1) C\left(\khalf-j_1,j_3,j_4 \right) }
modulo a $k$ dependent factor. Here we used \defofbtes .

As shown in \first\ and reviewed in the last subsection,
the spectral flow operator is given by the operator $\Phi_{{k\over
2}}$ in the $m$ basis. Thus we need to perform the integral
transform \basis\ on  \fpt\ and set $m_2 = -\khalf $.
As in the case of two point function, setting this value
of $m_2$ generates a pole in the amplitude so
 the spectral flow operator is defined by
\eqn\spectflow{
e^{- i \sqrt{  k \over 2} \varphi} \sim  {1\over V_{conf}}~
 \Phi_{\khalf,-\khalf ; \khalf, -\khalf }
}
where the operator  $ \Phi_{\khalf ,- \khalf ; \khalf,
-\khalf }$ is normalized as in
\basis . The factor $1/V_{conf}$ is there to remind us
 that we have to extract a pole residue at $m = -
\khalf$. This residue can be evaluated
by taking the limit $x_2 \to \infty $ of
$|x_2|^{2k}$ times \fpt . After performing the $x_i$ integrals,
we find  \zam ,
\eqn\zamres{\eqalign{
& \int \prod_{i=1,3,4} d^2 x_i x_i^{j_i-m_i -1} {\bar x_i}^{j_i - \bar
m_i-1}
 \left\{ \lim_{x_2 \to \infty} |x_2|^{2k} \fpt \right\} =
 \cr
& = \tilde C(j_1,j_3,j_4)
\delta^2 ( -\khalf +  m_1 + m_3 + m_4,
-\khalf +  \bar m_1 + \bar m_3 + \bar  m_4 ) \times
\cr
 & ~~~
\prod_{i=1,3,4} (z_2-z_i)^{m_i}
 z_{13}^{\Delta_4 - \Delta_1 -\Delta_3 + {k \over 4}  +m_4}
 z_{34}^{\Delta_1 - \Delta_3 -\Delta_4 + {k \over 4} + m_1}
 z_{41}^{\Delta_3 - \Delta_4 -\Delta_1 + {k \over 4} + m_3}
 \times
\cr
&~~~
\prod_{i=1,3,4} (\bar z_2-\bar z_i)^{\bar m_i}
 \bar z_{13}^{\Delta_4 - \Delta_1 -\Delta_3 + {k \over 4}
  +\bar m_4}
 \bar z_{34}^{\Delta_1 - \Delta_3 -\Delta_4 + {k \over 4}
+\bar  m_1}
 \bar z_{41}^{\Delta_3 - \Delta_4 -\Delta_1 + {k \over 4}
+ \bar m_3}
 \times \cr
& ~~~~~~  { 1 \over \gamma(j_1+ j_3 + j_4 - \khalf )}
\prod_{i=1,3,4} { \Gamma(j_i - m_i)
\over \Gamma(1 - j_i + \bar m_i),  }
}}
 where ``cyclic'' means a cyclic permutation of the
labels $134$.
The $z_2$ dependence is what we expect for the operator \spectflow .
 We can now extract
the action of the spectral flow operator on $\Phi_{j_1}$.
This is done by taking
the limit of $z_2 \to z_1$ and extracting the coefficient of the term
which goes like $ z_{12}^{m_1} {\bar z}_{12}^{\bar m_1} $.
This performs spectral flow on the operator inserted at $z_1$ by $-1$
unit.\foot{We have $-1$ unit of spectral flow because we
extracted the $m_2 = -\khalf$
component of the spectral flow operator in \zamres .
The resulting operator represents an outgoing state carrying away
one unit of winding number.} According to the rules \sfagain\
of the spectral
flow, the new spacetime quantum numbers of the operator at $z_1$ are
$M = m_1 -\khalf  $ and $\bar M = \bar m_1 -  \khalf $, and its
global $SL(2,C)$ left and right conformal weights are
$J=|M|$ and $\bar J= |\bar M| $.
Finally we find
\eqn\zamresm{\eqalign{
&\langle \Phi^{w=-1, \ j_1}_{J,M,\bar J,\bar M }(z_1)
\Phi_{j_3,m_3,\bar m_3}(z_3) \Phi_{j_4,m_4 \bar m_4 }(z_4)  \rangle =
 \cr
&= \tilde C(j_1,j_3,j_4)
~\delta^2( -\khalf +  m_1 + m_3 + m_4,
-\khalf +  \bar m_1 + \bar m_3 + \bar  m_4 ) \times
\cr
&  z_{13}^{\Delta_4 - \hat \Delta_1 -\Delta_3}
 z_{34}^{\hat \Delta_1 - \Delta_3 -\Delta_4}
 z_{41}^{\Delta_3-\Delta_4 - \hat \Delta_1}
\times \cr
&  \bar z_{13}^{\Delta_4 - \overline{\hat \Delta}_1 -\Delta_3}
 \bar z_{34}^{\overline{\hat \Delta}_1 - \Delta_3 -\Delta_4}
 \bar z_{41}^{\Delta_3-\Delta_4 - \overline{\hat \Delta}_1}
\times  \cr
& { 1 \over \gamma(j_1+ j_3 + j_4 - \khalf)}
 { \Gamma(j_1 - m_1)
\over \Gamma(1 - j_1 + \bar m_1)  }
 { \Gamma(j_3 - m_3)
\over \Gamma(1 - j_3 + \bar m_3)  }
 { \Gamma(j_4 - \bar m_4)
\over \Gamma(1 - j_4 + m_4)  } ,}}
with
\eqn\with{\eqalign{
&  \hat \Delta_1 = \Delta(j_1) + m_1 -{k\over 4} ~,~~~~~~~~
\overline { \hat \Delta}_1 = \Delta(j_1) + {\bar m}_1 -{k\over 4}
\cr
&J = -M = -m_1 + { k \over 2} ~,~~~~~~~~~~
\bar J = -\bar M = - {\bar m}_1 + { k \over 2},
 }}
where we used the $\delta$ function in $m_i$ to go from \zamres\ to
\zamresm .\foot{ We also used properties of the Gamma function to
absorb the sign  $(-1)^{m_4 - \bar m_4}$
that came from the powers of $z_{14}$ in going from
\zamres\ to \zamresm .}
This indeed has the expected $z$ dependence for the correlation
function of one spectral flowed operator with two unflowed operators.


\subsec{ Three point functions in $x$ basis}

In this subsection we will discuss how to go from the $m$ basis
to the $x$ basis for three point functions. We want to rewrite
\zamresm\ in the $x$ basis. This is similar to what we did for
the two point functions.

We start with a general three point function in the $x$-basis
\eqn\threept{\eqalign{
\langle \Phi_{J_1,\bar J_1} (x_1) & \Phi_{J_2 \bar J_2}(x_2)
\Phi_{J_3 \bar J_3}(x_3)\rangle  =
\cr
 & = { D(J_1,J_2,J_3) \over x_{12}^{J_1+ J_2 - J_3 }
 x_{13}^{J_1+ J_3 - J_2 }  x_{23}^{J_2+ J_3 - J_1 }
\bar x_{12}^{\bar J_1+ \bar J_2 - \bar J_3 }
 \bar x_{13}^{\bar J_1+ \bar J_3 - \bar J_2 }
\bar x_{23}^{\bar J_2+ \bar J_3 - \bar J_1 }  } ,}}
where the $J_i,~\bar J_i$ label
 the conformal weight under global $SL(2,C)$.
We can compute the integral transform of
this expression to go to  the $m$ basis. The integral can be written
using the Barnes hypergeometric function \SatohBI .
For our purposes
 we do not need to compute the most general expression since the
three point function \zamresm\ is really the residue of the pole
 at $J_1=-M_1 ~,\bar J_1 = - \bar M_1$
in the $x$ integral of \threept .
This pole comes  from the region where
$x_1$ is very large.
We  are interested in the coefficient of this pole.
This
is obtained by  taking the $x_1 \to \infty$
limit of $ x_1^{2J_1} {\bar x_1}^{2 \bar J_1 }$
times \threept\ and then
performing the integral transform with respect to $x_2$ and $x_3$.
We obtain
\eqn\corrmbasis{\eqalign{
&\langle \Phi_{J_1=-M_1,\bar J_1= -\bar M_1}
\Phi_{J_2 \bar J_2 ,M_2,\bar M_2}
\Phi_{J_3 \bar J_3 ,M_3,\bar M_3} \rangle \sim
\cr
& \sim V_{conf}~ \delta^2 (M_1 + M_2 + M_3)  D(J_1,J_2,J_3)\times
\cr &~~~~~ {\Gamma(\bar J_3- \bar M_3) \over \Gamma(1 - J_3 + M_3)}
  {\Gamma( J_2- M_2) \over \Gamma(1 - \bar J_2+ \bar M_2)}
{ \Gamma( 1 + \bar J_1 - \bar J_2- \bar J_3 )  \over
\Gamma( J_2+ J_3 - J_1 ),}
}}
where the $V_{conf}$ is there
to remind us that the rest is the residue of a pole.
Notice that only properties under global $SL(2,C)$ have been used to
derive this formula.

By comparing \zamresm\ to  \corrmbasis\
(and changing labels $(2,3) \to (3,4)$
in \corrmbasis\ in the obvious way), we find that
 the three point function in $x$ space is given by
\eqn\threepx{\eqalign{
&\langle \Phi^{w=1,j_1}_{J,\bar J }(x_1)
\Phi_{j_3} (x_3) \Phi_{j_4}(x_4)\rangle \sim
 {1\over V_{conf}}~B(j_1) C\left(\khalf -j_1 , j_3,j_4\right) \times \cr
&~~~~~~~
 {\Gamma( j_3 + j_4 - J) \over \Gamma( 1 + \bar J - j_3 - j_4)}
 {\Gamma( j_1  + J - \khalf  )
\over\Gamma(1- j_1  - \bar J +  \khalf  ) }
{1 \over\gamma(j_3 + j_4  + j_1 - \khalf ) }
\times
\cr
&~~~~~~~{ 1 \over  x_{13}^{J + j_3 - j_4 }
 x_{14}^{J + j_4 - j_3 }  x_{34}^{j_3+ j_4- J }
\bar x_{13}^{\bar J + j_3 - j_4 }
\bar x_{14}^{\bar J + j_4 - j_3 }  \bar x_{34}^{j_3+ j_4-\bar J }
}
 }}
where
\eqn\parame{\eqalign{
&J_1 =- m_1 + \khalf ~,~~\bar J_1 = -\bar m_1 + \khalf ~,\cr
&J_{3,4}= \bar J_{3,4} = j_{3.4}.}
}
In the case of $j = \half + is$, when the first operator
corresponds to a long string, this factor of $1/V_{conf}$
is cancelled since the long string operator $\hat{\Phi}$ comes with
the extra factor of $V_{conf}$ as in \correcont .
Thus we conclude that the three point function of
two short strings with $w=0$ and one long string with
$w=1$ is non-zero. In the following subsection, we
compare the expression \threepx\ with the factorization
of the four point function.

In Appendix D, we will show, using the representation theory
of the $SL(2,R)$ current algebra, that two short strings
with $w=0$ can only be mixed with short strings with
$w=0, 1$ or long strings with $w=1$. One may ask why
we did not see short strings with $w=1$ in
the factorization of the four point function.
In fact there is an additional reason for the vanishing
of the three point amplitude with two short strings
with $w=0$ and one short string with $w=1$.
If $j_1$ is real and $m_1, \bar m_1 <0$ ,
the operator $\Phi^{w=1, j_1}_{J,\bar J}$ in \threepx\ corresponds to
a short string with $w=1$. For this operator, the two point function
is finite as we saw in \floweddis, and we do not have to rescale
the operator as we did for the long string.
Thus we interpret the factor of $1/V_{conf}$ in \threepx\ as saying that the
three point function vanishes. This gives the additional
constraint on the winding
number violation stating that two short strings with $w=0$ cannot
produce a short string with $w=1$.

As a check that we are interpreting this factor of $1/V_{conf}$
correctly and
as a further application of \threepx , let us consider the case that
$j_1$ is real and $m_1 , \bar m_1>0$, $ m_1 = j_1 + p , ~
\bar m_1 = j_1 + \bar p $.
This can be interpreted as doing the spectral flow of a discrete
representation by $-1$ unit, thus producing the operator described at
\statef\ with $j = \khalf - j_1 $. This state is just a descendant
in a discrete representation with $w=0$ . Thus, in this case,
we do not expect the three point function to
vanish. Indeed we find that, as we set $m_1 = j_1 + p$,
one of the Gamma functions
in \threepx\ develops a pole, thereby cancelling the factor
$1/V_{conf}$ in \threepx . Finally we obtain
\eqn\threepxus{\eqalign{
&\langle {\hat \Phi}^{jp\bar p}_{J \bar J }(x_1)
\Phi_{j_3} (x_3) \Phi_{j_4}(x_4)\rangle\sim
\cr
 & \sim (-1)^{p + \bar p} C(j , j_3,j_4)
 {\Gamma( j_3 + j_4 -j  +p) \over p! \Gamma( j_3 + j_4 -j)}
 {\Gamma( j_3 + j_4 -j  + \bar p) \over
\bar p! \Gamma( j_3 + j_4 -j)}\times
\cr
&~~~~~~ { 1 \over x_{13}^{J+ j_3 - j_4 }
 x_{14}^{J+ j_4 - j_3 }  x_{34}^{j_3+ j_4- J }
\bar x_{13}^{\bar J+ j_3 - j_4 }
\bar x_{14}^{\bar J+ j_4 - j_3 }
\bar x_{34}^{j_3+ j_4- \bar J } }.
}}
Note that $j = \khalf - j_1$, where $j_1$ is the label appearing
in \threepx , and $J = j- p$, $\bar J = j- \bar p$. We have also
normalized
the operator as in \newnorm .
If we set $p=\bar p =0$, we indeed find that this is the same as the
correlation function of three $w=0$ discrete states. This is an
interesting consistency check of what we are doing.
Moreover we will see that the expression \threepxus\
exactly matches with the factorization of the four point function
in the target spacetime.

\newsec{Factorization of four point functions}

In the last section, we computed the two and
three point functions including spectral flowed
operators. In this section, we will use these results
to show  that the coefficients of the powers of
$x$ appearing in the spacetime operator product expansion
computed in section 4 are precisely what
are expected, $i.e.$, each of them is a product of two
three point functions
involving the intermediate state divided by the two point function
of that intermediate state.

\subsec{Factorization on long strings}

Let us first examine the coefficient for the continuous representations
appearing in \whatrone .
In the expression \whatrone , the integration contour runs along
$j_c = k/2 -1/2 + is = k/2 -  j$ where $ j = 1/2 - is$.
We are denoting the $SL(2,C)$ spin along the contour
by $j_c$, and $j$ is introduced for convenience.
Then we define
\eqn\ener{
J = j_c + d(j_c) = { s^2 + 1/4 \over k-2} + h -1~,~~~~~
\bar J = j_c + \bar d(j_c) = { s^2 + 1/4 \over k-2} + \bar h -1
}
{}From the power of $x$ in \whatrone , we conclude that $J$ is the
spacetime conformal weight of the intermediate state. The
coefficient of this power of $x$ is \whatrone
\eqn\coeff{\eqalign{
& {\Gamma( J - \khalf + j) \over \Gamma( 1 - \bar J + \khalf  - j) }
{\Gamma( j_1 + j_2 - \bar J ) \over \Gamma(1 -  j_1 - j_2 +  J )}
{\Gamma( j_3 + j_4 - \bar J ) \over \Gamma(1 -  j_3 - j_4 + J )}
{\Gamma(1 + J -j -\khalf) \over \Gamma( \khalf + j - \bar J) } \times
\cr
& ~~~
{ \gamma(2j) \over \gamma(j_1 + j_2 + j - \khalf)
\gamma(j_3 + j_4 + j - \khalf)}
{C(j_1,j_2,\khalf - j) C(\khalf - j,j_3,j_4) \over B(\khalf - j) }
}}
It can be shown that this coefficient is given by the product
of two of the three point functions divided by the two point function.
This can be seen explicitly by writing \coeff\ as
\eqn\coeffmex{\eqalign{
&
B(j)C\left(\khalf - j,j_1,j_2\right)
{\Gamma( j_1 + j_2 - J ) \over \Gamma(1 -  j_1 - j_2 + \bar J )}
 {\Gamma( j + J - \khalf ) \over \Gamma( 1 -j -  \bar J + \khalf  ) }
{1 \over \gamma(j_1 + j_2 + j - \khalf)}
\times \cr
& ~~~~~~~
{ \gamma(2 j) \over B(j) }
{\Gamma(1  -j -\khalf + J) \over \Gamma( j + \khalf  - \bar J) }
{ \Gamma( 1 -j  + \khalf  - \bar J ) \over \Gamma( j  - \khalf + J ) }
\times \cr
&~~~ B(j) C\left(\khalf - j, j_3, j_4\right)
{\Gamma( j_3 + j_4 - J ) \over \Gamma(1 -  j_3 - j_4 + \bar J )}
 {\Gamma( j + J - \khalf ) \over \Gamma( 1 -j -  \bar J + \khalf  ) }
{1 \over \gamma(j_3 + j_4 + j - \khalf)}.
}}
Note that we used that $B(\khalf -j) \sim B(j)^{-1} $.
We see that this has the form of the product of two three point
functions \threepx\ divided by the two point function  \correcont\
with $w=1$.
Note that, in the \correcont , there are two terms, one proportional to
$\delta(s+s')$ and the second proportional to $\delta(s-s')$.
Physically $s$ and $-s$ describe the same operator. So when we
consider the inverse of the two point function it is convenient to
restrict the integral over $s$ so that $s\geq 0$. This is
possible in \whatrone\ since the expression is symmetric under
$s \to -s$. In this prescription,
 only the term proportional to $\delta(s-s')$
in the two point function needs to be inverted.
We have checked that,
 if we took the other term in the
two point function, the one proportional to $\delta(s+s')$,
we will find the same result provided we switch the
sign of $s$ in one of the three point functions  in \coeffmex , precisely
as required.


\subsec{Factorization on short strings}

We now consider
\intadd\ where, as we  explained before, we should shift the $j$
contour of integration. This picks up some poles explicitly
displayed in \expandF .
These poles are at $d =-n$, $\bar d = -\bar n$. From their
$x$ dependence we conclude that the spacetime conformal weight
of the intermediate operator is $J = j-n$, $\bar J = j - \bar n$.
The residue of the pole is
\eqn\polesinj{\eqalign{
& { 1 \over \partial d / \partial j}
 { C(j_1,j_2, j)   C(j,j_3,j_4 ) \over B(j) } \times \cr
&~~~ {\Gamma(j_1+ j_2 - j + n) \over n! \Gamma(j_1+ j_2 - j) }
{\Gamma(j_3+ j_4 - j + n) \over n! \Gamma(j_3+ j_4 - j) }
{n! \Gamma(k-2j) \over  \Gamma(k-2j +n )} \times\cr
&~~~ {\Gamma(j_1+ j_2 - j + \bar n) \over \bar n! \Gamma(j_1+ j_2 - j) }
{\Gamma(j_3+ j_4 - j + \bar n) \over \bar n! \Gamma(j_3+ j_4 - j) }
{\bar n! \Gamma(k-2j) \over  \Gamma(k-2j +\bar n )}.}}
The factor $(\partial d(j) / \partial j)^{-1}$
 appears here since the pole we picked up
in \expandF\ is of the form $\sim (d(j)+ n)^{-1}$
and we are evaluating residues in the $j$ integral in \intadd .
We see that this
has precisely the expected form for a state like
\statef\ propagating in the intermediate channel.
Indeed we can write \polesinj\ as the product of two three point
functions \threepxus\ divided by the coefficient of the
two point function \twoneg , including
the factor involving $\partial d / \partial j \sim  (2 j-1) $,
which we discussed at \twoptcorr\ as
\eqn\polesinjtwo{\eqalign{
&(-1)^{n+\bar n}
C(j,j_1,j_2) {\Gamma(j_1+ j_2 - j + n) \over n! \Gamma(j_1+ j_2 - j) }
{\Gamma(j_1+ j_2 - j + \bar n) \over \bar n! \Gamma(j_1+ j_2 - j) }
 \times\cr
&~~~~~~~~{ 1 \over  (2j-1) B(j) }
{n! \Gamma(k-2j) \over  \Gamma(k-2j +n )}
{\bar n! \Gamma(k-2j) \over  \Gamma(k-2j +\bar n )}
\times \cr
&~~(-1)^{n+\bar n} C(j, j_3, j_4)
{\Gamma(j_3+ j_4 - j + n) \over n! \Gamma(j_3+ j_4 - j) }
{\Gamma(j_3+ j_4 - j + \bar n) \over \bar n! \Gamma(j_3+ j_4 - j) }
.}}
In other words, we need to
correct the two point function by the factor $(2j-1)$
as in \twoptcorr\ in order
to get the right spacetime factorization properties. This completes
the test of the factorization of the four point function.

\newsec{ Final remarks}

Most of what
 we said in this paper refered to the Euclidean theory, both on the
worldsheet and on target space.
These computations can also be interpreted as describing
string theory on a Lorentzian target space.
Note that string theory in Lorentzian $AdS_3$  can be thought
of in terms of the usual S-matrix formulation, where the asymptotic
states are the long strings. Short strings appear as
poles in the long string amplitudes. We did not compute this precisely
but this is the expected picture. It would be interesting to
expand the four point function for two long strings with $w=1$ and
two with $w=-1$, and see that indeed we produce only long and short
strings in accord to the winding violation rule described in the
appendix D. In this way,
the theory on the Lorentzian $AdS_3$ can be interpreted either in terms
of an S-matrix or in terms of a BCFT, albeit one with a non-compact
target space. The S-matrix computation of long strings is
in fact describing scatterings in the Lorentzian BCFT. It
is amusing to note that
this singular BCFT is reproducing some features which seem characteristic
to strings in flat space, such as
 having an S-matrix description. This may give us a
hint on how to construct a holographic description of flat
space physics.

This BCFT is rather peculiar due to the non-compactness of its
target space. All the computations we have been defining were for the
case that the BCFT is on $S^2$. These computations are well-defined when
properly interpreted, as we discussed in this paper. The only peculiarity
is that we cannot insert too many discrete state operators, but this
should not be surprising since we also saw simple quantum mechanical
models where that is true.
If we put the BCFT on a torus, we will find divergences in one
loop computations as we have shown explicitly in \second .
In \second\ these divergences were regulated by adding a volume cutoff
near the boundary, but strictly speaking the one loop free energy is
infinite. We would find a similar result in the quantum mechanical example
we discussed in section 3.2.
This BCFT would not be well-defined on a higher genus Riemann
surface.

The $SL(2,R)$ WZW model has an interesting algebraic structure which
should be explored further.

\bigskip
\bigskip

\centerline{{\bf Acknowledgments }}

\bigskip

We would like to thank
D. Kutasov,
 J. Teschner, N. Seiberg, S. Shenker,  and A. Zamolodchikov for discussions.

\smallskip

We thank the Institute for Theoretical Physics at the University
of California, Santa Barbara, for hospitality.
HO would also like
to thank Harvard University and the Aspen Center for
Physics. The research of JM
was supported in part by DOE grant DE-FGO2-91ER40654,
NSF grants PHY-9513835 and PHY99-07949, the Sloan Foundation
and the David and Lucile Packard Foundation.
The research of HO was supported in part by
DOE grant DE-AC03-76SF00098 and by Caltech Discovery Fund.

\vfill

\appendix{A}{Target space two point function of short string
with $w=0$}

In section 5, we computed the target space two point function
starting with the worldsheet two point function and dividing it
by the volume of worldsheet conformal symmetry which fixes the
two points. This process involved some subtlety since we have
to cancel two divergent factors, leaving the finite coefficient
$|2j-1+(k-2)w|$ for short string.\foot{There is no such factor for
long string.} In this appendix, we present an alternative
derivation of the target space two point function in the case
of $w=0$.

The idea is to use the target space Ward identity.
We assume that there is some current algebra symmetry
in the BCFT and use it to relate the three point function
including the conserved current to the two point function
that we want to compute.\foot{If there is no current algebra
symmetry, one can use the energy-momentum tensor, which exists
in any CFT. It is
straightforward to generalize the following computation
with the energy-momentum tensor, and one obtains the
same normalization for the target space two point function.}
One may view this as the string theory
version of the computation in \freedman\ where
a similar factor  for the two point function was
derived using the Ward identity in the supergravity
approximation.

The global symmetry of the BCFT comes from a current
algebra symmetry in the internal CFT on the worldsheet.
According to \gk, the vertex operator for the target
space current is given by $J(x,z) \bar{L}(\bar z)
\Phi_{j=1}(x,z)$, where
\eqn\currentform{
J(x,z)
= -J^-(z) + 2x J^3(z) - x^2 J^+(z),}
and $\bar{L}(\bar z)$ is the
current algebra generator in the internal CFT.
Thus, to compute the Ward identity for the target
space two point function,
we need to evaluate a three point function
$\langle \Phi_{j_1} \Phi_{j_2} \Phi_{j=1}\rangle$.
Due to the fact that
two point function of the internal CFT is non-zero
only between operators with the same conformal
dimensions, the on-shell condition requires
$j_1=j_2$ and we can focus our attention to this case.
We then find that the $AdS_3$ part
of the correlation function is of the form,
\eqn\threewithjone{\eqalign{
  &\langle \Phi_{j}(x_1,z_1) \Phi_{j}(x_2,z_2)
  \Phi_{1}(x_3,z_3)\rangle
\cr
& = {G(-2j)\over 2\pi^2 \nu^{2j} \gamma\left({k-1\over k-2}\right)
G(1-2j)}  {1\over
|z_{12}|^{2\Delta}
|x_{12}|^{2(2j-1)}|x_{23}|^2 |x_{31}|^2}
\cr
&= {1\over
2\pi^2 \nu^{2j} \gamma\left({k-1\over k-2}\right)
\gamma\left({2j-1\over k-2}\right)}
{1\over
|z_{12}|^{4\Delta}
|x_{12}|^{2(2j-1)}|x_{23}|^2 |x_{31}|^2}
\cr
&= {1\over \nu \cdot 2\pi  (k-2)
\gamma\left({k-1\over k-2}\right)}
{B(j)\over
|z_{12}|^{4\Delta}
|x_{12}|^{2(2j-1)}|x_{23}|^2 |x_{31}|^2}
,}}
where $\Delta=-j(j-1)/(k-2)$.
We then multiply the current generator $J(x_3,z_3)$ on
$\Phi_1(x_3,z_3)$. Using the operator product
expansion,
\eqn\currentope{
 J(x,z) \Phi_j(y,w)
\sim {1\over z-w}\left( (x-y)^2 {\partial \over \partial y}
 - 2j (x-y)\right) \Phi_j(y,w),}
we find
\eqn\spacetimeope{
 \eqalign{
 &\langle \Phi_{j}(x_1,z_1) \Phi_{j}(x_2,z_2)
\left( J(x_3,z_3) \Phi_{1}(x_3,z_3)\right)\rangle
\cr
&\sim {(2j-1)B(j) \over
|z_{12}|^{4\Delta} |x_{12}|^{4j}}
\left( {1\over \bar x_3-\bar x_1}
- {1 \over \bar x_3-\bar x_2}\right)
\left({1 \over z_3-z_1} - {1\over
z_3-z_2}\right),}}
where we ignored a constant independent of $j$.
To obtain the spacetime three point function,
we choose an operator $\phi_h$ of dimension $h$
in the internal CFT so that $\Delta + h = 1$
and multiply it to $\Phi_j$. Similarly we
multiply the current generator
$\bar{L}$ of the internal CFT to $\Phi_1$. We find
\eqn\finalwardid{
\eqalign{&
\langle \left(\Phi_{j}(x_1,z_1)\phi_h(z_1)\right)
\left( \Phi_{j}(x_2,z_2)
\phi_h'(z_2)\right)\left(
J(x_3,z_3)\bar{L}^a(\bar{z}_3) \Phi_{1}(x_3,z_3)
\right)\rangle
\cr&\sim {1\over |z_{12}|^2 |z_{23}|^2
|z_{31}|^2}\left( {q_1 \over \bar x_3 -\bar x_1}
+ {q_2 \over \bar x_3-\bar x_2}\right)
{(2j-1) B(j)  \over |x_{12}|^{4j}},}}
where $q_1$ and $q_2$ are the R-charges
$\phi_h(z_1)$ and $\phi_h'(z_2)$ respectively,
and we used the charge conservation,
$q_1 + q_2=0$. Comparing
this with what we expect for the target space Ward identity,
we find that the spacetime
two point function is given by
\eqn\targettwopointagain{
\langle \Phi_j(x_1) \Phi(x_2) \rangle
\sim  {(2j-1) B(j) \over |x_{12}|^{4j}},}
reproducing the result we have obtained
using the heuristic argument in section 5.1.

It is easy to see that if we insert in \targettwopointagain\ the
operator $J\bar J \Phi_1(x_3)$ we obtain \targettwopointagain\ times
an extra factor of $(2j-1)$ in agreement with the arguments in
\gklast .

\appendix{B}{ Some properties of the conformal blocks}

In this appendix, we will prove that the conformal block
${\cal F}_j(z,x)$
of the four point function has no poles in $j$ when  ${1\over 2}
 \leq Re~j \leq {k-1 \over 2}$.
We also argue that the integral
over $j$ in \expres\ is convergent.

\subsec{ Proof of no poles in ${\cal F}_j$ in ${1\over 2}
 \leq Re~j \leq {k-1 \over 2}$}

This has been shown in \teschnersecond\ using properties
of the Kac-Kazhdan determinant. Here we present a direct
proof of the absence of poles.

We use the expansion
\regionone\  as
\eqn\regionone{
\cf_j(z,x) = x^{\Delta(j)-\Delta(j_1)-\Delta(j_2) +j-j_1-j_2 }
             ~ u^{\Delta(j)-\Delta(j_1)-\Delta(j_2)}
 \sum_{n=0}^\infty g_n(u) x^n,}
where $u = z/x$. As we discussed in section 4,
the KZ equation and the boundary condition for small $z$
determines that $g_0(u)$ is given by the hypergeometric function,
\eqn\gzero{ g_0(x) = F(j_1+j_2-j,
j_3+j_4-j,k-2j;u).}
The standard Taylor expansion of the hypergeometric function
shows that
the $u$ expansion of $g_0(u)$ has
no poles in the region \deformregion.

 Let us write
\eqn\whatr{r\equiv {k-1 \over 2} - j,}
and
\eqn\whata{ a(r) \equiv  - {r^2 \over k-2},}
which is defined so that
\eqn\whatadoes{ \Delta(j) + j  =
a(r) + {k \over 4} + {1 \over 4(k-2)}.}
We then look for a solution to the KZ equation
in the power series expansion of the form,
\eqn\expansionwithb{ \eqalign{
 F(x,u) &= x^{a(r) +{k\over 4} + {1 \over 4(k-2)}
    -\Delta(j_1) -\Delta(j_2) -j_1-j_2}
  u^{b(r)   -{k-2\over 4} + {1 \over 4(k-2)}-
 \Delta(j_1) - \Delta(j_2)}\times \cr
&~~~~~~ \sum_{m,n=0}^\infty
  \tilde{c}_{m,n} u^m x^n,}}
where $b$ is some constant which will be determined below.
We have chosen $m, n = 0, 1, \cdots$
so that the expansion is consistent with  \smallz \whatfn \regionone .
The fact that $g_0(u)$ has no poles means that
$\tilde{c}_{m,n=0}$ has no poles. We will then show
inductively that this is also the case for all $\tilde{c}_{m,n}$
with $n\geq 1$.

Substituting this into the KZ equation, we find the recursive equation
for the coefficient $c_{n,m}$ of the form,
\eqn\recurse{ P(a(r)+n, b+m) \tilde{c}_{m,n} =
  ({\rm linear~combination~of~} \tilde{c}_{m',n'}, ~m'< m,~ n'\leq n ),}
for some function $P(a,b)$, which is quadratic in $b$.
The right hand side contains no
poles in $j$. For $n=0, m=0$, this
gives the condition $P(a(r), b)=0$. This is nothing but the characteristic
equation for the hypergeometric equation on $g_0(u)$, and
we know that this determines $b$ to be $b=b_\pm(r) = a(r)
\pm r$. In \gzero, we have chosen the $+$ root in order
to fit it with the boundary condition \smallz. With this
choice of $b$, we want to show that $P(a(r)+n, b_+(r) + m)$
is non-zero for any $n\geq 1$ and $m \geq 0$
in the region \deformregion , or equivalently
$0 \leq Re~r \leq {k-2\over 2}$. If this is true,
by recursive application of \recurse , we can show that
$\tilde{c}_{m,n}$ has no pole.

Our strategy is to look for a solution to $P(a(r)+n, b')=0$
for $n\geq 1$
and show that it can never be of the form $b'=b_+(r) + m$
for any $m \geq 0$. Let us write $a(r) + n = a(r')$ for
some $r'$. We know that the zeros of $P(a(r'),b')=0$
are given by $b'= b_\pm(r')$. Let us first consider the solution
of $b'=b_+(r')$. Since
\eqn\equalities{
 b_+(r') = a(r')+r' = a(r) + r' + n,}
then $b'$ could be  equal to $b_+(r) + m=a(r) + r + m$
if and only if $r' = r + m-n$. On the other hand,
$r'$ was defined by
$a(r)+n = a(r')$. Eliminating $r'$, we find the
condition
\eqn\notpossible{ -2mr - (n-m)^2 =
 n(k-2-2r).}
This cannot be satisfied by $r$ in the range
$0 \leq  Re~r \leq {k-2\over 2}$ for $(n,m) \not = (0,0)$
since the real part of the left-hand side is negative
while it is positive in the right-hand side.
For the other solution $b'=b_-(r')$, we also
find the same equation \notpossible.
Thus we have shown that $P(a(r)+n, b_+(r)+m)$
never vanishes for $n \geq 1, m \geq 0$ if $r$ is
in this range. This proves
that ${\cal F}_j(z,x)$ has no pole in the region of
our interest.

\subsec{Convergence of the expression for the four point function}

To see that the $j$ integral \expres\ is indeed convergent, we note
that, as a function of $j={1\over 2} + is$,
the coefficient $|{\cal C}(j)|$
behaves as $\sim e^{\alpha s}$ for large $s$ where $\alpha$ is
some constant. This can be deduced from the expression for the
two and three point functions, \defofbtes , \threecoeff ,
and \whatg , using the asymptotic formulae of the Gamma function
and the Barnes Gamma function.
Due to the factor
$u^{a(r)}$ in \expansionwithb\
each term in the $u$ expansion of ${\cal C}(j)|{\cal F}_j|^2$
decays as $e^{-\beta s^2}$ for
$s \rightarrow \pm \infty$ as long as $|u| < 1$.
We can see, using \recurse\ that the coefficients in the
$u$ and $x$ expansion do not grow more than polynomially in $n,m$,
so that these sums will converge if $|u|<1$, $|x|<1$.
For other values of $u$, $x$,
 \expres\ is defined by analytic continuation.

\appendix{C}{A useful formula}

In this appendix we derive the formula
\eqn\integraluse{
\eqalign{
I(a,b,c,d, \bar d ) & =
\int d^2 u u^{d-1}\bar{u}^{\bar{d}-1} \left(
|F(a,b,c;u)|^2
+ \lambda |u^{1-c} F(1+b-c,1+a-c,2-c;u)|^2\right) \cr
%
&=\pi
{\Gamma(d)\Gamma(a-\bar{d}) \Gamma(b-\bar{d})
\Gamma(1-c+d)
\over \Gamma(1-\bar{d})\Gamma(1-a+d)\Gamma(1-b+d)
\Gamma(c-\bar{d})}
{\gamma(c)\over \gamma(a)\gamma(b)}.
}}
where
\eqn\whatlambda{
\lambda = - {\gamma(c)^2 \gamma(a-c+1)\gamma(b-c+1)
\over (1-c)^2 \gamma(a)\gamma(b)},}
and $\gamma(x)$ is defined in \defofsmallgamma .
The formula \integraluse\ is obtained as follows.
Let us first prove the following identity,
\eqn\identity{
\eqalign{
&|F(a,b,c;u)|^2
+ \lambda |u^{1-c} F(1+b-c,1+a-c,2-c;u)|^2 = \cr
&=
{\gamma(c) \over \pi \gamma(b) \gamma(c-b)}
  |u^{1-c}|^2 \int d^2 t ~|t^{b-1}(u-t)^{c-b-1}(1-t)^{-a}|^2.}}
This is based on the following formula.
\eqn\dotsenko{
\eqalign{\int d^2 t |t^a(u-t)^c(1-t)^b|^2
 =& {\sin(\pi a) \sin(\pi c) \over \sin(\pi(a+c))}
\left| \int_0^u dt ~t^a(u-t)^c(1-t)^b\right|^2 + \cr
 &+ {\sin(\pi b) \sin(\pi (a+b+c)) \over \sin(\pi(a+c))}
\left| \int_1^\infty dt ~t^a(u-t)^c(1-t)^b\right|^2.}}
A derivation of this formula can be found, for example,
in \ref\dotsenkonote{
V. S. Dotsenko, ``Lectures on Conformal Field Theory,''
Proceedings of RIMS Workshop, {\it Conformal Field Theory
and Solvable Lattice Models}, Kyoto, 1986.}, where
it appears in the context of the free boson
realization of the $c<1$ conformal field theory. There the variable $t$
corresponds to the location of the screening operator.
Using the fact that the $t$ integrals in the right-hand side of
\dotsenko\ can be expressed in term of the hypergeometric
function, we obtain \identity.
The integral $I(a,b,c,d,\bar d)$ of the hypergeometric
functions can then be expressed as the following double-integral,
\eqn\newin{
I(a,b,c;d,\bar{d}) =
{\gamma(c) \over \pi \gamma(b) \gamma(c-b)}\int d^2u d^2t
 ~u^{d-1} \bar{u}^{\bar{d}-1}|t^{b-1}(u-t)^{c-b-1}(1-t)^{-a}|^2.
}
It turns out that both $u$ and $t$ integrals can be carried out
using the formula,
\eqn\integralformula{
\int d^x |x|^{2a} |1-x|^{2b} x^n (1-x)^m
= \pi {\Gamma(a+n+1)\Gamma(b+m+1) \Gamma(-a-b-1)
\over \Gamma(-a)\Gamma(-b)\Gamma(a+b+m+n+2)}.}
A derivation of this formula can be found, for example, in
section 7.2 of \ref\gsw{M. B. Green, J. H. Schwarz, and
E. Witten, ``Superstring Theory: 1,''
Cambridge University Press, 1987.}. Thus we
have proven the formula \integraluse .

\appendix{D}{ Constraints on winding number violation }

We have seen in \first\ that representations of the
$SL(2,R)$ current algebra
are parametrized in terms of an integer $w$. For long strings this integer
could be interpreted as the winding number of the long string. For short
string it is just a parameter of states with no obvious semi-classical
interpretation.

Let us clarify the meaning of $w$ for the short string.
The short string wave-function, when expanded at large $\rho$, has components
on all winding numbers. An explicit discussion of this in an expansion
around $\rho=0$ can be found in \first . By an abuse of notation, we
will still call $w$ as the winding number of short string, but it
should be
kept in mind that it is {\it not} the winding number in the semi-classical
sense. It is not even the
winding number of the largest component of the wave-function at infinity.
For example, when $k$ is large, the wave-function for a $w=0$ state
can be expanded large $\rho$ as
\eqn\wafun{
\Psi = e^{- 2 j \rho} \psi_0 + e^{ -2(k/2-j)\rho} \psi_1 + \cdots
}
where we separated the radial dependence and the indices on
$\psi_0, \psi_1, \cdots$ indicate the actual
winding numbers at $\rho=\infty$.
As $j\to k/2$ we see that the second term with winding number $1$
becomes more dominant even though we are still studying the wave
function with $w=0$. This second component of the wave-function
is responsible for giving the divergences in the two and
three point functions, which we discussed in section 2.
The winding number has a semi-classical meaning at the infinity.
However, since the circle
is contractible, we do not expect that it should be conserved.
In fact, it is not. We find, however, that there is an
interesting pattern in winding number violations.
It essentially says that the possible amounts of winding violation
are restricted by the number of operators in a way that we will make
precise below. This was first observed  in \zam .
Below
we will make a precise statement, and we will prove it using the
properties of the representations of the $SL(2,R)$ current algebra.

Let us work in the $m$ basis. The states are labeled by
$|d,\tilde j , w \rangle$
and $|c,\tilde j , w \rangle$,  as well as
some $m$ that we do not indicate since it
will not be important in what follows. Here the letter $d,c$ indicate
discrete or continuous representations. We will think of $d$ as $d^+$ and
we construct $d^-$ by considering $d^+$ with $w<0$. The winding
number $w$ can have any sign.
The sign of $w$ distinguishes
an incoming states and an outgoing state
 in the Lorentzian picture. The sign of
$m$ is correlated with the sign of $w$.\foot{ This is true
in our case, but it might not be true in some quotients of $AdS_3$ \son .}
These representation are such that there is ``lowest weight'' state that
obeys the conditions
\eqn\condit{\eqalign{
 J^+_{w + n} |d,\tilde j , w \rangle &=
J^-_{ -w + n-1} |d,\tilde j , w \rangle =0
\cr
 J^+_{w+ n} |c,\tilde j , w \rangle &=
J^-_{ -w + n} |c,\tilde j , w \rangle =0
\cr
{ n \geq 1 }
}}
All states in the representation can be generated by acting with the
generators that do not annihilate the states. Furthermore, for operators
with $j$ in the physical ranges for continuous and discrete representations,
there are no null states in the representation.

Now we will consider the following state
\eqn\state{
 \prod_{i=1}^{n_d} \Phi^d_{w_i}(z_i) \prod_{j=1}^{n_c}
 \Phi^c_{w_j}(z_j) | 0\rangle
}
where $n_d,~n_c$ is the number of continuous and discrete representations.
The state $|0\rangle $ does not quite make sense, but after we act with
any of the operators we get a state that does make sense.
Now we want to consider the state \state\ and decompose it into
representations of $SL(2,R)$.
For this we pick a circle $|z| = A$ sufficiently large
so that all
all the points where the operators are inserted are
left inside the circle.
We consider
$SL(2,R)$ generators that are defined by integrating the $SL(2,R)$ currents on
this contour times appropriate powers of $z$. In other words
$J^\pm_n \sim \oint dz J^\pm (z) z^n$.
Now let us show that some combination of the form
$J^P = J^+_{a} + c_1 J^+_{a-1} + \cdots $
annihilates the state \state .
The precise combination is
\eqn\precise{
 J^P = \oint dz \prod_{i=1}^{n_t} (z-z_i)^{w_i +1} J^+(z)
}
where $n_t = n_c + n_d$. We see that $a= \sum w_i + n_t $
We see that this  combination annihilates the state \state\
after using \condit . We can now decompose \state\ into
$SL(2,R)$ representations with definite $w$. This implies that
\precise\ will annihilate each of the states with definite
winding number independently.
Now we will show that this
implies that the state will carry winding
number less than or equal to $a-1 = \sum w_i + n_t -1$.
Suppose that there was  a state with winding number $a$. Then \precise\
would annihilate it. But on the other hand we know that all operators
in \precise\ act as creation operators on the Fock space due to \condit .
Since there are no null states in the representation we conclude that
this cannot happen. To be more precise let us expand the hypothetical
 state with winding
number $w\geq a$ in such a way that we fix $J_0^3$ and we look at the state
with fixed $J_0^3$ with minimum value of $L_0$ (though $L_0$ is not bounded
below, it is bounded below if we consider fixed $J_0^3$), let us denote
this state by $|h \rangle $ it is clear that $J^+_a |h \rangle =0$ since
there is no other state with which it could mix. This is inconsistent with
the idea that there are no null vectors. Therefore the state must have
winding number less or equal to $a-1$.

Now we can similarly form the combination
$J^N = J^-_b + c_1 J^-_{b-1} + \cdots $ annihilates the state.
The precise combination is
\eqn\negative{ J^N = \oint dz \prod_{i=1}^{n_d} (z-z_i)^{-w_i }
\prod_{j=1}^{n_c} (z-z_j)^{-w_j +1 } J^-(z)
}
so that $b = - \sum w_i + n_c $. We
see using \condit\ that \negative\
annihilates \state .
Now we show that the
total winding number of the state should be bigger than $-b$.
Suppose to the contrary that the winding number of the state is smaller
or equal than $-b$. Then \condit\ implies as above that $J_b^-$ will
annihilate at least one state. Actually the precise statement will
depend if the state we consider is discrete or continuous. If the state
is discrete, then the statement a bit weaker, $w$ should
be bigger than $-b - 1$.

If we expand \state\ in irreducible representations of the
$SL(2,R)$ current algebra it becomes a sum of
discrete and continuous states whose winding numbers are
restricted as
\eqn\windvio{\eqalign{
  - n_c +1 \leq &
  w - \sum w_i \leq n_t -1 ~,~~~~~~~{\rm continuous}
\cr
  - n_c  \leq &
  w - \sum w_i \leq n_t -1 ~,~~~~~~~{\rm discrete} .
}}

In terms of correlation functions of operators we need to take the
inner product of \state with $\langle 0| \Phi(z)$ where $\Phi$ could
be a discrete or continuous representation. Notice that,
in our conventions, when we take the adjoint of
a discrete representation we take $w \to -1-w$ while
for a continuous representation we take $w \to -w$.
We conclude
that correlators will obey the winding number violation rule
\eqn\windviocor{\eqalign{
 &- N_t +2 \leq
  \sum w_i \leq N_c -2 ~,~~~~~~{\rm at~least~one~continuous}
\cr
&- N_d +1 \leq
  \sum w_i \leq -1  ~,~~~~~~{\rm all~discrete }
}}
where now $N_t=N_c + N_d $ and $N_c, N_d$ is the total number of
 operators in the continuous and the discrete representations appearing
in the  correlation
function. Note that throughout this discussion we were thinking
of the correlators in the $m$-basis and the discrete states
where taken with $\tilde m=\pm \tilde j$.

Now let us consider the operators in the $x$ basis, then  the
labels $w_i$ of all operators can be taken to be non-negative.
In that case it is easy to show that in an $N$ point function
the  winding numbers  should obey
\eqn\easy{
 w_i - \sum_{j\not = i } w_j  \leq N-2
}
Note that an operator ${\cal O}^w(x,z)$ obeys simple
OPE expansion rules for the currents $J^a(x,z) = e^{x J_0^+} J^a(z)
e^{- x J_0^+}$ (see \ks ). Since
$J^+(x,z) = J^+(z) $  the analysis
done with the operator \precise\ goes through as before and leads
to \easy\ if we put  the $i$th operator at $z=x=\infty$.
This shows that for a three point function the winding violation
is only by one unit, so that the correlation function of
two discrete $w=0$ states with any  state with $w>1$
vanishes in $x$-space.

\appendix{E}{Another definition of the spectral flowed
operators}

In section 5, we defined the operator corresponding to
the spectral flowed representation by:

(1) starting with the
operator $\Phi_{j,\bar j}(x,\bar x)$ in the regular
representation in the $x$ basis,

(2) going to the $m$ basis by the integral transform \basis ,

(3) multiplying $e^{w{k\over 2} \sqrt{2\over k}
\varphi}$ with $J^3 = i \sqrt{k\over 2}\partial \varphi$
as in \spectflowact , and

(4) going back to the $x$ basis to obtain expressions such
as in \floweddis\ and \threepxus .

\noindent
Here we will describe a way to define the spectral flowed
operator $\Phi_{J,\bar J}^{w,j}(x)$ without going
through the $m$ basis. This approach has an advantage
that we do not have to deal with the infinite factor $V_{conf}$
as we did in section 5.
We will compute the two and
three point functions including $\Phi_{J,\bar J}^{w,j}(x)$,
and show that they agree with the results in section 5
when $w=1$.

\subsec{Definition in the $x$ basis}
The definition, in the case of $w=1$,
is given by the fusion of $\Phi_j$ with the spectral flow
operator $\Phi_{k/2}$ as
\eqn\defflow{
\Phi^{w=1 , j}_{J, \bar J} (x, z)
 \equiv  \lim_{\epsilon\to 0} {\epsilon}^{m} {\bar \epsilon}^{\bar m}
\int d^2 y y^{j - m -1}\bar y^{j - \bar m - 1}
 \Phi_{j}(x + y, z + \epsilon ) \Phi_{k/2}
(x,z) ,
}
where $J = m + {k\over 2}$ and $\bar J = \bar m + {k \over 2}$.
This equality is understood
to hold inside of any correlation functions.

First we need to show that the limit $\epsilon \rightarrow 0$
in \defflow\  exists, $i.e.$, the result of the $y$ integral
scales as $\epsilon^{-m}\bar{\epsilon}^{-\bar m}$ for
small $\epsilon$. We will prove this
for a correlation
function where there are
at least two more operators besides $\Phi_{J,\bar J}^{w,j}(x)$.
There is a subtlety with the argument when there is
only one additional operator in the correlation
function, $i.e.$, when we consider
a two point function including $\Phi^{w=1, j}_{J, \bar J}$.
This does not cause a problem since $\Phi^{w=1, j}_{J, \bar J}$
has a non-zero two point function only with another
operator with $w=1$, which actually is a composite of two
operators as in \defflow . In fact, we will be able
to compute the two point function using \defflow .

For simplicity, we set $x=0$ and $z=0$ and
consider a correlation function
\eqn\defoff{
{\cal F} = \langle \Phi_{j_1}(x_1,z_1)
\cdots \Phi_{j_N}(x_N,z_N) \Phi_j(y,\epsilon)
\Phi_{k/2}(0,0) \rangle.}
For $|y| \ll |x_i|$ and $|\epsilon| \ll |z_i|$ ($i=1,\cdots, N$),
with finite $\epsilon/y$, we can show that this behaves as
\eqn\smallyepsilon{
{\cal F} \sim \epsilon^j (\epsilon -
y)^{- 2j-{\cal D}} y^{{\cal D}}f(x_3,\cdots,x_N;
z_3,\cdots,z_N),}
where ${\cal D}$ is a differential operator acting
on $x_3,\cdots, x_N$. We have set $(x_1,z_1)=(1,1)$
and $(x_2,z_2)=(\infty,\infty)$ by using the $SL(2,C)$
symmetries on both worldsheet and the target space.
For $N=2$, we can check this explicitly by using
the formula \solutiontwo\ for the four point function
with the spectral flow operator. (In this case,
${\cal D}$ is a number depending on $j, j_1, j_2$.)
This can be generalized
for any correlation function with $N \geq 2$ as follows.
The spectral flow operator $\Phi_{k/2}$ obeys the null state condition
\eqn\nullagain{ J^-(z) \Phi_{k/2}(x,z) = 0.}
Using this, the KZ equation is simplified as
\eqn\kzagain{ {\partial\over \partial z}
 \Phi_{k/2}(x,z) = - J^3(z)\Phi_{k/2}(x,z).}
Let us evaluate the KZ equation in the correlation function
\defoff .
When $|\epsilon| \ll |z_1|, \cdots , |z_N|$, we can ignore
the operator product singularities of $J^3(z)$ at $z=0$
with the operators at $z_1,\cdots,z_N$, and we only have to
consider the contribution from $\Phi_j(y,\epsilon)$.
We then find that the KZ equation \kzagain\ leads to
\eqn\simplifiedkz{
{\partial \over \partial \epsilon} {\cal F}
= -{1\over \epsilon}\left( y{\partial \over \partial y}
+ j \right) {\cal F}.}
To evaluate the null state condition \nullagain\ in the limit of
our interest, we need to use the global $SL(2,C)$
invariance of ${\cal F}$ to turn derivatives
with respect to $x_i$, for example $\partial_{x_1}$
and $\partial_{x_2}$, into a derivative with
respect to $y$. This is where we need to assume that
there are at least two more operators in the correlation
function. Setting $(x_1,z_1)=(1,1)$ and $(x_2, z_2)=(\infty, \infty)$
after this procedure, and taking the limit $\epsilon, y
\rightarrow 0$ keeping $\epsilon/y$ finite, we find
that the null state condition \nullagain\ leads to the equation,
\eqn\simplifiednull{
\left\{ {y \over \epsilon}\left(-(\epsilon-y)
{\partial \over \partial y} + 2j\right)+ {\cal D} \right\}
{\cal F}=0,}
with some differential operator ${\cal D}$ acting
on $x_3,\cdots, x_n$. Here
$\epsilon^{-1} y^2  \partial_y$ acting on ${\cal F}$
comes from the operator product expansion of $J^{-}(0,0)$
with $\Phi_j(y,\epsilon)$, and the other terms
are obtained from $J^-(0,0)$ with
$\Phi_{j_1}(x_1,z_1) \cdots \Phi_{j_N}(x_N,z_N)$
and by converting $\partial_{x_i}$'s into
$\partial_y$ by using the $SL(2,C)$ invariance
in the target space.
We can then show
that a general solution to \simplifiedkz\ and
\simplifiednull\ is given by \smallyepsilon\ (besides the
contact term solution discussed in the footnote later).

Now we can estimate the $y$ integral
in \defflow . From the discussion in the above
paragraph, we see that
the product of the operators $\Phi_j(y, \epsilon)
\Phi_{k/2}(0,0)$ can be expanded, in the leading
order in $\epsilon \rightarrow 0$, as
\eqn\smallepsilonexpansion{
\langle \Phi_j(y, \epsilon)
\Phi_{k/2}(0,0) \cdots\rangle \sim
|\epsilon^j (\epsilon-y)^{- 2j-{\cal D}}y^{\cal D}|^2~
\sum_{n,\bar n=0}^\infty
f_{n,\bar n}(x_3,\cdots,x_N)
y^{n}\bar{y}^{\bar n},}
for some operators ${\cal O}_{n, \bar n}$.
The $y$ integral for each term in the expansion
\smallepsilonexpansion\ can then be estimated
as
\eqn\yintegral{
 |\epsilon|^{2j} \int d^2 y y^{j - m -1+{\cal D}+n}\bar
y^{j - \bar m-1+{\cal D} +\bar n} |\epsilon-y|^{-2(2j + {\cal D})}
\sim \epsilon^{-m + n} \bar\epsilon^{-\bar m + \bar n},
}
where we assumed $m-\bar m \in {\bf Z}$.
Thus the limit $\epsilon \rightarrow 0$
in \defflow\ is well-defined.
Only the $n=\bar n=0$
survives in the limit. Note that, although the
differential operator ${\cal D}$ has dropped out
from the exponent of $\epsilon$, there is a
product of Gamma functions whose arguments
include ${\cal D}$. When this operator acts on the finite term left
over it modifies its $z_i$ and $x_i$ dependence for $i=3,\cdots, N$,
but does give rise to additional $\epsilon$ dependence.
%

Next we need to show that the operator
defined by \defflow\ is indeed
in the flowed representation. We do
this by checking that it has the correct
OPE with the $SL(2,R)$ currents. To show this,
we start with the standard
operator product expansion for operators with $w=0$,
\eqn\standardope{
\eqalign{ &
J(x',z')\Phi_j(x+y, z+\epsilon) \Phi_{k/2}(x,z)\cr
&=\left\{ {1 \over z'-z-\epsilon} \left[
 (x+y-x')^2 {\partial \over \partial y}
 + 2j (x+y-x') \right]\right. +\cr
&~~~~~~~~~~~\left.
 + {1 \over z'-z}
\left[ (x-x')^2 \left(
{\partial \over \partial x} - {\partial \over \partial
y} \right) + k
(x-x') \right]\right\}
\Phi_j(x+y, z+\epsilon) \Phi_{k/2}(x,z).}}
Apply this to \defflow\ and performing the integration
by parts in $y$, we obtain
\eqn\flowedope{
\eqalign{ & J(x',z')  \Phi_{J,\bar J}^{w=1,j}(x,z)\cr
& = \lim_{\epsilon, \bar \epsilon
\rightarrow 0}
\epsilon^m \bar \epsilon^{\bar m}
\int d^2 y y^{j-m-1}\bar y^{j-\bar m -1}\times
\cr &~~~~~~
\left\{ {1 \over z'-z-\epsilon} \left[
 -(j-m-1)y^{-1} (x+y-x')^2
 + (2j-2) (x+y-x') \right]\right. +\cr
&~~~~~~~~~~~\left.
 + {1 \over z'-z}
\left[ (x-x')^2
{\partial \over \partial x}+ (j-m-1)y^{-1}(x-x')^2 + k
(x-x') \right]\right\} \times \cr &~~~~~~~~~~~~~~~~~\times
\Phi_j(x+y, z+\epsilon) \Phi_{k/2}(x,z)\cr
&=
\lim_{\epsilon, \bar \epsilon
\rightarrow 0}
\epsilon^m \bar \epsilon^{\bar m}
\int d^2 y y^{j-m-1}\bar y^{j-\bar m -1}\times
\cr &~~~~~~
\left\{- {j-m-1\over (z'-z)^2} {\epsilon \over y}
 (x-x')^2 +
{1\over z'-z}\left[
(x-x')^2 {\partial \over \partial x}
 + 2\left(m + {k \over 2}\right)(x-x')\right]\right\} \cr
&~~~~~~~~~~~~~~~~~\times
\Phi_j(x+y, z+\epsilon) \Phi_{k/2}(x,z)\cr
&=-(j-m-1){(x-x')^2 \over (z'-z)^2} \Phi^{w=1,j}_{J+1,\bar J}(x,z)
+ \cr
&~~~~~~+{1\over z'-z}
\left[ (x-x')^2 {\partial \over \partial x}
 + 2\left( m + {k \over 2} \right)
(x-x') \right] \Phi^{w=1,j}_{J, \bar J}(x,z).}}
This means that the corresponding state,
\eqn\correspondingstate{
| w=1,j; J, \bar J \rangle =
\Phi_{J, \bar J}^{w=1, j}(x=0, z=0) | 0 \rangle,}
obeys
\eqn\correctward{
\eqalign{&
 J_0^3 | w=1,j; J, \bar J \rangle
= \left( m + {k \over 2} \right)
 | w=1,j; J, \bar J \rangle,\cr
& J_n^3  | w=1,j; J, \bar J \rangle = 0,
J_{n\pm 1}^\pm  | w=1,j; J, \bar J \rangle
=0, ~~~~(n=1,2,\cdots).}}
This is the correct highest weight condition for
a state with $w=1$.

\subsec{Three point function}

Now we can use the definition \defflow\ to compute
correlation functions with spectral flowed states.
First let us study the three point function.
We start with the four point function with a spectral flow operator,
\eqn\fourpt{
 \eqalign{
& \langle \Phi_{j_1}(x_1) \Phi_{k/2}(x_2) \Phi_{j_3}(x_3)
\Phi_{j_4}(x_4) \rangle =\cr
=& |z_{43}|^{2(\Delta_2 + \Delta_1 - \Delta_4 - \Delta_3)}
|z_{42}|^{-4 \Delta_2}
| z_{41}|^{2(\Delta+ \Delta_2 - \Delta_4 - \Delta_1)}
|z_{31}|^{2(\Delta_4 - \Delta_1 - \Delta_2 - \Delta_3)}
|z|^{2j_1} |1-z|^{ 2 j_3}  \times
\cr
&
|x_{43}|^{2(k/2+j_1-j_4-j_3)}
|x_{42}|^{-2k}
|x_{41}|^{2(j_3+k/2-j_4-j_1)}
|x_{31}|^{2(j_4-j_1-k/2-j_3)}\times\cr
& B(j_1) C(k/2-j_1,j_3,j_4)
|z-x|^{2(-j_1-j_3-j_4+k/2)}
|x|^{2(-j_1+j_3+j_4-k/2)}
|x-1|^{2(j_1-j_3+j_4-k/2)}.}}
Setting $x_1 = x_2+w$,
\eqn\whatx{\eqalign{
&x = {x_{21}x_{43} \over x_{31}x_{42}}
 = {wx_{43} \over (w-x_{32})x_{42}} \cr
&1-x =
{(w-x_{42})x_{32} \over (w-x_{32})x_{42}}\cr
&z-x = {(zx_{42}-x_{43})w-zx_{32}x_{42} \over (w-x_{32})x_{42}}.}}
Substituting this into \fourpt , we find
\eqn\morefourpt{
 \eqalign{
& \langle \Phi_{j_1}(x_1) \Phi_{k/2}(x_2) \Phi_{j_3}(x_3)
\Phi_{j_4}(x_4) \rangle =\cr
=& |z_{43}|^{2(\Delta_2 + \Delta_1 - \Delta_4 - \Delta_3)}
|z_{42}|^{-4 \Delta_2}
| z_{41}|^{2(\Delta_3+ \Delta_2 - \Delta_4 - \Delta_1)}
|z_{31}|^{2(\Delta_4 - \Delta_1 - \Delta_2 - \Delta_3)}
  \times
\cr
 &|z|^{2j_1} |1-z|^{ 2 j_3} B(j_1) C(k/2-j_1,j_3,j_4)
|x_{42}|^{2(j_1+j_3-j_4-k/2)}
|x_{32}|^{2(j_1-j_3+j_4-k/2)}\times\cr
&
|w|^{2(-j_1+j_3+j_4-k/2)}
|(zx_{42}-x_{43})w-z x_{32}x_{42}|^{-j_1-j_3-j_4+k/2}.}}
We then multiply the factor $|w|^{2(j_1-m_1-1)}$
and integrate over $w$. We find
\eqn\integral{
\eqalign{&
\int dw^2 |w|^{2(j_1-m_1-1)}
\langle \Phi_{j_1}(x_1) \Phi_{k/2}(x_2) \Phi_{j_3}(x_3)
\Phi_{j_4}(x_4) \rangle =  ({\rm standard~powers ~of~} z_i ) \times \cr
&  B(j_1) C(k/2-j_1,j_3,j_4) |x_{42}|^{2(j_1+j_3-j_4-k/2)}
|x_{32}|^{2(j_1-j_3+j_4-k/2)}\times \cr
& \int d^2 w |w|^{2(j_3+j_4-m_1-k/2-1)}
 |(zx_{42}-x_{43})w-x_{32}x_{42}|^{-j_1-j_3-j_4+k/2} \cr
=& |z_{43}|^{2(\Delta_2 + \Delta_1 - \Delta_4 - \Delta_3)}
|z_{42}|^{-4 \Delta_2}
| z_{41}|^{2(\Delta_3+ \Delta_2 - \Delta_4 - \Delta_1)}
|z_{31}|^{2(\Delta_4 - \Delta_1 - \Delta_2 - \Delta_3)}
  \times
\cr
 &~~~ |1-z|^{ 2 j_3} |z|^{-2 m_1}
 B(j_1) C(k/2-j_1,j_3,j_4)\times\cr
&~~~ |x_{42}|^{2(j_3-j_4-m_1-k/2)}
|x_{32}|^{2(-j_3+j_4-m_1-k/2)}
|zx_{42}-x_{43}|^{2(-j_3-j_4+m_1+k/2)}\times \cr
& \pi { \Gamma(j_3+j_4-m_1-k/2) \Gamma(-j_1-j_3-j_4+k/2+1)
   \Gamma(j_1+m_1) \over
 \Gamma(1-j_3-j_3+m_1+k/2)\Gamma(j_1+j_3+j_4-k/2)
\Gamma(1-j_1-m_1) }.}}
Now we multiply by $|z_{21}|^{2 m_1}$ and
send $z_{21} \to 0$. We find
\eqn\final{
\eqalign{&
\lim_{z_{21} \to 0}  |z_{21}|^{2 m_1} \int dw^2 |w|^{2(j_1-m_1-1)}
\langle \Phi_{j_1}(x_1) \Phi_{k/2}(x_2) \Phi_{j_3}(x_3)
\Phi_{j_4}(x_4) \rangle = \cr
= & B(j_1) C(k/2-j_1,j_3,j_4)
\pi { \Gamma(j_3+j_4-J) \over \Gamma(1-j_3-j_4+J)}
   {\Gamma(j_1+J-k/2) \over\Gamma(1-j_1-J+k/2) }
 {1\over \gamma(j_3+j_4+j_1-k/2)}
 \times \cr
& |x_{42}|^{2(j_3-j_4-J)}
|x_{32}|^{2(j_4-j_3-J)}|x_{43}|^{2(J-j_3-j_4)}
|z_{43}^{\hat \Delta_1 - \Delta_3 - \Delta_4} z_{42}^{ \Delta_3 -
\hat \Delta_1 - \Delta_4} z_{32}^{\Delta_4 - \hat \Delta_1 -\Delta_3}
|^{2}
}}
where
\eqn\defdeltahat{
\hat \Delta_1 = \Delta(j_1) - m_1 - {k\over 4},~~J = m_1 + {k \over 2}.
}
Due to the limit in \final , we can neglect higher powers of $z$
appearing at various places.

The result \final\ is in agreement with
\threepx , which we computed by going through the $m$ basis\foot{
Note that $m$ in \defdeltahat\ is $-m$ in \parame .}.
We should point out that the factor $1/V_{conf}$ in
\threepx\ is absent in \final. Thus
the definition \defflow\ includes the rescaling
$\Phi \rightarrow \widehat{\Phi} = V_{conf} \Phi$
that we performed for the long string.

\subsec{Worldsheet two point function}

To compute
 the two point function with spectral flowed operators, we start
with the four point function with two insertions of  spectral
flow operators, say $j_2 = j_4 = k/2$. The KZ equation and the
null state conditions imply that
$j_1=j_3$. To see this we notice that the four point function
should be symmetric under $2 \leftrightarrow 4$, leaving 1 and 3
unchanged. This changes $z \to 1-z$, $x \to 1-x$. Taking into
account also the prefactors, we find
\eqn\ratio{
{ {\rm 4pt}(1,2,3,4) \over  {\rm 4pt}(1,4,3,2)}
=\left|   \left( {z \over 1-z} \right)^{\Delta_1-\Delta_3+j_1-j_3}
\left( {1-x \over x} \right)^{j_1-j_3}  \right|^2
}
Demanding that this is $1$, we find $j_1=j_3$.\foot{
A solution with $j_1 = 1-j_3$ appears to come from a contact
term for the four point function. In fact, the function
$ z^{-j_1}  \delta^2(x-z)$ is a solution to \simplifiedkz\
and \simplifiednull , with
$j_3=1-j_1$, and ${\cal D}  = 1 -2 j_1$, which is the value that
appears in the four point function equation when $j_2 = j_4 = k/2$.
(Note that $\delta^2(x-z)$ is not a standard contact term,
for $x=z$ is not a coincidence limit of two operators.
If we use the relation between the four point function
${\cal F}_{SL(2)}(z,x)$ in the $SL(2,C)/SU(2)$ coset model
and a five point function in the Liouville model, recently
pointed out in \teschnernew , one can interpret $\delta^2(x-z)$
as a contact term coming from the coincidence limit involving
the extra operator one inserts in the Liouville model.
It would be interesting to find a direct interpretation
of such a contact term in the $SL(2,C)/SU(2)$ model.)
Inserting  $ z^{-j_1}  \delta^2(x-z)$ into the $x$ integral we
describe below and doing the same change of variables, we see
that we recover the term proportional to $\delta(j_1 +j_2 -1)$
in the two point function. }

Now let us apply \defflow\ to extract the two point function
of the spectral flowed state. As we explained in the above,
we expect $j_1 = j_3$ from the null vector equations.
In fact, the factor $C(k/2-j_1, j_3, j_4)$ in \fourpt\
with $j_4=k/2$ vanishes for $j_1\neq j_3$ and is infinite
at $j_1=j_3$. We can regularize the infinity by slightly
modifying the spectral flow operator as
$k/2 \rightarrow k/2 + i \epsilon$. Indeed, in the limit
$\epsilon \rightarrow 0$, we recover the
delta function enforcing $j_1=j_3$,
\eqn\regularized{
 C\left({k\over 2} - j_1, j_3, {k\over 2}\right)
 = (k{\rm ~ dependent~coefficient}) \times \delta(j_1-j_3). }
Thus the four point function in this case reduces to
\eqn\fptfl{\eqalign{&
\langle \Phi_{j}(x_1,z_1)
\Phi_{k/2}(x_2,z_2) \Phi_{j'}(x_3,z_3)
\Phi_{k/2}(x_4,z_4) \rangle \cr
&=
\left|   z_{42}^{k\over 2} z_{31}^{-2 \Delta}  z^j (1-z)^j
x_{42}^{-k} x_{31}^{-2 j} (z-x)^{-2 j} \right|^2
B(j) \delta(j-j'),}
}
where we ignored a $k$ dependent overall coefficient.
Now we set $x_1 = w_1 + x_2$, $x_3= w_3 + x_4$, multiply
by $|w_1|^{2(j-m_1 -1)} |w_3|^{2(j-m_3-1)}$, and
integrate over $w_1$ and $w_3$.
It is convenient to introduce new variables $u_1$ and
$u_3$ defined by $w_i = x_{42} u_i$, $i=1,3$.
The integrated correlation function becomes
\eqn\inte{
 |z|^{2j}
\int d^2 u_1 d^2 u_2 \left| u_1^{j-m_1-1} u_2^{j-m_3 -1}
\left( u_1 u_3 - z ( u_{31} + 1) \right)^{-2j} \right|^2,
}
where we set $z=0$ in the term with $(1-z)$ since we are going
to be interested in the small $z$ behavior of \fptfl . Here
we omitted the standard factors of $x_{42}$ and $z_{42}$.
It is convenient to change of the integration variables as
$u_1 = \sqrt{s} y$ , $u_3 = \sqrt{s}y^{-1}$.
After rescaling $s = z t $, we  find that \inte\ goes as
\eqn\intenew{
 | z^{ - {1\over 2}(m_1 + m_3)} |^2
\int d^2 y  d^2 s~|y^{m_3 - m_1-1}|^2     \left|
s^{ j-{1\over 2}(m_1+ m_3) -1}
\left( s + \sqrt{sz} (y-y^{-1})  -1 \right)^{-2j} \right|^2.
}
Since we are interested in the leading term in the $z$ expansion,
we set $z=0$ in the last factor.
The integral over $y$ then gives $\delta^2(m_1-m_3)$, and
the integral over $s$ gives a  combination of
Gamma functions,
\eqn\gammacombi{
2\pi {\Gamma(j-\bar m_1) \Gamma(j+m_1)\over
\gamma(2j) \Gamma(1-j+\bar m_1)
\Gamma(1-j-m_1)}.}
Combining this with the factor $B(j)\delta(j_1-j_3)$ in
\fptfl , we have reproduced the expression
   for the two point function in \flowedx  .

Finally, let us note that, instead of the definition \defflow ,
we could also define the spectral flowed operator via
\eqn\defflowz{\Phi^{w=1 , j}_{J, \bar J} (x, z)
 \equiv  \lim_{y\to 0} {y}^{j-m} {\bar y}^{j- \bar m}  \int d^2 \epsilon
\epsilon^{m -1}\bar{\epsilon}^{\bar m-1}
 \Phi_{j}(x+y, z + \epsilon ) \Phi_{k/2} (x,z),
}
for $J = m + {k\over 2}$. Instead of integrating over $y$,
here we are taking an integral over $\epsilon$.
In this definition of the flowed operator, the expression is
manifestly local in $x$. On the other hand, the definition
\defflow\ is manifestly local in $z$.
In order to show that the two definitions are equivalent, we
note that the relevant part of the correlation function behaves
as \yintegral .
Then, with the previous definition in \defflow ,
we  find that the spectral flowed correlator goes as
\eqn\xlimit{
\int d^2 w  |  w^{j - m-1+{\cal D}} (1-w)^{ - 2 j -{\cal D}}|^2
}
after we rescale $w \to z w $ and taking the $z \to 0 $ limit.
Similarly from \defflowz , we obtain
\eqn\zlimit{
\int d^2 t | t^{j + m -1} ( 1-t)^{-2 j -{\cal D} } |^2
}
after rescaling $z = xt$ and and taking the $x \to 0 $ limit.
We see that after the change of variables $ t = 1/w$, the two integrals
becomes the same. This shows that the two definitions
\defflow\ and \defflowz\ give the same results in general.

\subsec{Target space two point function}

Let us turn to the target space two point function
for the state with $w=1$. We apply the method used
in Appendix A for $w=0$ and use the Ward identity
to determine the normalization of the two point function.
We start with the following identity for the three point-function,
\eqn\threeptwardid{
\eqalign{
& \langle \Phi^{w=1,j_1}_{J_1,\bar J_1}(x_1,z_1)
\Phi^{w=1,j_2}_{J_2,\bar J_2}(x_2,z_2)
J(x_3,z_3) \Phi_1(x_3,z_3) \rangle \cr
& =-(j_1-m_1-1)
{(x_3-x_1)^2 \over (z_3-z_1)^2}
 \langle \Phi^{w=1,j_1}_{J_1+1,\bar J_1}(x_1,z_1)
\Phi^{w=1,j_2}_{J_2,\bar J_2}(x_2,z_2)
\Phi_1(x_3,z_3) \rangle - \cr
&~~~- (j_2-m_2-1)
{(x_3-x_2)^2 \over (z_3-z_2)^2}
 \langle \Phi^{w=1,j_1}_{J_1,\bar J_1}(x_1,z_1)
\Phi^{w=1,j_2}_{J_2+1,\bar J_2}(x_2,z_2)
\Phi_1(x_3,z_3) \rangle + \cr
&~~~ +\left\{
{1 \over z_3-z_1}\left[ (x_1-x_3)^2 {\partial
\over \partial x_1} + 2\left(m_1 + {k\over 2}\right)
\right] + \right.\cr
&~~~~~~~~~\left .+
{1 \over z_3-z_2}\left[ (x_2-x_3)^2 {\partial
\over \partial x_2} + 2\left(m_2 + {k\over 2}\right)
\right] \right\} \times \cr
&~~~~~~~~~~~~~~~~~~~
 \langle \Phi^{w=1,j_1}_{J_1,\bar J_1}(x_1,z_1)
\Phi^{w=1,j_2}_{J_2,\bar J_2}(x_2,z_2)
\Phi_1(x_3,z_3) \rangle .}}
We then have to compute the
three point functions in the right-hand side of this equation.
We start with the expression for the 5 point function with
two spectral flow operators, obtained in \zam ,
\eqn\zamformula{
\eqalign{
&\langle \Phi_{k/2}(x_1,z_1) \Phi_{k/2}(x_2,z_2)
 \Phi_{j_1}(y_1,\zeta_1) \Phi_{j_2}(y_2,\zeta_2)
 \Phi_{1}(y_3,\zeta_3) \rangle  \cr
& = B(j_1) B(j_2) C\left( {k \over 2} -j_1,
{k \over 2} - j_2, 1\right) \times\cr
& ~~~~~
|(x_1-x_2)^{j_1+j_2+1-k}
 \mu_1^{j_1-j_2-1} \mu_2^{j_2-j_1-1} \mu_3^{1-j_1-j_2}|^2,}}
where
\eqn\whatmu{
 \mu_i = {(x_1-y_{i+1})(x_2-y_{i+2})
\over (z_1-\zeta_{i+1})(z_2-\zeta_{i+2})}
-{(x_1-y_{i+2})(x_2-y_{i+1})
\over (z_1-\zeta_{i+2})(z_2-\zeta_{i+1})}.}
We have neglected the  $z$ and $\zeta$ dependent
factors.
When $j_1=j_2$, the factor $B^2 C$ in \zamformula\ is equal to
$B(j_1)$ up to a $k$ dependent factor as
\eqn\fateofc{
C\left( {k\over 2} - j, {k\over 2} - j, 1\right)
= {G(2j-k) \over 2\pi^2 \nu^{k-2j} \gamma\left(
{k-1 \over k-2}\right) G(1+2j-k)}
= \left({k-2 \over 2\pi}\right)^2 \nu^{2-k} {1 \over B(j)}
.}

We apply \defflow\ to \zamformula\ and integrate
over $\zeta_1$ and $\zeta_2$. It is convenient to
set $x_1=z_1=0$, $x_2=z_2=1$, $y_3=\zeta_3=\infty$,
$y_1=u \zeta_1$, $y_2-1 = v(\zeta_2-1)$ and take
$\zeta_1$, $1-\zeta_2$ to be small. In this limit we find
\eqn\mulimit{
\mu_1 = 1-v~,~~~~\mu_2 = 1-u~,~~~~\mu_3 = uv -1
}
The integral we need to evaluate, in order to compute
$\langle \Phi_{J_1,\bar{J}_1}^{w=1,j_1}
\Phi_{J_2,\bar J_2}^{w=1,j_2} \Phi_1 \rangle$ is then
\eqn\integraluv{\eqalign{
\int du^2 dv^2&
u^{j_1-m_1-1}\bar{u}^{j_1-\bar m_1 - 1}
v^{j_2-m_2-1} \bar{v}^{j_2-\bar m_2-1} \times
\cr
& ~~~|(u-1)^{j_1-j_2-1}
(v-1)^{j_2-j_1-1} (uv-1)^{1-j_1-j_2}|^2 .}}

Let us consider the case of the long string. We then have
$j_a = 1/2 + is_2$, and
the integral gives a delta-function singularity at $s_1=s_2$ coming
from the region of the integral of $u\sim 1$ or $v\sim 1$.
The term proportional to the delta-function can be evaluated as
\eqn\moreintegral{
\eqalign{& \delta(j_1-j_2) \left(
 \int du^2
u^{j_1-m_1-1} \bar{u}^{j_1-\bar m_1-1}
|(u-1)^{-2j_1}|^2 + (m_1, \bar m_1 \rightarrow m_2,
\bar m_2) \right) \cr
&= \delta(j_1-j_2){\pi \over \gamma(2j_1)}\left(
 {\Gamma(j_1-m_1)\Gamma(j_1+\bar m_1)
\over\Gamma(1-j_1-m_1) \Gamma(1-j_1+\bar m_1) }
+ (m_1,\bar m_1 \rightarrow m_2, \bar m_2) \right).}}
Since
\eqn\whatlargej{
J = {k \over 4} + {{1\over 4} + s^2 \over k-2} + h -1,}
the delta-function $\delta(s_1-s_2)$ together with
the condition $h_1=h_2$ in the internal CFT implies
$J_1=J_2$ and therefore $m_1=m_2$.
The correlator multiplying the
double pole term in \threeptwardid\ then gives
\eqn\remainingintegral{
\eqalign{
\int du^2 dv^2&
u^{j_1-m-2}\bar{u}^{j_1-\bar m - 1}
v^{j_2-m-1} \bar{v}^{j_2-\bar m-1} \times \cr
&~~~~|(u-1)^{j_1-j_2-1}
(v-1)^{j_2-j_1-1} (uv-1)^{1-j_1-j_2}|^2 \cr
&\sim  \delta(j_1-j_2){1 \over \gamma(j)}
\left( {-j_1-m\over j_1-m-1} + 1 \right)
 {\Gamma(j_1-m)\Gamma(j_1+\bar m)
\over\Gamma(1-j_1-m) \Gamma(1-j_1+\bar m)}\cr
& \sim  -\delta(j_1-j_2){1 \over \gamma(j_1)}
{2m+1 \over j_1-m-1}
{\Gamma(j_1-m)\Gamma(j_1+\bar m)
\over\Gamma(1-j_1-m) \Gamma(1-j_1+\bar m)}
. }}
On the other hand, the correlator multiplying the
single pole term in \threeptwardid\ gives
\eqn\singlepoleterm{\delta(j_1-j_2)
{2m_1+k \over \gamma(j_1)}
{\Gamma(j_1-m_1)\Gamma(j_1+\bar m_1)
\over\Gamma(1-j_1-m_1) \Gamma(1-j_1+\bar m_1)}.}
We combine them with \fateofc\ and
 multiply by the correlation function
 $\langle \phi_h(z_1) \phi_h'(z_2) \bar{L}(\bar{z}_3)
\rangle $  in
the internal CFT, as we did in Appendix A,
to compute the on-shell three point function involving
the target space R-current. We find
\eqn\moreonthreewardid{
 \eqalign{&\langle (\Phi^{w=1,j_1}_{J,\bar J}(x_1,z_1)
\phi_h(z_1))(\Phi^{w=1,j_2}_{J,\bar J}(x_2,z_2)
\phi_h(z_2))(J(x_3,z_3) \bar L(\bar z_3)
\Phi_1(x_3,z_3)) \rangle \cr
&\sim {1\over |z_{12}|^2 |z_{23}|^2
|z_{31}|^2} \left({q_1 \over \bar x_3 -\bar x_1}
+ {q_2 \over \bar x_3 - \bar x_2}\right) \times \cr
&~~~~~~~~\delta(j_1-j_2)
{\Gamma(j_1-m)\Gamma(j_1+\bar m)
\over \gamma(j_1) \Gamma(1-j_1-m)
\Gamma(1-j_1-\bar m)} {B(j_1)
\over |x_{12}|^{4J}},}}
where $q_1$ and $q_2$ are the R-charges of the two
operators.
{}From this, we find that the spacetime
two point function of $\Phi_{J,\bar J}^{w=1,j} \phi_h$
is ${\Gamma(j_1-m_1)\Gamma(j_1+\bar m_1)
\over \gamma(j_1) \Gamma(1-j_1-m_1)
\Gamma(1-j_1-\bar m_1)} B(j_1)$. We do not have
the extra factor of $(2j-1)$ for the long string.

\listrefs

\bye